\documentclass[journal]{IEEEtran}
\usepackage[T1]{fontenc}

\usepackage{cite}
\usepackage{amsmath}
\usepackage{algorithmic}
\usepackage{array}
\usepackage{bm}
\usepackage{bbm}
\usepackage{amssymb}
\usepackage{amsthm}
\usepackage{mathrsfs}
\usepackage{empheq}	
\usepackage[mathcal]{euscript}
\usepackage{xcolor}
\usepackage{floatrow}

\newtheorem{theorem}{Theorem}
\newtheorem{lemma}{Lemma}
\newtheorem{corollary}{Corollary}
\newtheorem{definition}{Definition}
\newtheorem{proposition}{Proposition}
\newtheorem{property}{Property}

\newtheorem{assumption}{Assumption}

\renewcommand{\P}{\mathbb{P}}
\newcommand{\E}{\mathbb{E}}
\newcommand{\thetashare}{\theta_{\sf TX}}
\newcommand{\thetasharec}{\bar{\theta}_{\sf TX}}
\newcommand{\thetatruec}{\bar{\theta}_{0}}
\newcommand{\thetatrue}{\theta_0}
\newcommand{\T}{^\top}
\newcommand{\kappagood}{k^{\star}}

\newcommand*{\QED}{\hfill\ensuremath{\square}}%

\ifCLASSINFOpdf
   \usepackage[pdftex]{graphicx}

\else
\fi
\ifCLASSOPTIONcompsoc
  \usepackage[caption=false,font=normalsize,labelfont=sf,textfont=sf]{subfig}
\else
  \usepackage[caption=false,font=footnotesize]{subfig}
\fi
\allowdisplaybreaks
\begin{document}

\title{Partial Information Sharing over Social Learning Networks}

\author{Virginia~Bordignon,
        Vincenzo~Matta
        and~Ali~H.~Sayed%
\thanks{Virginia Bordignon and Ali H. Sayed are with the School of Engineering, EPFL, CH1015 Lausanne, Switzerland. E-mails: \{virginia.bordignon, ali.sayed\}@epfl.ch.}% 
\thanks{Vincenzo Matta is with the Department of Information and Electrical Engineering and Applied Mathematics (DIEM), University of Salerno, 84084 Fisciano (SA), Italy, and also with the National Inter-University Consortium for Telecommunications (CNIT), Italy. E-mail: vmatta@unisa.it.}
\thanks{An early version with partial results from this paper was presented at ICASSP 2020 \cite{2020partialinfo}.}\thanks{ This work was supported in part by grant 205121-184999 from the Swiss National Science Foundation.}%
}

%\markboth{IEEE Transactions on Information Theory,~Vol.~00, No.~0, Month~Year}%
%{Shell \MakeLowercase{\textit{et al.}}: Social Learning with Partial Information}

\maketitle

\begin{abstract}
This work addresses the problem of sharing partial information within social learning strategies. In social learning, agents solve a distributed multiple hypothesis testing problem by performing two operations at each instant: first, agents incorporate information from private observations to form their \emph{beliefs} over a set of hypotheses; second, agents combine the \emph{entirety} of their beliefs locally among neighbors. Within a sufficiently informative environment and as long as the connectivity of the network allows information to diffuse across agents, these algorithms enable agents to learn the true hypothesis. Instead of sharing the entirety of their beliefs, this work considers the case in which agents will only share their beliefs regarding one \emph{hypothesis of interest}, with the purpose of evaluating its validity, and draws conditions under which this policy does not affect truth learning. We propose two approaches for sharing partial information, depending on whether agents behave in a self-aware manner or not. The results show how different learning regimes arise, depending on the approach employed and on the inherent characteristics of the inference problem. Furthermore, the analysis interestingly points to the possibility of deceiving the network, as long as the evaluated hypothesis of interest is close enough to the truth.
\end{abstract}


\IEEEpeerreviewmaketitle

\section{Introduction}
\IEEEPARstart{D}{istributed} decision-making is a relevant research domain that has attracted great interest over the last decades, encompassing several useful paradigms and strategies --- see~\cite{VarshneyBook,ViswanathanVarshney,BlumKassamPoor,ChamberlandVeeravalli,ChamberlandVeeravalliSPMag,ChenTongVarshneySPMag,Saligrama,ScutariBarbarossaPescosolidoTSP2008,BarbarossaAcademicPress,CattivelliSayedDetection,MattaSayedCoopGraphSP2018,MouraLDGauss,MouraLDnoisy,Zoubir2018} for a non-comprehensive list of works on the subject.  
One relatively recent paradigm of distributed decision-making is {\em social learning}. In a nutshell, social learning is the process through which a collection of networked agents can update their opinions on a set of possible hypotheses. The models used for carrying out these updates are several in the literature \cite{chamley2004rational, golub2010naive, acemoglu2011opinion, chamley2013models, jadbabaie2013information, krishnamurthy2013social,molavi2013reaching, krishnamurthy2014interactive,golub2017learning}, ranging from \emph{consensus}-based strategies inspired by the classical DeGroot model \cite{degroot1974reaching},  to solutions with online elements, such as the use of exogenous streaming observations that influence the evolution of opinions. Falling into the latter class of models, we find approaches that range from \emph{fully Bayesian} \cite{banerjee1992asimple,bikhchandani1998learning,smith2000path,acemoglu2011bayesian} to \emph{locally Bayesian} (usually referred to as non-Bayesian) approaches\cite{jadbabaie2012non,molavi2018theory}, all of which share the common attribute of processing information in a decentralized manner. 

We focus on non-Bayesian social learning strategies \cite{zhao2012learning,shahrampour2015distributed, nedic2017fast, lalitha2018social, bordignon2020adaptive}, which are computationally more tractable and have been shown to enable truth learning. In this context, each agent observes individual signals, describing some phenomenon or event of interest and arising from possibly different distributions, all of which depending on a \emph{true state of nature}. The true state of nature is assumed to be part of a set of possible hypotheses and, for each hypothesis, agents will consolidate over time a confidence level. The collection of confidence levels across all hypotheses then form the agent's \emph{belief vector}. To allow information to diffuse across the network, agents will share their beliefs with neighboring agents. For example, consider a network of meteorological stations at different locations monitoring weather conditions. Each of the stations (or agents) will observe measurements such as temperature, humidity, atmospheric pressure and wind speed, which are functions of the underlying weather condition (or state of nature) that the agents are trying to learn such as declaring that the conditions are sunny, rainy, cloudy, or snowing. 

One classical assumption adopted in social learning is that information propagates over a strongly-connected network. Under this assumption, several existing social learning implementations successfully drive the agents to identify the true state of nature with full confidence. Recently, alternative network settings have been considered, including weakly-connected networks \cite{ying2016information, salami2017social, matta2019interplay} and social learning in the presence of forceful or stubborn agents \cite{acemoglu2010spread, yildiz2013binary}. Under these settings, new relevant phenomena arise, such as mind control, propagation of misinformation, emergence of contrasting opinions.

In this work, we examine the scenario in which agents within a strongly-connected network do not share their full belief vectors but only the confidence they have in a particular hypothesis of interest (such as their opinion about whether the weather conditions are rainy or not). This amounts to requiring a scenario where agents are only sharing \emph{partial information}. The best learning outcome agents could hope for with the sharing of such minimal information is to infer whether the hypothesis of interest corresponds to the truth or not. We will discover in this work that this process will give rise to a rich set of convergence regimes. The main contributions of this work consist in the characterization of the learning and mislearning regimes, under the social learning process with partial information. We will propose two approaches for diffusing partial information: a first approach without self-awareness and a second approach with this feature, where self-awareness endows each agent with the ability to combine neighbors' {\em partial} information to its own {\em full} belief vector. We will reveal, supported by some non-trivial analysis, under which conditions one can attain learning or even {\em mislearning} for both approaches. The theoretical results will highlight some interesting phenomena in this context: one of them being that truth sharing preserves truth learning; but also that, when the hypothesis of interest is false, a sufficient distance between this hypothesis and the truth must exist in order for agents to make a clear distinction between both and be able to correctly discard the presumed hypothesis. 

\emph{Notation:} We will use boldface fonts to indicate random variables. The symbols $\stackrel{\textnormal{a.s.}}{\longrightarrow}$, $\stackrel{\textnormal{p}}{\longrightarrow}$ and $\stackrel{\textnormal{d}}{\longrightarrow}$ indicate respectively almost sure convergence, convergence in probability and convergence in distribution \cite{billingsley2008probability}, as the time index $i$ goes to infinity.
\section{Background}
\subsection{Inference Problem}
In social learning problems, we consider a network of $K$ agents, indexed by $k\in\{1,2,\dots,K\}$, trying to identify the state of nature $\theta$ from a set of $H$ possible hypotheses $\Theta=\{1,2,\dots,H\}$. To accomplish this task, each agent $k$, at each time instant $i=1,2,\dots$, relies on the observation of an exogenous signal $\bm{\xi}_{k,i}$, which belongs to agent $k$'s signal space $\mathcal{X}_{k}$. Each agent $k$ possesses a private family of likelihood functions $L_k(\xi|\theta)$ for all $\xi\in \mathcal{X}_k$ and $\theta\in\Theta$. 

The hypothesis $\thetatrue \in\Theta$, corresponding to the true state of nature, is the same for all agents. In this case, the observed signal at each instant $i$ for each agent $k$ is generated according to the marginal likelihood function $L_k(\xi|\thetatrue)$:
\begin{equation}
\bm{\xi}_{k,i}\sim L_k(\xi|\thetatrue), \quad \quad i=1,2,\dots\,.
\end{equation}
Signals are assumed independent and identically distributed (i.i.d.) over time, but they can be dependent across agents.\footnote{It is typically unrealistic to assume that an individual agent has sufficient knowledge to build a $K$-dimensional multivariate model embodying dependencies across all the agents. For this reason, in social learning it is assumed that each agent $k$ uses only a {\em marginal} likelihood $L_k(\xi|\theta)$ that pertains to the signals $\bm{\xi}_{k,i}$ observed locally. All probabilities and expectations will be evaluated under the true likelihood functions $L_k(\xi|\thetatrue)$ for every $k$, which will represent probability mass functions or probability density functions, if the random variables $\bm{\xi}_{k,i}$ are respectively discrete or continuous.} We will sometimes refer to the likelihood functions $L_k(\xi|\theta)$ as $L_k(\theta)$ instead, in order to simplify the notation.

To characterize the identifiability of the inference problem, we will make use of the KL divergence metric. We denote the Kullback-Leibler (KL) divergence for agent $k$ between its true likelihood and the likelihood corresponding to some arbitrary hypothesis $\theta$ by:
\begin{equation}
d_k(\theta)\triangleq \mathbb{E}\left(\log \frac{L_k(\bm{\xi}_{k,i}|\thetatrue)}{L_k(\bm{\xi}_{k,i}|\theta)}\right),
\label{eq:kldiv}
\end{equation} 
where, as said, the expectation in \eqref{eq:kldiv} is computed assuming that $\bm{\xi}_{k,i}$ is distributed according to $L_k(\xi|\thetatrue)$. The KL divergence metric quantifies agent $k$'s ability to distinguish hypothesis $\theta$ from the true hypothesis $\thetatrue$.

To avoid pathological cases, we consider a first assumption that ensures the regularity of the likelihood ratios, namely, we assume that all KL divergences \cite{cover2012elements} between any two likelihood functions $L_k(\xi|\theta)$ and $L_k(\xi|\theta')$ for $\theta, \theta'\in\Theta$ are finite. 

\begin{assumption}[{\bf Finiteness of KL divergences}]
	\label{as:integ}
	For each $k=1,2,\ldots,K$ and each pair of distinct hypotheses $\theta$ and $\theta'$, the Kullback-Leibler divergence between $L_k(\xi|\theta)$ and $L_k(\xi|\theta')$ is finite.
	\QED
\end{assumption}
%Note that Assumption~\ref{as:integ} implies that all likelihood functions have common support.
We further define, for each agent $k$, a set of {\em locally indistinguishable hypotheses}, namely,
\begin{equation}
\Theta_k \triangleq \left\{\theta: d_{k}(\theta)=0 \right\},\label{eq:kl_divergence}
\end{equation}
which is the set of hypotheses whose KL divergence with respect to the true hypothesis is zero. The complementary set made of {\em locally distinguishable} hypotheses is:
\begin{equation}
\bar{\Theta}_k \triangleq \Theta\setminus \Theta_k.\label{eq:distset}
\end{equation}
In many practical situations, the limited knowledge available {\em locally} at the {\em individual} agents precludes them from identifying the true state of nature. When this happens, we say that the problem is not locally identifiable, which formally means that all local indistinguishable sets would have cardinality larger than one, i.e.,
\begin{equation}
|\Theta_k|>1,\quad \forall k=1,2,\dots,K.
\end{equation}
When an agent is not able to learn locally, it is stimulated to cooperate with other agents. Even when local identifiability is possible, cooperation can greatly improve the learning performance since the entire network has generally a significantly larger amount of data.  

As we will see in the following treatment, under certain conditions, the agents can overcome their local limitations through social interaction and eventually learn the truth. In particular, we remark that all our theorems contemplate the possibility that the problem is not locally identifiable. In these social interactions, the agents form their individual opinions about the hypotheses of interest by integrating the signals observed locally with some proper information exchanged with their neighbors. The next section introduces the network model relevant to such information diffusion.
\subsection{Network Model}
\begin{figure}[tb]
	\centering
	\includegraphics[width=2.5in]{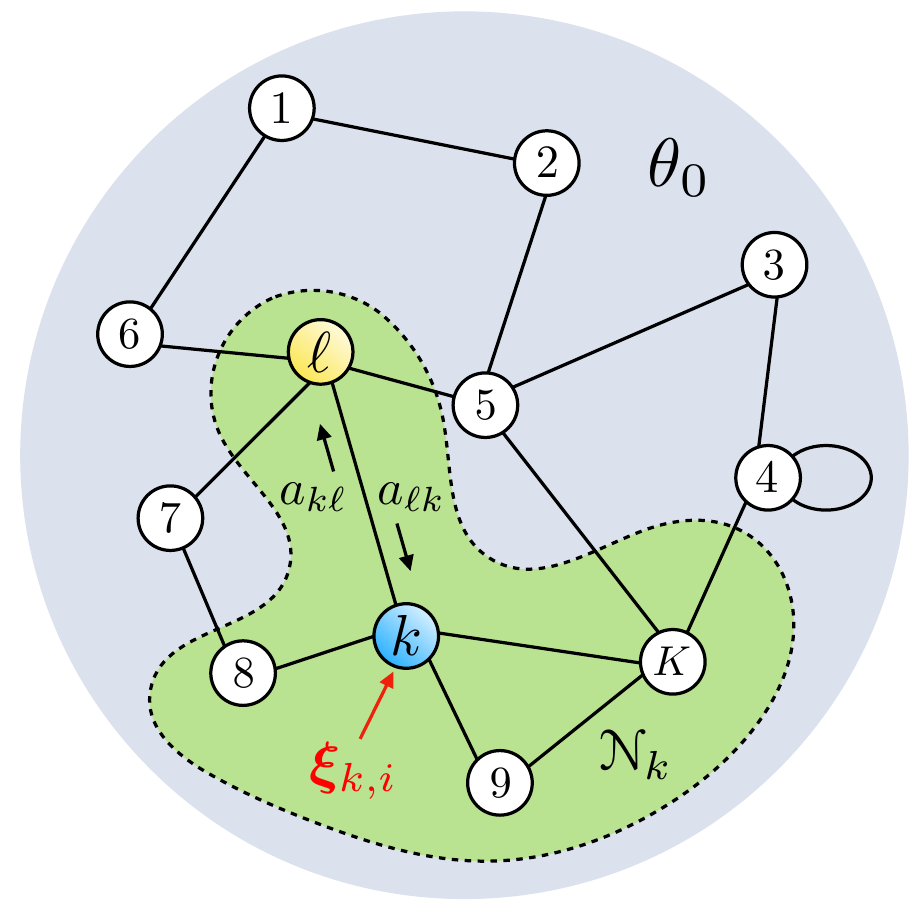}
	\caption{Network topology diagram. Agent $k$ weights the information received from neighboring agent $\ell$ using $a_{\ell k}$.}
	\label{fig:network}
\end{figure}

We assume the $K$ agents are linked by a (weighted) strongly-connected graph defined by the entities $(\mathcal{N},\mathcal{E}, A)$, where $\mathcal{N}$ represents the set of nodes (agents), $\mathcal{E}$ is the set of edges (communication links), and $A$ is a combination matrix that encodes the weights assigned to the graph edges --- see Fig.~\ref{fig:network}.

We assume that $A$ is a left-stochastic matrix, where each element, namely $a_{\ell k}$ for $\ell=1,2,\dots,K$ and $k=1,2,\dots,K$, contains the nonnegative weight agent $k$ assigns to the information\footnote{The term ``information'' is deliberately vague at this stage. Depending on the particular social learning implementation, different types of information will be shared among agents. The necessary details will be given when we introduce the social learning algorithms.} received from agent $\ell$. The weight $a_{\ell k}$ can be interpreted as the credibility assigned by agent $k$ to the information propagated by its neighbor $\ell$. If the weight $a_{\ell k}$ is exactly equal to zero, we say that agent $k$ is not receiving information from agent $\ell$, or yet, we say that $\ell$ does not belong to the set of neighbors of $k$, denoted by $\mathcal{N}_k$. In other words, the combination weights satisfy:
\begin{equation}
\mathbbm{1}\T A=\mathbbm{1}\T,\qquad a_{\ell k}\geq 0, \qquad a_{\ell k}=0\text{ if }\ell\notin \mathcal{N}_k,\label{eq:combweights}
\end{equation}
where $\mathbbm{1}$ represents a vector of all ones.

In a strongly-connected network, there exists at least one path with nonzero weights connecting any two nodes and, moreover, at least one $a_{kk}$ is nonzero~\cite{sayed2014adaptation}. Both conditions imply that $A$ is a primitive matrix, and from the Perron-Frobenius theorem \cite{horn2012matrix} we conclude that there exists a vector $v$ (referred to as the Perron vector) that satisfies the following conditions:
\begin{equation}
Av=v,\qquad \mathbbm{1}\T v=1,\qquad v\succ 0,\label{eq:perronv}
\end{equation}
where the symbol $\succ$ indicates an element-wise strict inequality. In other words, $v$ is a right-eigenvector for $A$ associated with the eigenvalue at one, and all its entries are positive and add up to one. Using the Perron entries, we further define the \textit{network average of divergences} for all $\theta\neq\thetatrue$ as
\begin{equation}
d_{\sf ave}(\theta)\triangleq \sum_{\ell =1}^Kv_\ell d_\ell (\theta),\label{eq:avediv}
\end{equation}
which, as shown in previous works~\cite{nedic2017fast,lalitha2018social}, plays an important role in traditional social learning as the asymptotic rate of convergence, and which will appear as well in the results that follow.

\subsection{Beliefs}
	In social learning, the agents form their opinions about the possible hypotheses by using their own private signals as well as by exchanging information with their neighbors. 
	In this process of opinion formation, each agent constructs its {\em private belief} about each hypothesis $\theta$, giving rise to a {\em belief vector}:
	\begin{equation}
	\bm{\mu}_{k,i}=\left[\bm{\mu}_{k,i}(1),\bm{\mu}_{k,i}(2),\ldots,\bm{\mu}_{k,i}(H)\right]\T,
	\end{equation}
which is a random vector (bold notation) because it depends on the random signals collected by the agent. The element $\bm{\mu}_{k,i}(\theta)$ is a scalar value between $0$ and $1$, which represents the confidence agent $k$ has at instant $i$ that hypothesis $\theta$ is the true one. The sum of these values along $\theta$ is equal to $1$, implying that the {\em belief vector} lies in the probability simplex in $\mathbb{R}^H$, denoted by $\Delta^H$.   

To ensure that no hypothesis is preliminarily discarded by any agent, an assumption on the positivity of initial beliefs is considered. We model the initial belief vectors $\mu_{k,0}$ as deterministic (normal font notation).
\begin{assumption}[\textbf{Positive Initial Beliefs}]\label{as:posit}
	We have $\mu_{k,0}(\theta)>0$ for each agent $k$ and all $\theta\in\Theta$, i.e., all agents start with a strictly positive belief for all hypotheses. 
	\QED
\end{assumption}

In order to construct and update continuously its belief, each agent $k$ at every instant $i$ uses its current signal $\bm{\xi}_{k,i}$ as well as some information exchanged with its neighbors.
The particular social learning strategy will be determined by several factors, such as the update rule, the type of information exchanged by neighboring agents, or the combination rule. In this article we will consider different strategies. We start by examining the non-Bayesian social learning setting.

\subsection{Non-Bayesian Social Learning}\label{sec:sl}
Social learning refers to a large family of methods that address the decision-making problem in a multi-agent environment. In these methods, agents interact by exchanging some information to discover the true state of nature. Approaches for social learning are typically categorized as Bayesian (a.k.a. fully Bayesian) or non-Bayesian. In the former case, the agents update their beliefs by computing {\em exact} posterior distributions accounting for dependencies involving the data of {\em all other agents} and keeping memory of {\em all the past history}. Such
approach is deemed unrealistic for different reasons. From a
system-design perspective, a Bayesian approach requires too
much distributed knowledge and complexity to be pursued
in many practical applications~\cite{hazla2021}. From a behavioral perspective,
it has been demonstrated in previous studies that social interactions
seldom take place in a fully-Bayesian way~\cite{demarzo2003,golub2010naive}. In Economics, these observations translate into the concept of bounded rationality, which considers that human cognition is limited, and that agents cannot be fully rational (i.e., fully Bayesian). They act instead in a {\em non-Bayesian} way, adapting their rational behavior in view of the complexity and cost of the decision process~\cite{simon1990bounded,conlisk1996why}. For example, in traditional non-Bayesian social learning, the agents perform {\em locally} a Bayesian update to incorporate their individual information into their own beliefs. Then, they implement some form of {\em combination policy} to aggregate information received from their neighbors. 
	
Among these works we can distinguish two groups of strategies: the ones that consider that agents share beliefs (or opinions)\cite{jadbabaie2012non,zhao2012learning, molavi2018theory} and those which consider the sharing of actions, i.e., functions of the beliefs~\cite{chamley2004rational,krishnamurthy2014interactive,mossel2017opi}. The latter, commonly found in the context of Economics, considers that agents update their beliefs with the goal of choosing an optimal instantaneous action and that agents know the likelihood functions relating actions and beliefs of other agents. In this work, we will focus on the former group of methods, in which agents can communicate in a synchronous (as opposed to sequential) and decentralized manner. We will often refer to these methods as non-Bayesian or traditional social learning.

More specifically, non-Bayesian social learning strategies can be described as follows. First, to emulate rational behavior, each agent $k$ performs a local Bayesian update to generate an \emph{intermediate belief vector} $\bm{\psi}_{k,i}$ using the new observation $\bm{\xi}_{k,i}$ and the previous \emph{private belief vector} $\bm{\mu}_{k,i-1}$. Second, we depart from a fully-Bayesian paradigm, by having the intermediate beliefs shared among neighbors. The received information is combined by each agent using some functional form, denoted by the function $g_k(\cdot)$.
	\begin{align}
	\bm{\psi}_{k,i}(\theta) &= \text{BU}(\bm{\mu}_{k,i-1};\bm{\xi}_{k,i}),\\
	\bm{\mu}_{k,i}(\theta) &=g_k(\bm{\psi}_{k}^i),
	\end{align}
	where $\bm{\psi}^i_k$ denotes the full history of intermediate beliefs received by agent $k$, i.e., $\bm{\psi}^i_k\triangleq \{ \bm{\psi}_{\ell,j}\}$ for $\ell\in\mathcal{N}_k$ and $j=1,2,\dots, i$. The Bayesian update, i.e., operator $\textnormal{BU}(\cdot\,;\,\cdot)$, uses knowledge of the private likelihood functions $L_k(\xi|\theta)$ to update the intermediate belief given the new observations $\bm{\xi}_{k,i}$ in the following manner for any $\theta\in\Theta$:
	\begin{equation}
	\bm{\psi}_{k,i}(\theta)=\frac{L_k(\bm{\xi}_{k,i}|\theta)\bm{\mu}_{k,i-1}(\theta)}{\sum_{\theta'\in\Theta}L_k(\bm{\xi}_{k,i}|\theta')\bm{\mu}_{k,i-1}(\theta')}.\label{eq:bayesup}
	\end{equation}
In principle, the only certainty in non-Bayesian social learning is that agents implement a local Bayesian update step. Then, the way they aggregate information is left unspecified, and several possibilities are found in the literature to choose the functions $g_k(\cdot)$. 

One structured approach to finding combination policies that correspond to rational agents' behavior, can be pursued by taking into account the fundamental constraints imposed by the distributed nature of the problem. This approach is pursued in~\cite{molavi2018theory}, where the shape of $g_k(\cdot)$ is established by imposing desired behavioral axioms to the output of the combination function: imperfect recall (IR), label neutrality (LN), independence of irrelevant alternatives (IIA), and monotonicity\footnote{Actually, in \cite{molavi2018theory} the authors focus on the Combine-Then-Adapt approach~\cite{sayed2014adaptation} when introducing the social learning framework, i.e., where the first step is the combination step. Similar conclusions can be obtained for the Adapt-Then-Combine form considered here, whose first step is the local Bayesian update. The framework in~\cite{molavi2018theory} admits more generally a time-varying combination rule. In our work, we focus on a stationary rule.}. Each of these axioms captures a distinct aspect found in the context of social interactions. For example, IR establishes that the output $g_k(\bm{\psi}_{k}^i)$ should only depend on present intermediate beliefs, as opposed to the full history of beliefs. The authors show that the only functional form satisfying IR, LN, IIA and monotonicity is the log-linear combination rule, otherwise written for any $\theta\in\Theta$ as\footnote{The result in~\cite{molavi2018theory} requires furthermore that $|\Theta|>3$ and $g_k(\cdot)$ to be continuous.}:
	\begin{equation}
	\bm{\mu}_{k,i}(\theta) =\frac{\displaystyle\exp\left\{\sum_{\ell \in\mathcal{N}_k}c_{\ell k}\log\bm{\psi}_{\ell,i}(\theta)\right\}}{\displaystyle\sum_{\theta'\in\Theta}\exp\left\{\sum_{\ell\in\mathcal{N}_k}c_{\ell k}\log\bm{\psi}_{\ell,i}(\theta')\right\}},\label{eq:combsl}
	\end{equation}
	for arbitrary positive constants $c_{\ell k}$. Under the additional unanimity axiom, i.e., agent $k$ agrees with its neighbors if all their beliefs are the same, $c_{\ell k}$ should sum up to 1 within the neighborhood of $k$. This condition is notably satisfied by the set of combination weights $a_{\ell k}$ introduced in \eqref{eq:combweights}. The combination step in~\eqref{eq:combsl} is not the only acceptable non-Bayesian strategy: additional or distinct assumptions may yield different algorithms that provably allow for truth learning~\cite{jadbabaie2012non,molavi2018theory,zhao2012learning,bordignon2020adaptive,salami2017social,KaiLaiChungBook}. For example, the authors in~\cite{molavi2018theory} show that replacing IIA for a separability property yields the linear combination rule found in~\cite{jadbabaie2012non,zhao2012learning,salami2017social} and based in deGroot's learning model~\cite{degroot1974reaching}. Another element to compare different combination rules is their learning performance, and it was shown in~\cite{nedic2017fast,lalitha2018social} that the log-linear combination rule in \eqref{eq:combsl} exhibits superior exponential convergence to the true state. In view of these results, additional aspects of log-linear combination rule have been explored in the literature, e.g., non-asymptotic evolution of beliefs, topology learning and adaptive capabilities in face of a non-stationary environment~\cite{shahrampour2015distributed,nedic2017fast,matta2019interplay, bordignon2020adaptive}. 

In most of the social learning literature, the concept of truth learning corresponds to the capacity of each agent to concentrate their beliefs around the true hypothesis, as described in the following definition.
\begin{definition}[\textbf{Truth Learning}]\label{def:truth}
	We say that agent $k$ learns the truth, within the traditional social learning setup, when the following convergence behavior is observed:
	\begin{equation}
	\bm{\mu}_{k,i}\stackrel{\textnormal{a.s.}}{\longrightarrow}e_{\thetatrue},
	\end{equation}
	where $e_{\thetatrue}$ represents a vector of zeros everywhere except for the $\thetatrue$-th element, which is equal to one. 
	\QED
\end{definition}

An important condition to enable truth learning under traditional social learning is {\em global identifiability}.

\begin{assumption}[\textbf{Global Identifiability}]\label{as:glob}
	For each $\theta\neq \thetatrue$, there exists at least one agent $\kappagood$ for which the KL divergence $d_{\kappagood}(\theta)$ is strictly positive.
	\QED
\end{assumption}

An agent $\kappagood$ that is able to distinguish some hypothesis $\theta$ from $\thetatrue$ will be referred to as a \emph{clear-sighted agent}. In order to satisfy Assumption~\ref{as:glob}, at least one clear-sighted agent must exist {\em for each} hypothesis $\theta\in\Theta\setminus\{\thetatrue\}$. 

Due to Assumption~\ref{as:glob}, the network as a whole has sufficient information to solve the inference problem. For strongly-connected networks, under Assumptions ~\ref{as:integ},~\ref{as:posit} and~\ref{as:glob}, references \cite{nedic2017fast,lalitha2018social} show that every agent is able to learn the truth exponentially fast, i.e., over time the belief vectors $\bm{\mu}_{k,i}$ collapse into the vector $e_{\thetatrue}$ almost surely for all $k=1,2,\dots, K$. 

In all the aforementioned works, a critical assumption is that each agent shares its entire intermediate belief, namely $\bm{\psi}_{k,i}$, with neighbors. This assumption can be unrealistic, for example, from a behavioral point of view, i.e., in social interactions, individuals oftentimes express their opinion only when provoked about a specific hypothesis. 
\section{Partial Information Sharing}\label{sec:partialinfo}
In the scenario described above, in order to learn the true state of nature out of a set of $H$ hypotheses, agents share the full extension of their intermediate belief vector with their respective neighbors. We consider now that agents are interested in answering a different question. For instance, in our example of the network of meteorological stations, consider that these stations want to answer the question ``is it sunny?", do they still need to share their entire belief vectors repeatedly to find out whether it is sunny or not? If we devise a cooperation scheme where agents share only, at every iteration, the confidence they have regarding the ``sunny" condition, can agents still learn? 

In this work, we adapt the non-Bayesian social learning framework by incorporating the following communication constraint (due, for example, to communication or regulation requirements): agents share a \emph{single} belief component, namely $\bm{\psi}_{\ell,i}(\thetashare)$, where $\thetashare$ denotes a \emph{hypothesis of interest} or the \textit{transmitted hypothesis}. 
This constraint reflects a situation in which agents possess a certain level of private knowledge, but, for reasons such as social dynamics, limited bandwidth, regulation, diffuse only certain aspects of it. For example, consider the following situation. A group of agents exchanges reviews concerning a product from brand $\theta_1$ that was recently released in the market. The information contained in these reviews is limited to the product of interest, i.e., the hypothesis of interest. The content of these reviews can be positive or negative according to the agent's perception of the product, i.e., the review conveys a soft decision. From these repeated interactions, agents would like to reach a conclusion on which product brand is best among brands $\{\theta_1, \theta_2, \theta_3\}$. In their reviews, agents do not share opinions on brands $\theta_2, \theta_3$, which correspond to the non-transmitted hypotheses. 

Besides the appeal of the partial information strategy from a behavioral standpoint, a second relevant aspect consists in taking into account {\em compressed information}. While technological advances allow for improved communication bandwidth capacity, therefore enabling many edge-intelligent solutions such as distributed and federated learning, we see a growing interest for communication efficiency~\cite{millsiot2019}. The problem of sharing partial or quantized information in the context of distributed learning arises in other applications. For example, in~\cite{sarwate2014distributed} nonparametric estimation of a discrete probability distribution is performed through a consensus protocol with quantized messages.

\begin{figure}[t]
	\centering
	\includegraphics[width=2.6in]{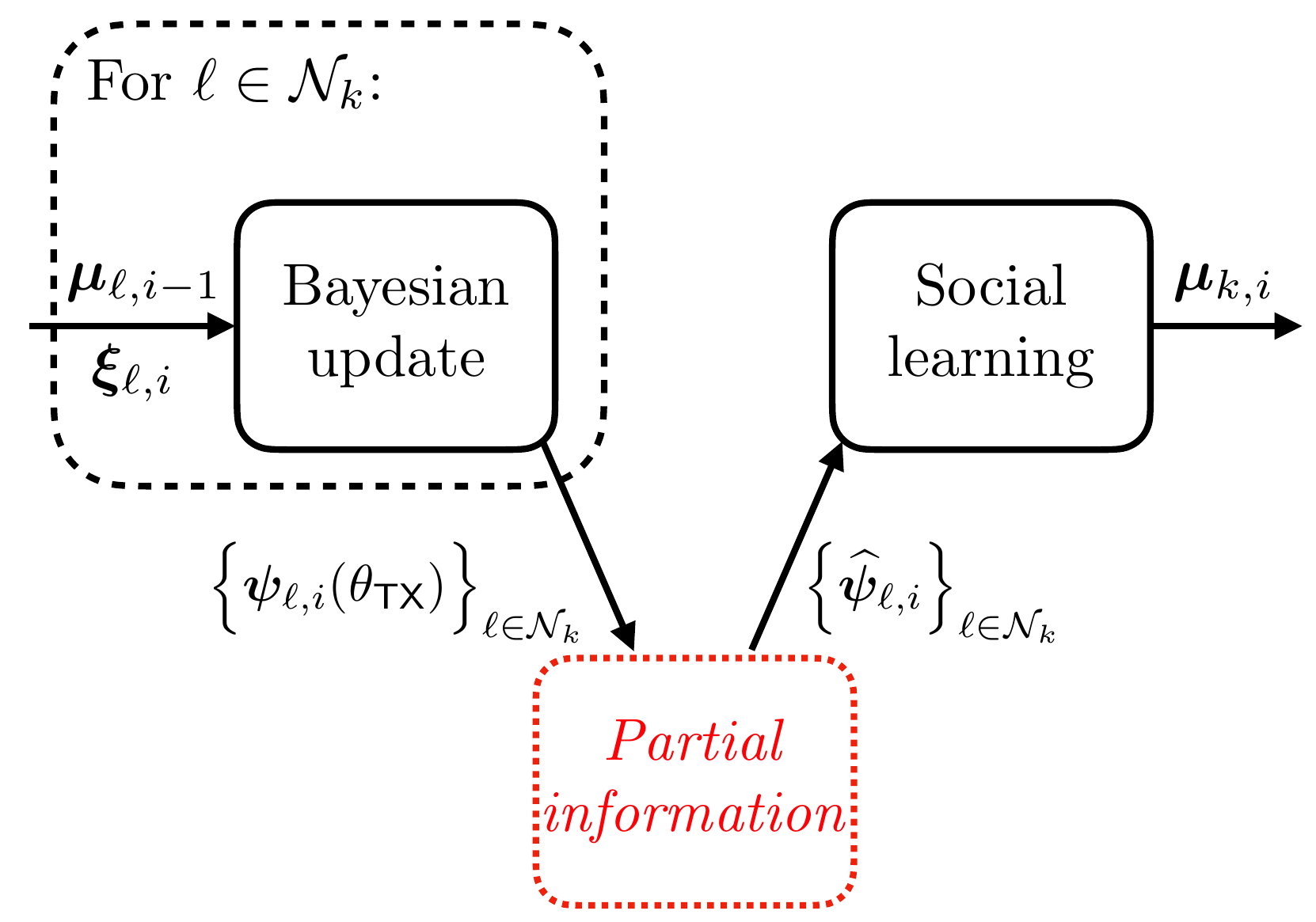}
	\caption{Diagram of the social learning strategy with a partial information mechanism.}	\label{fig:pidiag}
\end{figure}

In Fig.~\ref{fig:pidiag}, we see how the output of the Bayesian update step is limited to the transmitted component $\bm{\psi}_{\ell, i}(\theta_{\sf TX})$. We see also that an intermediate step is included between the individual learning and the social learning steps. This additional step is referred to as a \emph{partial information} mechanism and its role consists in transforming the received transmitted components into a valid belief vector to be used during the social learning step. To this end, the red block in Fig.~\ref{fig:pidiag} implements some transformation
\begin{equation}
\bm{\psi}_{\ell,i}(\thetashare)
\mapsto
\widehat{\bm{\psi}}_{\ell,i}
\end{equation}
to incorporate the information in the transmitted component into an estimate $\widehat{\bm{\psi}}_{\ell,i}$ of the locally-updated belief vector $
\bm{\psi}_{\ell,i}$. Different transformations correspond to different application scenarios and represent various types of behavior of the learning agents. 
Notably, once a particular transformation is specified, the other two blocks in Fig.~\ref{fig:pidiag} keep their original meaning, since $i)$ the Bayesian update was already shown to be perfectly well-supported at a local level, and, since it comes before the red block, it is unaffected by the specific transformation; and $ii)$ it was shown in~\cite{molavi2018theory} that, given the aforementioned axioms, and given certain belief vectors available to agent $k$, the log-linear combination of these beliefs is the only admissible rule. 
In summary, what is left unspecified in Fig. 2 is the particular transformation implemented by the red block. 

	Given that the only information shared by neighbors is the transmitted component of the intermediate beliefs, the transformation $\bm{\psi}_{\ell,i}(\theta_{\sf TX})\mapsto\bm{\widehat\psi}_{\ell,i}$, performed in the partial information mechanism, can be designed according to:
	\begin{equation}
	\widehat{\bm{\psi}}_{\ell,i}(\theta)=\begin{cases}\bm{\psi}_{\ell,i}(\theta_{\sf TX}), &\theta=\theta_{\sf TX},\\
	\frac{1}{H-1}(1-\bm{\psi}_{\ell,i}(\theta_{\sf TX})), &\theta\neq \theta_{\sf TX}.
	\end{cases}\label{eq:pimec}
	\end{equation}
	Intuitively the partial information mechanism in \eqref{eq:pimec} preserves the component of interest shared by agent $\ell$, i.e., $\bm{\psi}_{\ell,i}(\theta_{\sf TX})$, and redistributes the excess mass, i.e., $1-\bm{\psi}_{\ell,i}(\theta_{\sf TX})$, over the remaining hypotheses $\theta\neq\thetashare$ uniformly following a maximum entropy principle. We say that \eqref{eq:pimec} implements partial information using a {\em  memoryless} approach, that is, disregarding prior knowledge that might bias the agents to give more or less importance to the non-transmitted components. 

Alternatively, and in contrast to the memoryless choice of uniform weights, we can also consider a {\em memory-aware} strategy suggested by one of the anonymous reviewers to this paper, namely, for $\theta\neq \theta_{\sf TX}$: 
\begin{equation}
	\widehat{\bm{\psi}}_{\ell,i}(\theta)=\frac{\bm{\mu}_{k,i-1}(\theta)}{1-\bm{\mu}_{k,i-1}(\thetashare)}(1-\bm{\psi}_{\ell,i}(\theta_{\sf TX})).\label{eq:memoweig}
\end{equation}
From a behavioral perspective, the memoryless choice reflects well situations where the agents focus on the transmitted hypothesis (for example, they are discussing/sharing opinions on a particular candidate in an election process), and their learning mechanism does not allow them to care about the detailed update of the other components. From a design-oriented perspective, the memory-aware choice seems to carry more information, and we have preliminary results supporting this intuition, but it requires detailed analysis that is beyond the scope and page limitations of this work. In this paper, we will provide a detailed assessment of strategy \eqref{eq:pimec}, with the analysis of the latter strategy being left for future work.

We propose in this work two algorithms that include partial information sharing regarding one hypothesis of interest $\thetashare$ and examine how the constrained communication affects truth learning in this setup. The objective of the agents, in both approaches, is to verify whether the state of nature agrees with $\theta_{\sf TX}$ or not. We consider that agent $k$ succeeds in doing so whenever it learns the truth according to Definition~\ref{def:truthpartial}.

\begin{definition}[\textbf{Truth Learning with Partial Information}]\label{def:truthpartial}
	Within the partial information framework, the definition of truth learning depends on the choice of $\thetashare$:
	\begin{itemize}
		\item If $\thetashare=\thetatrue$, agent $k$ learns the truth when
		\begin{equation}
		\bm{\mu}_{k,i}(\thetashare)\stackrel{\textnormal{a.s.}}{\longrightarrow}1.
		\end{equation}
		\item If $\thetashare\neq\thetatrue$, agent $k$ learns the truth when
		\begin{equation}
		\bm{\mu}_{k,i}(\thetashare)\stackrel{\textnormal{a.s.}}{\longrightarrow}0.
		\end{equation}
	\end{itemize}
	Any other case is classified as a mislearning outcome.
	\QED
\end{definition}

\subsection{Social Learning under Partial Information}
In the first partial information approach, we propose the following modified version of the social learning algorithm \eqref{eq:bayesup}--\eqref{eq:combsl}, where at each instant $i=1,2,\dots$ each agent $k$ performs the following operations:
\begin{align}
&\bm{\psi}_{k,i}(\theta)=\frac{L_k(\bm{\xi}_{k,i}|\theta)\bm{\mu}_{k,i-1}(\theta)}{\sum_{\theta'\in\Theta}L_k(\bm{\xi}_{k,i}|\theta')\bm{\mu}_{k,i-1}(\theta')},\label{eq:Bayesupdate}\\
&\widehat{\bm{\psi}}_{\ell,i}(\theta)=\begin{cases}\bm{\psi}_{\ell,i}(\theta_{\sf TX}), &\theta=\theta_{\sf TX},\\
	\frac{1}{H-1}(1-\bm{\psi}_{\ell,i}(\theta_{\sf TX})), &\theta\neq \theta_{\sf TX},
	\end{cases}\label{eq:justone}\\
&\bm{\mu}_{k,i}(\theta) =\frac{\displaystyle\exp\left\{\sum_{\ell \in\mathcal{N}_k}a_{\ell k}\log\widehat{\bm{\psi}}_{\ell,i}(\theta)\right\}}{\displaystyle\sum_{\theta'\in\Theta}\exp\left\{\sum_{\ell\in\mathcal{N}_k}a_{\ell k}\log\widehat{\bm{\psi}}_{\ell,i}(\theta')\right\}}.\label{eq:combine}
\end{align}
In \eqref{eq:Bayesupdate}, agent $k$ performs a local Bayesian update to incorporate its new private observation $\bm{\xi}_{k,i}$. By doing so, the agent builds its intermediate belief vector $\bm{\psi}_{k,i}$, which in the traditional social learning implementation would have been the variable shared with the neighbors of $k$. In the partial information setting, however, agent $k$ will only share the component $\bm{\psi}_{k,i}(\thetashare)$ with its neighbors, which will then split the remaining mass $1-\bm{\psi}_{k,i}(\thetashare)$ uniformly across the hypotheses $\theta\neq \thetashare$. This process is shown in \eqref{eq:justone}, which gives origin to the \textit{modified belief vector} $\widehat{\bm{\psi}}_{k,i}$. The final belief vector $\bm{\mu}_{k,i}$ is obtained by locally aggregating the neighbors' modified belief vectors using the same log-linear combination rule as shown in \eqref{eq:combine}.

Note that the aggregation step in \eqref{eq:combine} implies for every pair $\theta, \theta'\in \Theta$ that
\begin{equation}
\log \frac{\bm{\mu}_{k,i}(\theta)}{\bm{\mu}_{k,i}(\theta')} =\sum_{\ell=1}^{K}a_{\ell k}\log \frac{\widehat{\bm{\psi}}_{\ell,i}(\theta)}{\widehat{\bm{\psi}}_{\ell,i}(\theta')}.
\label{eq:combine_2}
\end{equation}
The social learning strategy with partial information hereby proposed admits a useful relation to traditional social learning, which is described in Proposition~\ref{prop:binary}. 

\begin{proposition}[\textbf{Binary Hypotheses Test}]\label{prop:binary}
	The social learning algorithm with partial information presented in \eqref{eq:Bayesupdate}-\eqref{eq:combine} can be interpreted as solving a binary hypothesis test problem for the set $\Theta_b=\{\thetashare, \thetasharec\}$, with the likelihood $L_k(\xi|\thetasharec)$ associated to $\thetasharec$ defined as:
	\begin{equation}
	L_{k}(\xi|\thetasharec)\triangleq \sum_{\tau\neq\thetashare}\frac{L_{k}(\xi|\tau)}{H-1}.\label{eq:likelihoodthetac}
	\end{equation}
In other words, the belief $\bm{\mu}_{k,i}(\theta_{\sf TX})$ obtained from implementing \eqref{eq:Bayesupdate}--\eqref{eq:combine} with $H$ hypotheses in $\Theta$ is equivalent to the belief $\bm{\mu}_{k,i}(\theta_{\sf TX})$ obtained from implementing traditional social learning (i.e., \eqref{eq:bayesup}--\eqref{eq:combsl}) using a binary set of hypotheses $\Theta_b$ and the likelihood $L_k(\xi|\thetasharec)$ defined in \eqref{eq:likelihoodthetac}.
	%In other words, the original problem with $H$ hypotheses considered by the agents can be reformulated as a binary hypothesis test problem over $\Theta_b$, with a fictitious likelihood for the ``aggregate'' fictitious hypothesis $\bar{\theta}_{\sf TX}$, namely, $L_k(\xi|\thetasharec)$.
\end{proposition}
\begin{IEEEproof}
	See Appendix~\ref{ap:bin}.
\end{IEEEproof}
The proposition shows that the algorithm under partial information in \eqref{eq:Bayesupdate}--\eqref{eq:combine} can be reinterpreted in terms of a traditional social learning algorithm with a binary set of hypotheses $\Theta_b=\{\thetashare,\thetasharec\}$, and with likelihoods $L_k(\xi|\thetashare)$ and $L_k(\xi|\thetasharec)$. Intuitively, hypothesis $\thetasharec$ corresponds to an artificial hypothesis that representing any hypothesis distinct from $\thetashare$.

When $\thetashare\neq\thetatrue$, the algorithm under partial information without self-awareness is equivalent to a traditional (binary) social learning algorithm with mismatched distribution, i.e., with a distribution of the data that does not match the assumed likelihood. Under these conditions, the evolution of beliefs, particularly its asymptotic learning behavior, is known to depend on the KL divergence between the true likelihood $L_k(\xi|\thetatrue)$ and likelihoods $L_k(\xi|\thetasharec)$ and $L_k(\xi|\thetashare)$, an can be characterized using the theoretical results in~\cite{nedic2017fast}. In order to keep this article complete and self-contained, we establish the convergence behavior of beliefs in Lemma~\ref{lemma:rate} (see Appendix~\ref{ap:truthpartial}), Theorem~\ref{the:truthpartial} and Theorem~\ref{the:belcoll}.

In the second approach, we take into account the fact that each agent $k$ has full knowledge about its own intermediate belief vector $\bm{\psi}_{k,i}$. Agent $k$ will still perform the same Bayesian update seen in \eqref{eq:Bayesupdate} and share only its belief component corresponding to the hypothesis of interest $\thetashare$, reflected in \eqref{eq:justone}. However, now, we rewrite the combination step of the algorithm in such a way that agent $k$ combines its neighbors' modified beliefs $\{\widehat{\bm{\psi}}_{\ell,i}\}_{\ell\in\mathcal{N}_k\backslash k}$ with its own \emph{true} belief $\bm{\psi}_{k,i}$, using the following log-linear combination rule:
\begin{equation}
\bm{\mu}_{k,i}(\theta)=\frac{\exp\left\{a_{kk}\log \bm{\psi}_{k,i}(\theta)+\sum\limits_{\substack{\ell=1\\\ell\neq k}}^Ka_{\ell k}\log \widehat{\bm{\psi}}_{\ell,i}(\theta) \right\}}{\sum\limits_{\theta'\in\Theta}\exp\left\{a_{kk}\log \bm{\psi}_{k,i}(\theta')+\sum\limits_{\substack{\ell=1\\\ell\neq k}}^Ka_{\ell k}\log \widehat{\bm{\psi}}_{\ell,i}(\theta') \right\}}.\label{eq:combsa}
\end{equation}
Note that this combination step leads to:
\begin{equation}
\log \frac{\bm{\mu}_{k,i}(\theta)}{\bm{\mu}_{k,i}(\theta')}=a_{kk}\log \frac{\bm{\psi}_{k, i}(\theta)}{\bm{\psi}_{k, i}(\theta')}+\sum\limits_{\substack{\ell=1\\\ell\neq k}}^Ka_{\ell k} \log \frac{\widehat{\bm{\psi}}_{\ell, i}(\theta)}{\widehat{\bm{\psi}}_{\ell, i}(\theta')}, \label{eq:combsa_2}
\end{equation}
where we can distinguish two terms on the RHS of \eqref{eq:combsa_2}: a first term representing the \textit{self-awareness} of agent $k$ and a second term, which combines the neighbors' partial information contribution. In this formulation, it is necessary that $a_{kk}>0$ in order for the self-awareness of agent $k$ to count in the combination step. We will refer to it as \textit{self-awareness coefficient} and we will assume that $a_{kk}>0$ for all $k=1,2,\dots,K$ in this setup.

\subsection{Non-Transmitted Components}\label{sec:ntc}
Before presenting the theoretical results, it is useful to make a parallel between the evolution of non-transmitted belief components for both partial information approaches. For the algorithm without self-awareness, all non-transmitted components of the belief vector \emph{evolve equally} over time. To see that, replace \eqref{eq:justone} into \eqref{eq:combine_2} for any two non-transmitted components $\tau,\tau'\neq \thetashare$:
\begin{align}
&\log \frac{\bm{\mu}_{k,i}(\tau)}{\bm{\mu}_{k,i}(\tau')}=\sum_{\ell=1}^{K}a_{\ell k}\log \frac{\widehat{\bm{\psi}}_{\ell,i}(\tau)}{\widehat{\bm{\psi}}_{\ell,i}(\tau')}\nonumber\\&=\sum_{\ell=1}^{K}a_{\ell k}\log\frac{\frac{1-\bm{\psi}_{\ell,i}(\thetashare)}{H-1}}{\frac{1-\bm{\psi}_{\ell,i}(\thetashare)}{H-1}}=0\nonumber
\\&\implies \bm{\mu}_{k,i}(\tau)=\bm{\mu}_{k,i}(\tau').\label{eq:allmuequal}
\end{align}
Since the entries of the vector $\bm{\mu}_{k,i}$ sum up to one, it follows that we can write, for any non-transmitted hypothesis $\tau\neq\thetashare$:
\begin{align}
&\sum_{\tau\neq\thetashare}\bm{\mu}_{k,i}(\tau)=1-\bm{\mu}_{k,i}(\thetashare)\nonumber\\
&\implies \bm{\mu}_{k,i}(\tau)=\frac{1-\bm{\mu}_{k,i}(\thetashare)}{H-1}.
\label{eq:1}
\end{align} 
This equal evolution for all $\tau\neq\thetashare$ will have the following important effect on the learning behavior: If one non-transmitted hypothesis is rejected, then so are all the non-transmitted hypotheses.

For the approach with self-awareness, the beliefs regarding non-transmitted hypotheses will no longer evolve equally as is the case for the first partial information strategy. More precisely, for two non-transmitted hypotheses $\tau,\tau'\neq\thetashare$, considering \eqref{eq:justone}, the combination step in \eqref{eq:combsa_2} yields:
	\begin{align}
	&\log\frac{\bm{\mu}_{k,i}(\tau)}{\bm{\mu}_{k,i}(\tau')}=a_{kk}\log \frac{\bm{\psi}_{k,i}(\tau)}{\bm{\psi}_{k,i}(\tau')}+\sum_{\substack{\ell=1\\\ell\neq k}}^{K}a_{\ell k}\log\frac{\frac{1-\bm{\psi}_{\ell,i}(\thetashare)}{H-1}}{\frac{1-\bm{\psi}_{\ell,i}(\thetashare)}{H-1}}\nonumber\\&\stackrel{\text{(a)}}{=}a_{kk}\log \frac{\bm{\mu}_{k,i-1}(\tau)}{\bm{\mu}_{k,i-1}(\tau')}+a_{kk}\log \frac{L_k(\bm{\xi}_{k,i}|\tau)}{L_k(\bm{\xi}_{k,i}|\tau')},\label{eq:recursnontx}
	\end{align}
where in (a) we used \eqref{eq:Bayesupdate}.
In Lemma~\ref{lem:unshared} (Appendix~\ref{ap:alsa}), we show that the log-ratio of beliefs for non-transmitted hypotheses in the LHS of \eqref{eq:recursnontx} converges in distribution to some asymptotic random variable. In practice, this implies that the non-transmitted components will exhibit an oscillatory behavior over time. Lemma~\ref{lem:discardonediscardall} (Appendix~\ref{ap:alsa}) will nevertheless ensure the following stronger result: for the algorithm with self-awareness all non-transmitted components are \emph{rejected in parallel}. Although not as strong as the equal evolution seen in \eqref{eq:1}, this property will be essential to enable learning in the self-aware case.

\section{Performance Analysis}\label{sec:perfan}
Before delving upon the analysis of the learning performance, it is useful to introduce some auxiliary quantities. Recall the definition of $\thetasharec$, which corresponds to a ``fictitious'' hypothesis that represents occurrence of any hypothesis distinct from $\thetashare$. This fictitious hypothesis does not explicitly belong to $\Theta$, and therefore is not associated to any of the likelihood functions. To this end, we use instead the definition of the fictitious likelihood, seen in \eqref{eq:likelihoodthetac}, which embodies compressed information on all likelihoods relative to $\theta\neq \thetashare$. These two concepts allow us to extend the notation of the KL divergence introduced in \eqref{eq:kldiv} to likelihood $L_k(\xi|\thetasharec)$:
\begin{equation}
d_{k}(\thetasharec)\triangleq\E \left(\log \frac{L_k(\bm{\xi}_{k,i}|\thetatrue)}{L_k(\bm{\xi}_{k,i}|\thetasharec)}\right).\label{eq:klarithmean}
\end{equation}
We also introduce the corresponding network average of divergences, where the averaging weights are given by the entries of the Perron eigenvector in \eqref{eq:perronv}:\footnote{From the convexity of $-\log(\cdot)$ and using Jensen's inequality, we have that:
		\begin{equation}
		\log\frac{L_k(\bm{\xi}_{k,i}|\thetatrue)}{\frac{1}{H-1}\sum\limits_{\tau\neq\thetashare}L_k(\bm{\xi}_{k,i}|\tau)}\leq\frac{1}{H-1}\sum\limits_{\tau\neq\thetashare}\left(\log \frac{L_k(\bm{\xi}_{k,i}|\thetatrue)}{L_k(\bm{\xi}_{k,i}|\tau)}\right).\label{eq:ineqlike}
		\end{equation}
		Taking expectation of both sides in \eqref{eq:ineqlike} allows us to
		relate $d_{k}(\thetasharec)$ to the KL divergences relative to the non-transmitted hypotheses according to:
		\begin{equation}
		d_{k}(\thetasharec)\leq\sum\limits_{\tau\neq\thetashare}\frac{d_{k}(\tau)}{H-1}.\label{eq:klineq}
		\end{equation}
	From \eqref{eq:klineq}, we see that the finite KL divergence assumption (Assumption~\ref{as:integ}) extends naturally to $d_k(\thetasharec)$ for all $k=1,2,\dots,K$.}
\begin{equation}
d_{\sf ave}(\thetasharec)\triangleq\sum_{\ell =1}^Kv_\ell d_{\ell}(\thetasharec).\label{eq:klarithmeanave}
\end{equation} 
In the following sections, we are interested in determining for each of the algorithms, and for different choices of the transmitted hypothesis, the conditions for learning and mislearning. The convergence analysis will be split in two complementary cases: $i)$ when $\thetashare=\thetatrue$; and $ii)$ when $\thetashare\neq\thetatrue$.

\subsection{Truth Learning when $\thetashare=\thetatrue$}\label{sec:partialinfo_learning}
For both partial information strategies, we will show that truth sharing, i.e., choosing $\thetashare=\thetatrue$, results in truth learning. Consider first the approach without self-awareness, namely algorithm \eqref{eq:Bayesupdate}--\eqref{eq:combine}. Truth learning under truth sharing is guaranteed conditioned on the existence of at least one agent that is clear-sighted in the following sense (we use the notation $\thetatruec$ in place of $\thetasharec$ since we are focusing on the case $\thetashare=\thetatrue$).
\begin{assumption}[\textbf{Existence of a Clear-Sighted Agent: Approach without Self-Awareness}]
	\label{as:clear}
	There exists at least one agent $\kappagood$ that satisfies the following condition:
	\begin{equation}
	d_{\kappagood}(\thetatruec)>0.
	\label{eq:assumdistmin1}
	\end{equation}
\QED
\end{assumption}
From \eqref{eq:assumdistmin1}, we require that this clear-sighted agent is endowed with the ability of distinguishing the true likelihood $L_{\kappagood}(\xi|\thetatrue)$ from the fictitious likelihood $L_{\kappagood}(\xi|\thetatruec)$ defined in \eqref{eq:likelihoodthetac}. Note that Assumption~\ref{as:clear} implies that $|\bar{\Theta}_{\kappagood}|>0$. Actually, requiring that the true likelihood is not a combination, with weights $1/(H-1)$, of all the likelihoods for $\theta\neq\thetatrue$, is tantamount to requiring that the true likelihood is not a combination, with weights $1/|\bar{\Theta}_{k^{\star}}|$, of the  {\em distinguishable} hypotheses.  
This is not a strong assumption, since the case in which the true likelihood matches \emph{exactly} a mixture of the likelihoods relative to the distinguishable hypotheses with uniform weights is deemed to be an unlucky coincidence.
\begin{theorem}[\textbf{Truth Sharing Implies Truth Learning: Approach without Self-Awareness}]\label{the:truthpartial}
	Under Assumptions ~\ref{as:integ},~\ref{as:posit} and~\ref{as:clear}, if $\thetashare=\thetatrue$, then every agent $k$ learns the truth, i.e.,
	\begin{equation}
	\bm{\mu}_{k,i}(\thetashare)\stackrel{\textnormal{a.s.}}{\longrightarrow}1.
	\end{equation}
\end{theorem}
\begin{IEEEproof}
	See Appendix~\ref{ap:truthpartial}.
\end{IEEEproof}
Next, consider the partial information approach with self-awareness, whose algorithm can be seen in \eqref{eq:Bayesupdate}, \eqref{eq:justone} and \eqref{eq:combsa}. For this algorithm, truth learning under truth sharing requires another notion of clear-sighted agent.
\begin{assumption}[\textbf{Existence of a Clear-Sighted Agent: Approach with Self-Awareness}]
	\label{as:strongclear}
There exists at least one agent $\kappagood$ whose set of distinguishable hypotheses $\bar{\Theta}_{k^{\star}}$ is nonempty, and whose likelihood models satisfy the following condition. For any convex combination vector $\alpha\in \Delta^{|\bar{\Theta}_{\kappagood}|}$,
\begin{equation}
	\E\Bigg(\log\frac{L_{\kappagood}(\bm{\xi}_{{\kappagood},i}|\thetatrue)}{\sum\limits_{\tau\in\bar{\Theta}_{\kappagood}}\alpha(\tau) L_{\kappagood}(\bm{\xi}_{{\kappagood},i}|\tau)}\Bigg)\geq c>0.
	\label{eq:assumdistmin}
	\end{equation}
\QED
\end{assumption} 
This assumption is stronger than Assumption~\ref{as:clear} since it requires that the true likelihood $L_{\kappagood}(\xi|\thetatrue)$ is not an \emph{arbitrary} mixture of the likelihoods relative to distinguishable hypotheses. We will discuss these differences in due detail in Section~\ref{sec:assumpt}. 

\begin{theorem}[\textbf{Truth Sharing Implies Truth Learning: Approach with Self-Awareness}]\label{the:selfaware}
Under Assumptions~\ref{as:integ},~\ref{as:posit} and~\ref{as:strongclear}, when $\thetashare=\thetatrue$ we have: 
	\begin{equation}
	\bm{\mu}_{k,i}(\thetashare)\stackrel{\textnormal{a.s.}}{\longrightarrow} 1.
	\end{equation}
\end{theorem}
\begin{IEEEproof}
	See Appendix~\ref{ap:selfaware}.
\end{IEEEproof}

Theorems~\ref{the:truthpartial} and~\ref{the:selfaware} ensure, under some technical assumptions, that both partial information approaches drive agents to learn the truth when they share information relative only to the true hypothesis. These results motivate us to draw a parallel between the learning behaviors of the partial information and the traditional social learning strategies.
	
\begin{figure*}[tb]
	\centering
	\subfloat[Traditional social learning.]{\includegraphics[height=1.22in]{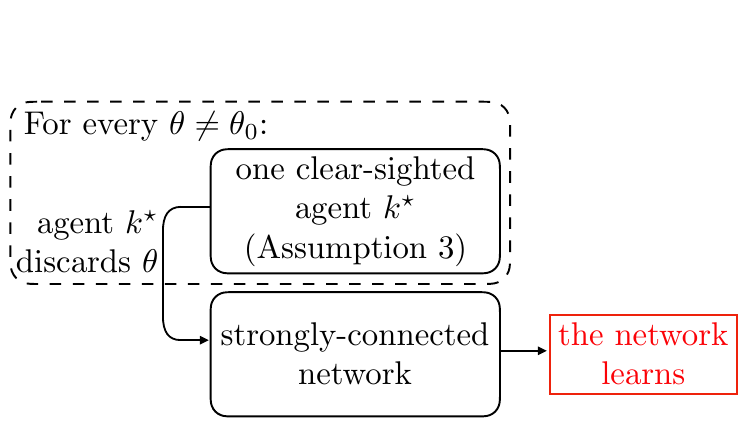}\label{fig:mech1}}
	\hspace{4pt}\subfloat[Partial information approach without self-awareness.]{\includegraphics[height=1.22in]{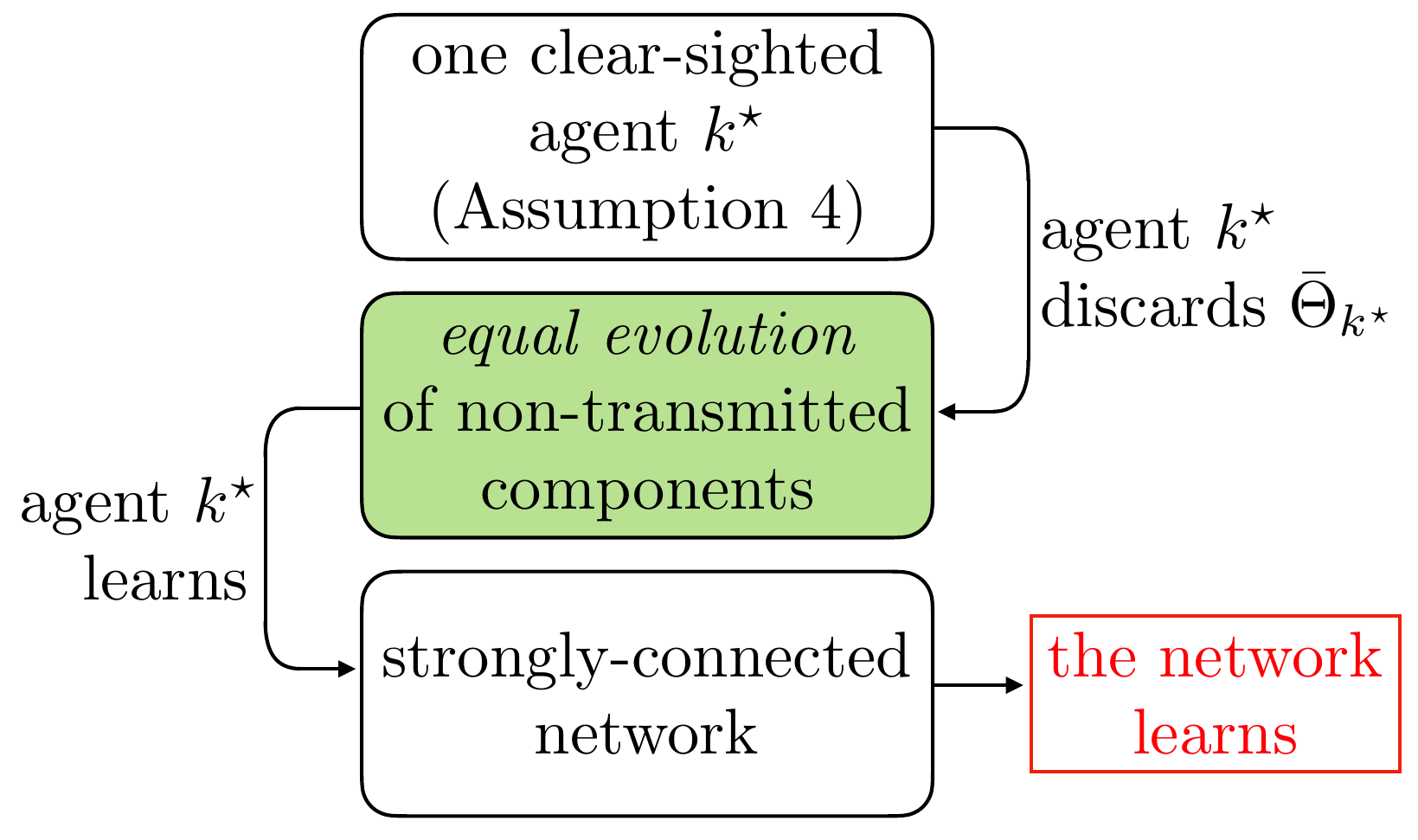}\label{fig:mech2}}
	\hspace{10pt}\subfloat[Partial information approach with self-awareness.]{\includegraphics[height=1.22in]{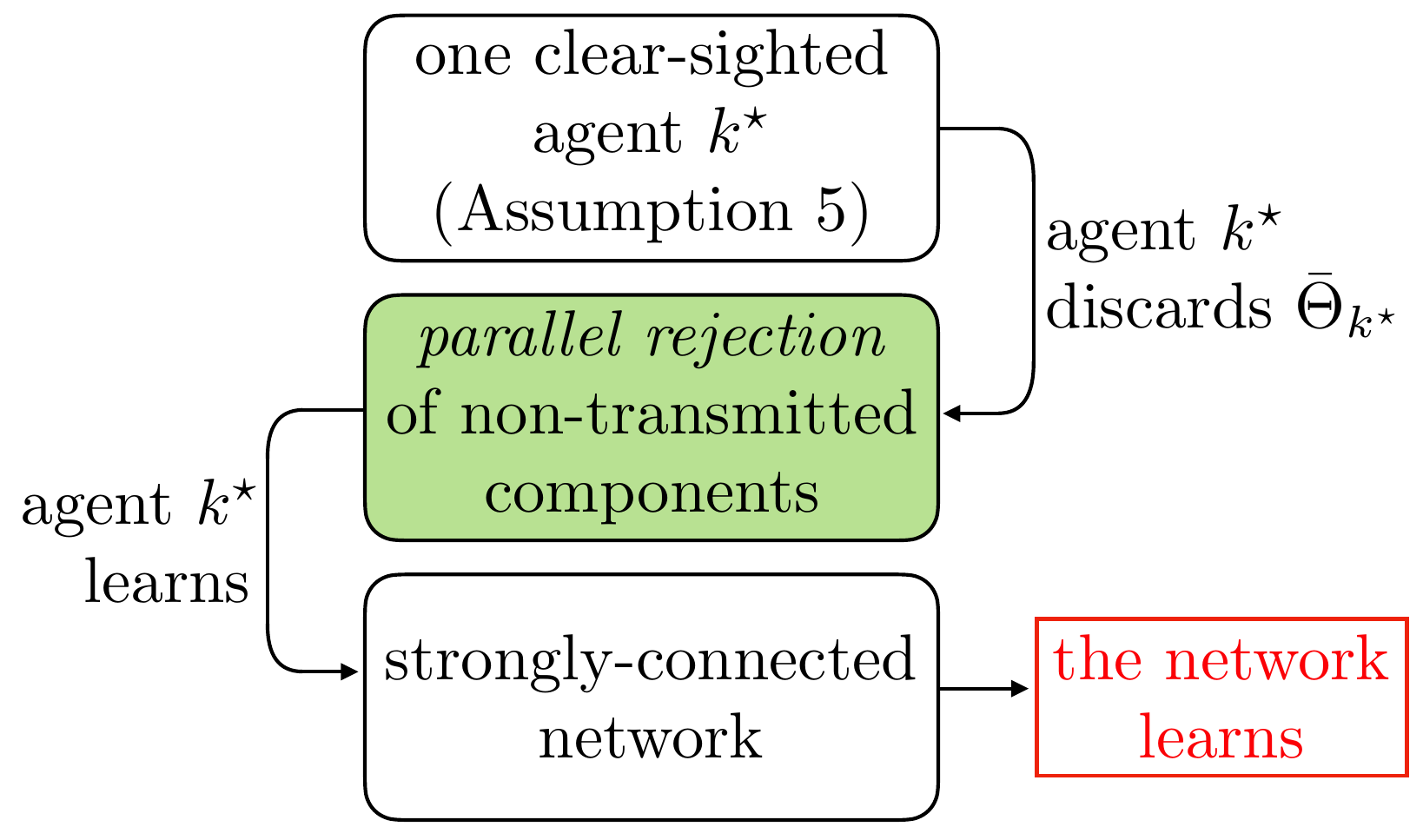}\label{fig:mech3}}
	\caption{Learning mechanism for the traditional social learning and the two partial information approaches under the truth-sharing regime.}\label{fig:learnmech}
\end{figure*}
\subsubsection{Learning with Traditional Social Learning (Fig.~\ref{fig:mech1})}
Consider a fixed hypothesis $\theta\neq\thetatrue$. In view of the global identifiability assumption, there is (at least) one {\em clear-sighted} agent (in the sense of Assumption~\ref{as:glob}) whose data and likelihoods allow it to distinguish $\theta$ from $\thetatrue$. Due to the propagation of information across the strongly-connected network, the other agents are eventually endowed with the same ability. As a result, all agents are able to discard $\theta$ from $\thetatrue$. Repeating the above process for every $\theta\neq\thetatrue$ leads the agents to choose finally $\thetatrue$.
\subsubsection{Learning without Self-Awareness (Fig.~\ref{fig:mech2})}
In this case, the meaning of the qualification ``clear-sighted'' changes. 
Let $k^\star$ be the index of a clear-sighted agent, and recall that the distinguishable set of this agent is denoted by $\bar{\Theta}_{\kappagood}$. 
In the context of partial information without self-awareness, the clear-sighted agent is required to distinguish $\thetatrue$ from some fictitious hypothesis ``aggregating'' the hypotheses in $\bar{\Theta}_{\kappagood}$ --- see Assumption~\ref{as:clear}. We showed that if this condition is satisfied, then all agents decide correctly. This result admits a useful interpretation. 
Assumption~\ref{as:clear} implies that the clear-sighted agent has some capability of discounting $\bar{\Theta}_{\kappagood}$.
Now, since we have shown in \eqref{eq:1} that under partial information without self-awareness the beliefs evaluated at $\theta\neq\thetashare$ {\em evolve equally in parallel} (i.e., $\bm{\mu}_{k,i}(\theta)$ takes the same value for all $\theta\neq\thetashare$ during the algorithm evolution), once the clear-sighted agent is able to discount the hypotheses in $\bar{\Theta}_{\kappagood}$, it is also able to discount {\em all} $\theta\neq\thetashare$. Finally, this possibility is extended to all the other agents by propagation of information across the strongly-connected network.
\subsubsection{Learning with Self-Awareness (Fig.~\ref{fig:mech3})}
As happens in the case without self-awareness, we need a clear-sighted agent, say agent $\kappagood$, that is required to distinguish $\thetatrue$ from some aggregate hypothesis involving $\bar{\Theta}_{\kappagood}$, but now in a different sense. 
In the self-aware strategy, the clear-sighted agent must be able to discern the likelihood at $\theta_0$ from {\em any convex combination} of likelihoods of the distinguishable hypotheses, as detailed in condition \eqref{eq:assumdistmin}.
This condition is stronger than \eqref{eq:assumdistmin1} since \eqref{eq:assumdistmin1} requires discriminability for a particular (uniform) combination. 
The reason for this stronger requirement is as follows.
In the Bayesian update rule, the social learning algorithms evaluate convex combinations of the likelihoods that use as weights the beliefs $\bm{\mu}_{k,i-1}(\theta)$ --- see the denominator in \eqref{eq:Bayesupdate}. 
In the social learning strategy without self-awareness, due to the equal evolution of the beliefs at the non-transmitted hypotheses, this convex combination ends up being a uniformly weighted convex combination of the likelihoods.
In contrast, as discussed in Section~\ref{sec:ntc}, in the strategy with self-awareness the beliefs at $\theta\neq\thetashare$ experience unpredictable mutual oscillations, and due to this unpredictability we require discriminability with respect to any convex combination.  

Now, as we show in Lemma~\ref{lem:oneagentlearns} (Appendix~\ref{ap:selfaware}), if condition \eqref{eq:assumdistmin} is satisfied the clear-sighted agent is able to discount the hypotheses in $\bar{\Theta}_{\kappagood}$. We will be able to show that also in this case the correct choice of the clear-sighted agent propagates across the other agents, albeit with a different learning mechanism, due to the self-awareness term. Comparing \eqref{eq:combine} against \eqref{eq:combsa}, we see that the self-awareness term introduces
a slight asymmetry in the social learning algorithm, since the self-loop term is treated differently from all the other terms. 

In traditional social learning~\cite{zhao2012learning,matta2019interplay,nedic2017fast,lalitha2018social} and in the approach without self-awareness \eqref{eq:Bayesupdate}--\eqref{eq:combine}, we do not have a different treatment among agents in the belief recursion. On the theoretical side, this allows us to characterize the learning strategy by expanding the recursion of logarithmic belief ratios over time and using concentration techniques such as the strong law of large numbers~\cite{billingsley2008probability} to establish the convergence of beliefs. In the approach with self-awareness, the slight asymmetry in the rule makes the exploitation of the belief recursion more demanding, and requires us to use different tools, such as martingales~\cite{billingsley2008probability} and useful results from the theory of stochastic convergence~\cite{ShaoBook}. The technical proofs that examine the belief convergence are thus significantly more complicated than in the traditional social learning or in the partial information case without self-awareness. On the practical side, the beliefs at the non-transmitted hypotheses do not  evolve equally in parallel as happens in
the case without self-awareness. Instead, as already mentioned, the beliefs will feature mutual oscillations among different entries $\theta\neq\thetashare$. Lemma~\ref{lem:discardonediscardall} (Appendix~\ref{ap:alsa}) is used to show that the oscillatory behavior of the beliefs does not impair the extension of this knowledge to the remaining $\theta\neq\thetashare$. 
As a result, despite the oscillatory behavior, the clear-sighted agent is able to discount all the hypotheses $\theta\neq\thetashare$.
Finally, this possibility is extended to all the other agents by propagation of information across the network (see Lemma~\ref{lem:oneall} in Appendix~\ref{ap:selfaware}).

\subsection{Truth Learning/Mislearning when $\thetashare\neq\thetatrue$}\label{sec:asaltbeh}
For both partial information approaches, we will establish conditions for obtaining truth learning and mislearning as an outcome of choosing $\thetashare\neq\thetatrue$. First, we introduce these results for the strategy without self-awareness.
\begin{theorem}[\textbf{Learning/Mislearning Regimes: Approach without Self-Awareness}]\label{the:belcoll}
Under Assumptions~\ref{as:integ} and~\ref{as:posit}, for every agent $k=1,2,\dots,K$, we observe two convergence behaviors:\footnote{We rule out the pathological case in which $d_{\sf ave}(\thetashare)=d_{\sf ave}(\thetasharec)$, which typically results in a (non-convergent) asymptotic oscillatory behavior of the belief components.}
\begin{enumerate}
	\item If $d_{\sf ave}(\thetashare)>d_{\sf ave}(\thetasharec)$, 
	\begin{equation}
	\bm{\mu}_{k,i}(\thetashare)\stackrel{\textnormal{a.s.}}{\longrightarrow}0\text{ and then }\bm{\mu}_{k,i}(\theta)\stackrel{\textnormal{a.s.}}{\longrightarrow}\frac{1}{H-1},\label{eq:th3}
	\end{equation}
	for all $\theta \neq \thetashare$.
	\item If $d_{\sf ave}(\thetashare)<d_{\sf ave}(\thetasharec)$,
	\begin{equation}
	\bm{\mu}_{k,i}(\thetashare)\stackrel{\textnormal{a.s.}}{\longrightarrow}1.\label{eq:th3a}
	\end{equation}
\end{enumerate}
\end{theorem}
\begin{IEEEproof}
	See Appendix~\ref{ap:thebelcoll}.
\end{IEEEproof}
Theorem~\ref{the:belcoll} shows two possible convergence behaviors for the beliefs across the network: asymptotically, either agents correctly discard $\thetashare$ or they mistakenly believe that $\thetashare$ is the true hypothesis. The former case takes place whenever the transmitted hypothesis is sufficiently distinct from the true hypothesis. The latter case happens whenever the transmitted hypothesis is more easily confounded with the true one than the fictitious complementary hypothesis $\thetasharec$. 

Before presenting similar results for the strategy with self-awareness, we introduce an extra assumption on the boundedness of the likelihood functions.
\begin{assumption}[\textbf{Bounded Likelihoods}]\label{as:boundedlik}
	Let there be a finite constant $B>0$ such that, for all $k$:
	\begin{equation}
	\left|\log \frac{L_k(\xi|\tau)}{L_k(\xi|\tau')}\right|\leq B, \label{eq:assboundedlike}
	\end{equation}
	for all $\tau, \tau'\in\Theta\setminus\{\thetashare\}$ and for all $\xi\in\mathcal{X}_k$.
	\QED
\end{assumption}

\begin{theorem}[\textbf{Learning/Mislearning Regimes: Approach with Self-Awareness}]\label{the:altsa}
	Under Assumptions~\ref{as:integ} and~\ref{as:posit}, when $\thetashare\neq\thetatrue$, for any agent $k$, we have:
	\begin{enumerate}
		\item  If $d_{\sf ave}(\thetashare)>\frac{1}{H-1} \sum\limits_{\tau \neq\thetashare}d_{\sf ave}(\tau),$
		\begin{equation}
		\bm{\mu}_{k,i}(\thetashare)\stackrel{\textnormal{a.s.}}{\longrightarrow}0.\label{eq:th4}
		\end{equation}
		\item Considering furthermore Assumption~\ref{as:boundedlik}, if $d_{\sf ave}(\thetashare)<d_{\textnormal{ave}}(\thetasharec)-\sum_{k=1}^K a_{kk}(d_k(\thetasharec)+v_k B ),$
		\begin{equation}
		\bm{\mu}_{k,i}(\thetashare)\stackrel{\textnormal{a.s.}}{\longrightarrow}1.\label{eq:th4a}
		\end{equation}
	\end{enumerate}
\end{theorem}
\begin{IEEEproof}
	See Appendix~\ref{ap:alt_sa}.
\end{IEEEproof}

Comparing the conditions for truth-learning when $\thetashare=\thetatrue$ (Theorems~\ref{the:truthpartial} and~\ref{the:selfaware}) against the conditions for truth-learning/mislearning when $\thetashare\neq\thetatrue$ (Theorems~\ref{the:belcoll} and~\ref{the:altsa}), we see that a fundamental difference arises. The conditions relative to the case $\thetashare=\thetatrue$ are formulated {\em at an individual agent level}, i.e., they depend on local characteristics of a clear-sighted agent. 
In contrast, the conditions relative to the case $\thetashare\neq\thetatrue$ are formulated at a {\em network} level, since they depend on average KL divergences and network parameters in a way that does not allow disentangling the individual agent contributions.

Let us now provide some interpretation of the results in Theorems~\ref{the:belcoll} and~\ref{the:altsa}. We will examine the two theorems separately. 
\subsubsection{Learning and Mislearning without Self-Awareness} 
\begin{figure*}[tb]
	\begin{minipage}{2.2in}
		\subfloat[Family of Gaussian likelihood functions.]{\includegraphics[width=\linewidth]{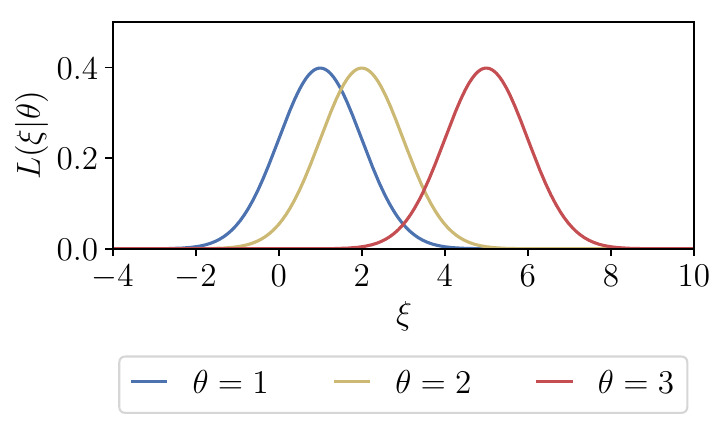}\label{fig:ex_likelihood1}}
	\end{minipage}
	\begin{minipage}{4.2in}
		\subfloat[Transmitted hypothesis $\thetashare=2$.]{\includegraphics[width=\linewidth]{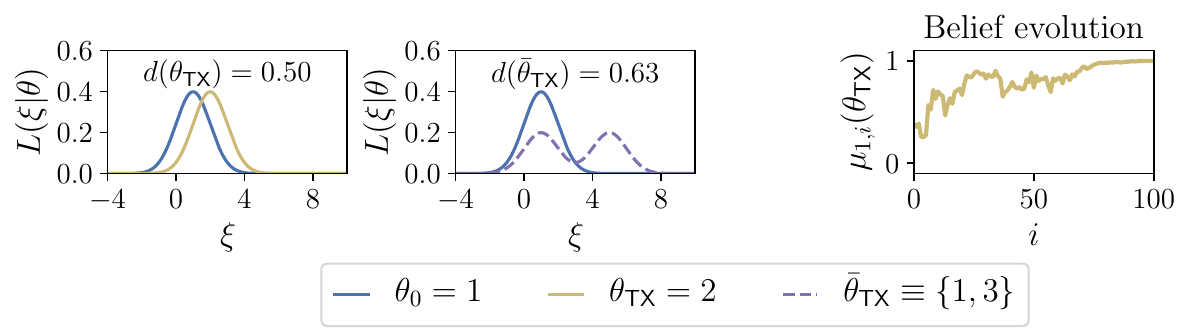}\label{fig:ex_likelihood2}}
		
		\vspace{-10pt}
		\subfloat[Transmitted hypothesis $\thetashare=3$.]{\includegraphics[width=\linewidth]{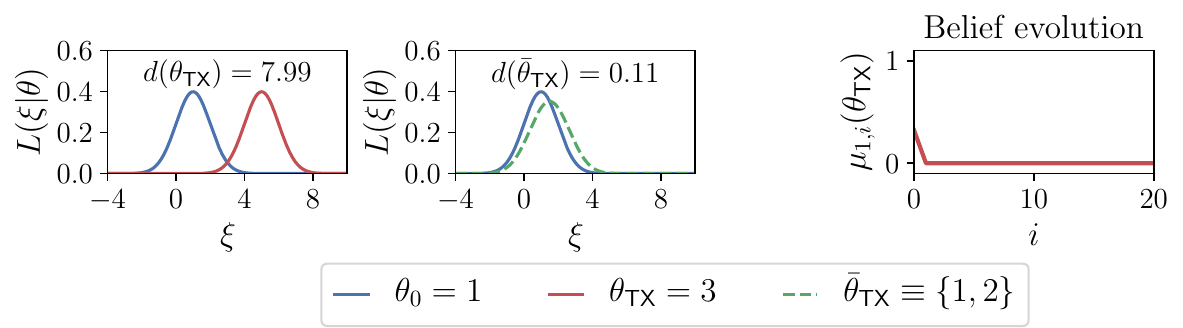}\label{fig:ex_likelihood3}}
	\end{minipage}
	\caption{Example of family of likelihood functions with $\thetatrue=1$. In the middle panels of (b) and (c), \emph{solid lines} represent the actual likelihood functions and \emph{dashed lines} depict the ``fictitious" likelihood functions associated with the complementary hypothesis $\thetasharec$, defined in \eqref{eq:likelihoodthetac}.}\label{fig:ex_likelihood}
\end{figure*}

To explain the intuition behind Theorem~\ref{the:belcoll}, we will introduce a numerical example. Let there be a strongly-connected network of $K=10$ agents solving a social learning problem under the partial information regime without self-awareness, i.e., under \eqref{eq:Bayesupdate}-\eqref{eq:combine}. The set of hypotheses is $\Theta=\{1,2,3\}$, where we assume the true hypothesis is $\thetatrue=1$. We consider that all agents possess the same family of Gaussian likelihood functions with same variance and distinct means, denoted by $L(\xi|\theta)$ for $\theta\in\Theta$, which are illustrated in Fig.~\ref{fig:ex_likelihood1}. 

Since the likelihood functions are the same across all agents, the Perron eigenvector does not play a role in the convergence behavior, and only the following two quantities of interest will determine the behavior of all agents:
\begin{equation}
d(\thetashare)~~\text{ and }~~d(\thetasharec),
\end{equation}
which, in the considered example, are the same across all agents, and which quantify, respectively, the KL divergence between the likelihood of the true hypothesis and hypothesis of interest $\thetashare$, and the KL divergence between the likelihood of the true hypothesis and the fictitious likelihood of the complementary hypothesis $\thetasharec$--- see \eqref{eq:likelihoodthetac}.

Consider first that the hypothesis of interest is chosen to be $\thetashare=2$. We see in Fig.~\ref{fig:ex_likelihood2} that the likelihood relative to the transmitted hypothesis is closer to the true likelihood in comparison with the likelihood relative to the non-transmitted hypothesis, i.e.,
\begin{equation}
d(\thetashare)<d(\thetasharec),
\end{equation}
which implies that condition 2) of Theorem~\ref{the:belcoll} is satisfied, and all agents are fooled into believing that $\thetashare$ is the true state. This behavior is confirmed by the experiment shown in the righmost panel of Fig.~\ref{fig:ex_likelihood2} for agent $1$.

When the hypothesis of interest is chosen as $\thetashare=3$, Fig.~\ref{fig:ex_likelihood3} shows that the likelihood relative to the transmitted hypothesis is farther from the true likelihood in comparison with the likelihood relative to the non-transmitted hypothesis, i.e.,
\begin{equation}
d(\thetashare)>d(\thetasharec),
\end{equation}
and agents can properly distinguish the transmitted hypothesis as being false, as seen in case 1) of Theorem~\ref{the:belcoll}. Therefore agents are able to discount hypothesis $\thetashare$, as shown in the belief evolution in the rightmost panel of Fig.~\ref{fig:ex_likelihood3}.

\subsubsection{Learning and Mislearning with Self-Awareness}
Let us comment on the result for the algorithm with self-awareness in the case $\thetashare\neq\thetatrue$. 
The addition of a self-awareness term is expected to improve the learning performance, and this behavior will be examined in the forthcoming section. However, as already noticed, the asymmetry introduced in the learning algorithm by the self-awareness term makes the theoretical analysis more complicated. For example, different from the case without self-awareness, the learning/mislearning bounds are not tight, and, as far as we can tell, do not suggest a neat physical interpretation of the learning/mislearning behavior. We notice furthermore that the RHS of condition 2) in Theorem~\ref{the:altsa}, can in principle be negative, particularly when the self-awareness coefficients approach $1$. In this case, the mislearning condition is never satisfied, and simulation results, detailed in the next section, suggest that higher self-weights can mitigate mislearning.

\begin{figure}[tb]
\begin{center}
\includegraphics[width=3.5in]{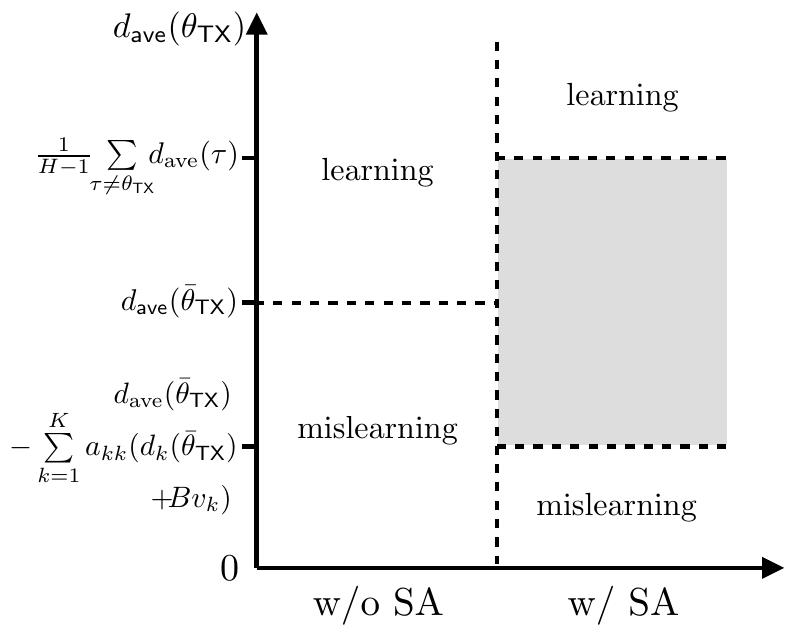}
\end{center}
\caption{Learning regions for $d_{\sf ave}(\thetashare)$ for the partial information algorithm without and with self awareness (denoted by ``w/o SA" and ``w/ SA" respectively) when $\thetashare\neq \thetatrue$. }\label{fig:regions}
\end{figure}

In summary, Theorems \ref{the:belcoll} and~\ref{the:altsa} show, for both partial information approaches, that when $\thetashare\neq\thetatrue$ there exist regions of $d_{\sf ave}(\thetashare)$ for which respectively truth learning and mislearning occur. These regions are illustrated in Fig.~\ref{fig:regions}. As a general comment applying to both algorithms, we see that if the transmitted hypothesis is more easily confounded with the true one (small $d_{\textnormal{ave}}(\thetashare)$) we have mislearning, while the converse behavior occurs for relatively high values of $d_{\textnormal{ave}}(\thetashare)$. However, a difference emerges between the results available from the two theorems.
For the algorithm without self-awareness, we can determine the learning/mislearning behavior for any value of $d_{\sf {ave}}(\thetashare)$, whereas for the algorithm with self-awareness, we cannot determine the behavior whenever $d_{\sf ave}(\thetashare)$ is found in the gray area of Fig.~\ref{fig:regions}. 

Exploiting the structure of the lower boundary in Fig.~\ref{fig:regions}, we can examine how this boundary is influenced by the self-weights $a_{kk}$. If the value of one or more self-terms decreases (i.e., if self-awareness decreases) the lower boundary moves  upward, and the region where mislearning occurs becomes wider, eventually approaching the threshold $d_{\textnormal{ave}}(\thetasharec)$ pertaining to the algorithm without self-awareness when the self-terms vanish. Conversely, if $a_{kk}$ increases the lower boundary moves downward. This implies that the gray area becomes wider, i.e., the region where we are sure to mislearn reduces. On the other hand, a wider gray area leaves open the possibility that correct learning occurs over an ampler range of cases. We will get confirmation of this behavior in the forthcoming section.
 
	\subsection{Comparative Discussion on Main Assumptions}\label{sec:assumpt}
As seen in Section~\ref{sec:sl}, traditional social learning requires global identifiability, i.e., for every $\theta\neq\thetatrue$, at least one agent should be able to distinguish $\theta$ from $\thetatrue$ (Assumption~\ref{as:glob}). In comparison, Assumptions~\ref{as:clear} and~\ref{as:strongclear} require the existence of one agent whose true likelihood is not equal to convex combinations of the other likelihoods, which are in some sense representative of the ``alternative'' w.r.t. the transmitted hypothesis. 
The situation that one likelihood is a convex combination of the other likelihoods is often an unlikely situation (for example, if we have Gaussian or exponential likelihoods, a convex combination thereof is not Gaussian or exponential). 
In summary, Assumptions~\ref{as:clear} and~\ref{as:strongclear} can be satisfied even if global identifiability is violated.
For example, if $\theta^{\star}$ is indistinguishable from $\thetatrue$ at all agents, our results imply that when $\thetashare=\thetatrue$ we can still guess the right hypothesis.
This might appear strange in view of traditional social learning, however we must not forget that the problem of truth learning contemplates also the case $\thetashare\neq\thetatrue$. In the latter case, the impact of indistinguishability becomes more relevant, since the partial information strategies learn well provided that condition 1) in Theorem~\ref{the:belcoll} (without self-awareness) or condition 1) in Theorem~\ref{the:altsa} (with self-awareness) holds.
Examining \eqref{eq:th3} and \eqref{eq:th4}, we see that one necessary condition for them to hold is that $d_{\textnormal{ave}}(\thetashare)>0$, which implies that some agent must be able to distinguish $\thetashare$ from $\thetatrue$. 
In particular, if we require truth learning for all $\thetashare\neq\thetatrue$, we need $d_{\textnormal{ave}}(\thetashare)>0$ for all $\thetashare\neq\thetatrue$, i.e., at least global identifiability is required. In summary, in the truth sharing regime, conditions for learning are weaker than in traditional social learning, whereas in the regime with $\thetashare\neq\theta_0$, global identifiability is necessary but not sufficient to guarantee truth learning. In other words, with partial information the agents concentrate their beliefs on $\thetashare$ more often than desired, i.e., not only when $\theta_{\sf TX}=\theta_0$, but also in some cases when $\theta_{\sf TX}\neq\theta_0$. 

\subsection{Main Issues in Social Learning with Partial Information}
In summary, the main questions to be answered in social learning with partial information sharing are overall ones like:
\begin{enumerate}
\item
In which instances the agents learn regardless of the true state?
\item
When agents mislearn, how does this happen?
\end{enumerate}
The answer to question $1$ is provided by Theorems~\ref{the:truthpartial}--\ref{the:altsa}. In particular, since Theorems~\ref{the:truthpartial} and~\ref{the:selfaware} reveal that, when $\thetashare=\thetatrue$, truth learning is guaranteed, both with and without self-awareness, the answer to question $1$ is contained in Eqs.~\eqref{eq:th3} and \eqref{eq:th4}, which provide conditions under which truth learning takes place regardless of the transmitted hypothesis.  Likewise, the answer to question $2$ is provided by \eqref{eq:th3a}  and \eqref{eq:th4a}, which in particular specify that when an agent mislearns, it gives full credit to the transmitted (wrong) hypothesis. 

\section{Simulation Results}
\begin{figure}[htb]
	\centering
	\includegraphics[width=3in]{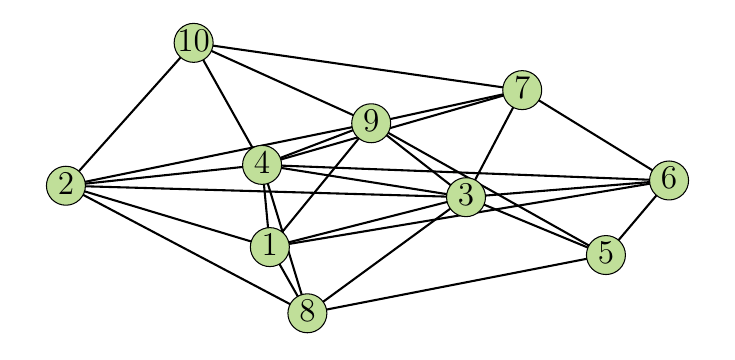}
	\caption{Strongly-connected network topology with $K = 10$ agents.}
	\label{fig:networksim}
\end{figure}

We now proceed to illustrate the results seen in Theorems~\ref{the:truthpartial}--\ref{the:altsa}. To do so, we set up an inference problem with ten hypotheses, i.e., $\Theta=\{1,2,\dots, 10\}$, from which $\thetatrue=1$ is the true state of nature.  We consider a strongly-connected network of $10$ agents, whose topology can be seen in Fig.~\ref{fig:networksim}, designed so that all agents have self-loops. 

Besides, the adjacency matrix is designed to be left-stochastic using a parametrized averaging rule \cite{sayed2014adaptation}:
\begin{equation}
a_{\ell k}=\begin{cases}
\lambda, & \text{ if }\ell=k,\\
(1-\lambda)/n_k,&\text{ if }\ell\neq k\text{ and }\ell\in\mathcal{N}_k,\\
0,& \text{ otherwise}.
\end{cases}\label{eq:combmatrix}
\end{equation}
where $n_\ell$ is the degree of node (agent) $\ell$, excluding node $\ell$ itself. 
Each agent is trying to determine whether some hypothesis $\thetashare\in\Theta$ corresponds to the true state of nature, by exchanging among neighbors partial information regarding the hypothesis of interest. In the following we consider two inference problems, one with continuous observations, the other with discrete observations.
\subsection{Continuous Observations}\label{sec:gaussiansim}
The first example considers a family of unit-variance Gaussian likelihood functions given by:
\begin{equation}
f_n(\xi)=\frac{1}{\sqrt{2\pi }}\exp\left\{-\frac{(\xi-0.5(n-1))^2}{2}\right\},
\end{equation}
for $n=1,2,\dots, 10$. From the above family of distributions, we will design the likelihood models for each agent in the following manner in order to guarantee that the inference problem is globally identifiable (see Assumption~\ref{as:glob}). To make the problem more challenging, we consider the following identifiability limitations. For each agent $k=1,2,\dots, 10$,
	\begin{equation}
	L_k(\xi|\theta)=\begin{cases}
	f_1(\xi), &\text{ for }\theta\leq k,\\
	f_\theta(\xi), &\text{ for }\theta>k. 
	\end{cases}\label{eq:likes}
	\end{equation}
	In this case, only agent 1 is able to solve the inference problem alone, that is, the indistinguishable set of hypotheses satisfies:
	\begin{equation}
	|\Theta_k|>1,\text{ for }k=2,\dots,10.
	\end{equation}
Under the aforementioned setup, we now examine both the partial information algorithm proposed in \eqref{eq:Bayesupdate}--\eqref{eq:combine} and the algorithm with self-awareness in \eqref{eq:Bayesupdate}, \eqref{eq:justone} and \eqref{eq:combsa}. We also wish to compare the performance of both algorithms with the performance of the traditional social learning algorithm (seen in \eqref{eq:bayesup}--\eqref{eq:combsl}), in which the agents share all elements of the belief vector.

\begin{figure}[tb]
	\centering
	\subfloat[Self-awareness parameter $\lambda =0.7$.	\label{fig:partial_07}]{\includegraphics[width=3.5in]{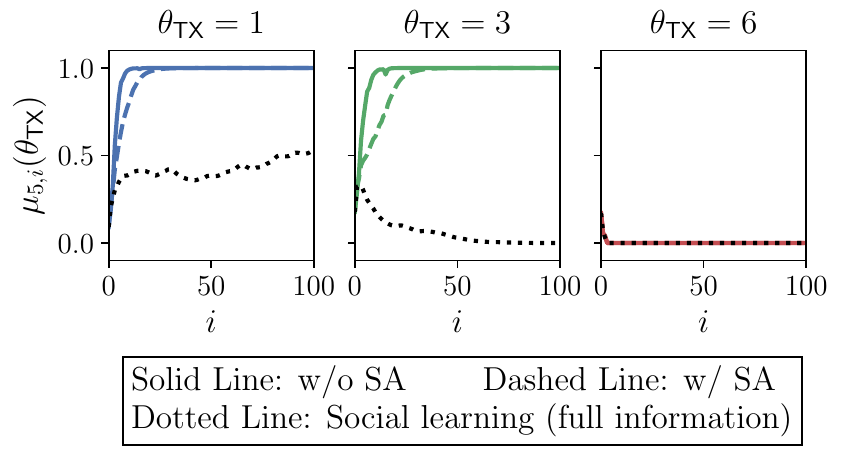}}%

	\subfloat[Self-awareness parameter $\lambda =0.95$.	\label{fig:partial_095}]{\includegraphics[width=3.5in]{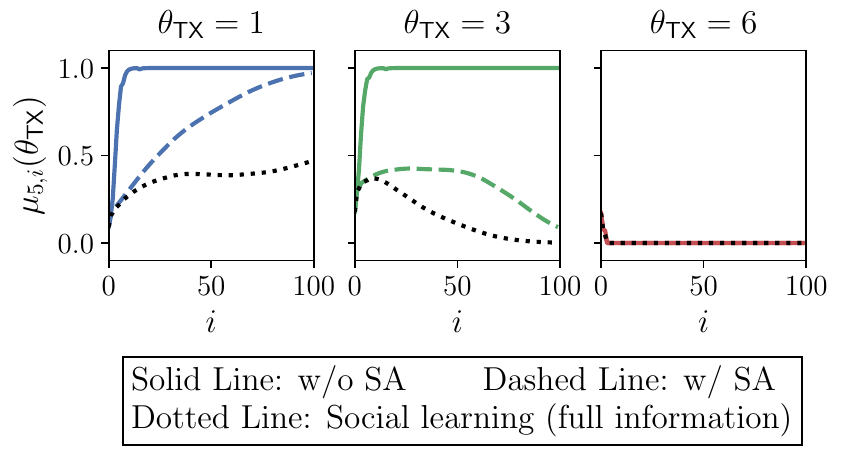}}%
	\caption{Convergence of the belief component regarding different transmitted hypotheses for agent $5$, where $\thetatrue=1$.}
	\label{fig:partial}
\end{figure}

At first, we consider that the combination matrix is parameterized according to \eqref{eq:combmatrix} with $\lambda=0.7$. In Fig.~\ref{fig:partial_07} we can see the evolution of belief at agent $5$ (similar behavior is observed for the other agents) for each different hypothesis of interest $\thetashare$. Colorful solid and dashed lines refer to the partial information algorithm without and with self-awareness, respectively. Black dotted lines refer to traditional social learning.

We start by examining the behavior of the algorithm under truth sharing, i.e., when $\thetashare=\thetatrue=1$ (leftmost panel in Fig.~\ref{fig:partial_07}). We see that all social learning algorithms are able to learn the true hypothesis, as predicted by Theorems~\ref{the:truthpartial} and~\ref{the:selfaware} for the partial information algorithms, and by the existing results on traditional social learning. We switch to the case $\thetashare\neq\thetatrue$ (middle and rightmost panel in Fig.~\ref{fig:partial_07}). As expected, traditional social learning learns well. The partial information algorithms behave instead in accordance with Theorems~\ref{the:belcoll} and~\ref{the:altsa}: when the hypothesis of interest is sufficiently ``close" to the true one, which is the case for $\thetashare=3$ (middle panel), the agent mistakenly learns that $\thetashare$ is the true hypothesis. Conversely, when the hypothesis of interest is far enough from the true one, which is the case for $\thetashare=6$ (rightmost panel), the agent learns well.

It is interesting to see what happens when all agents give more weight to their individual information by increasing the self-awareness parameter, setting it to parameter $\lambda$. In Fig.~\ref{fig:partial_095} we consider the case $\lambda=0.95$. The algorithm with self-awareness is now able to learn the truth for any of the three transmitted hypothesis, and its convergence curve is now closer to the curve of the traditional social learning algorithm. In a nutshell, concentrating the weights of the combination matrix $A$ around the self-loops entails a decrease in cooperation and hence a slower convergence. It also mitigates the effect of partial information received from neighbors, allowing for truth learning in all three cases. 

Another interesting phenomenon emerging from the simulations pertains to the learning rate. In the considered example, the algorithm without self-awareness can be faster\footnote{We have also noted that, for very small values of $\lambda$, it is possible for this convergence to be slightly slower instead.} than that with self-awareness, which can, in turn, be faster than traditional social learning. This can be counterintuitive, since one could expect that traditional social learning is the best one. However, in making this observation one should not forget the inherent trade-off of decision systems. Think of a decision system that always chooses $\thetashare$. This system learns {\em instantaneously} when $\thetashare=\thetatrue$, but fails invariably in the other cases.
In other words, the superiority of traditional social learning resides in the fact that it {\em always} allows correct learning. In contrast, the algorithms with partial information can learn faster {\em when they learn well}, but they can fail. Likewise, the fact that the algorithm without self-awareness can be faster than the algorithm with self-awareness when $\thetashare=\thetatrue$, is justified by the fact that the latter algorithm can perform better when $\thetashare\neq\thetatrue$.

\subsection{Discrete Observations}
\begin{figure}[htb]
	\centering
	\includegraphics[width=3.4in]{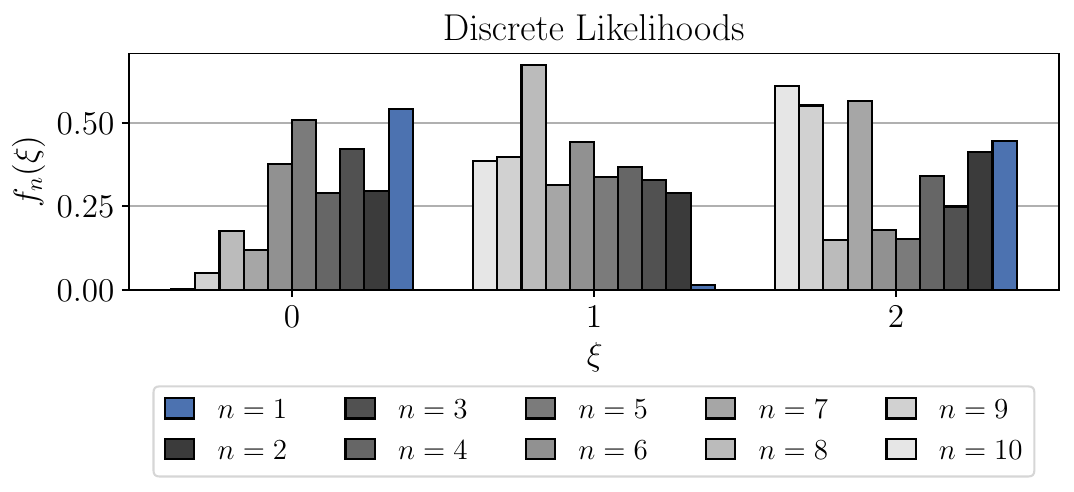}
	\caption{Family of discrete likelihood functions.}
	\label{fig:discretelike}
\end{figure}

Consider the same network topology seen in Fig.~\ref{fig:networksim} and combination matrix in \eqref{eq:combmatrix}. We now consider a family of discrete probability mass functions given by $f_n(\xi)$ for $n=1,2,\dots, 10$, defined over a space of signals $ \mathcal{X}\triangleq\{0,1,2\}$, which can be seen in Fig.~\ref{fig:discretelike}. From these distributions, we choose the likelihood models for each agent following the same identifiability constraints specified in \eqref{eq:likes}.
In this example, the true state is $\thetatrue=1$. 
According to \eqref{eq:likes}, we have $L_k(\xi|\thetatrue)=f_1(\xi)$ for all $k$, that is, the true distribution is equal to $f_1$ for all agents. In Fig.~\ref{fig:discretelike}, we highlight in blue the distribution $f_1$.

\begin{figure}[htb]
	\centering
	\subfloat[Self-awareness parameter $\lambda =0.7$.	\label{fig:partial_07d}]{\includegraphics[width=3.5in]{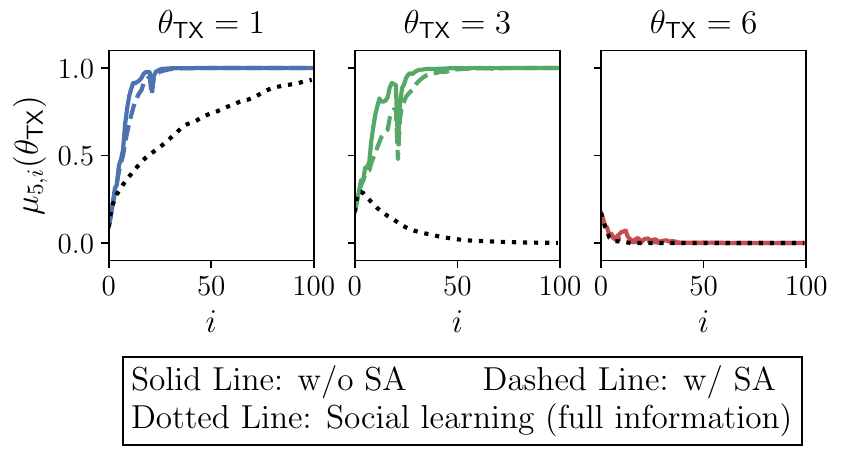}}%

	\subfloat[Self-awareness parameter $\lambda =0.95$.	\label{fig:partial_095d}]{\includegraphics[width=3.5in]{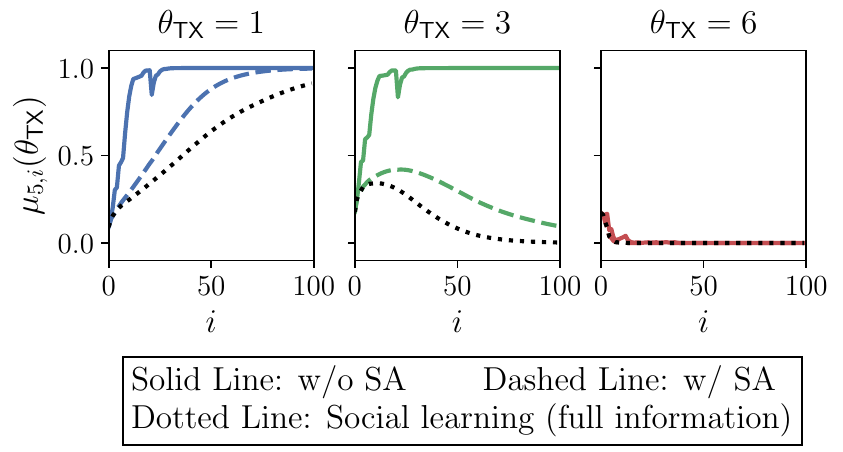}}%
	\caption{Convergence of the belief component regarding different transmitted hypotheses for agent $5$, where $\thetatrue=1$.}
	\label{fig:partiald}
\end{figure}

At first, we consider a self-awareness parameter $\lambda = 0.7$. We wish to compare the two partial information approaches for different transmitted hypotheses and the traditional social learning strategy with full information sharing. We can see in Fig.~\ref{fig:partial_07d} the evolution of belief at agent 5 for each transmitted hypothesis $\thetashare\in\Theta$ (similar behavior is observed for the other agents). As in the previous example, due to the likelihood functions setup, when $\thetashare=3$, the true hypothesis and $\thetashare=3$ are confounded by the algorithm with partial information, and the agent mislearns.

However, when the self-awareness parameter increases to $\lambda =0.95$, again a switch in the convergence behavior happens as can be seen in Fig.~\ref{fig:partial_095d} for $\thetashare=3$. As the agents are more self-aware, they are able to make correct decisions for all scenarios of $\thetashare$. 

Different from the previous example with Gaussian likelihoods, in the present example with discrete likelihoods Assumption~\ref{as:boundedlik} holds. This implies that we can exploit the lower boundary in Fig.~\ref{fig:regions} corresponding to the algorithm with self-awareness. Examining this lower boundary, we see that as  self-awareness grows (i.e., as $\lambda$ grows), the mislearning region shrinks and gives place to a gray area, where either learning or mislearning could possibly occur. We recall that getting a wider gray region does not allow to conclude that the algorithm with self-awareness would learn inside this region. However, a wider gray area reduces the region where we would be {\em sure to observe mislearning}. As a matter of fact, in the specific example we are dealing with, self-awareness can be used to tune the learning behavior in the case $\thetashare = 2$, and bring the network from mislearning to full learning.

\section{Concluding Remarks}
In this work, we introduced two approaches for taking into account partial information within a social learning framework, where agents communicate their belief about a single hypothesis of interest. We have proposed and examined in detail two social learning strategies aimed at operating under partial information. In the first strategy, agents consider only partial beliefs. In the second strategy, each individual agent becomes {\em self-aware}, in the sense that it exploits its own {\em full} belief (being still forced to use {\em partial} beliefs from its neighbors). We established the following main trends. While the traditional social learning algorithms that leverage full belief sharing are always able to learn correctly the true hypothesis, a richer behavior characterizes social learning under partial information. Both social learning algorithms with partial information proposed in this work learn correctly when the hypothesis of interest is the true hypothesis. When the transmitted hypothesis is false, however, mislearning can occur. Moreover, we showed that there are cases where the algorithm without self-awareness mislearns, while the algorithm with self-awareness can be led to the right conclusion by increasing the self-weights in the combination matrix.

Several important questions remain open. For example, the results we obtain for the algorithm with self-awareness lead to an undetermined region (the gray area in Fig.~\ref{fig:regions}) where we are not in a position to state whether learning or mislearning occurs. Filling this gap would constitute an important advance. Another extension is to characterize the beliefs in the non-asymptotic regime inspired by works~\cite{shahrampour2015distributed,nedic2017fast} and use large deviations analysis to characterize the probability of beliefs deviating from their asymptotic limit as performed in~\cite{lalitha2018social} for traditional social learning.

One future direction of research would be to consider partial information within a continuous version of social learning, i.e., for estimation/regression problems where the parameter space is continuous, and the concept of ``partial information'' under this setting will need to be properly formalized. Another interesting extension pertains to the design of an {\em adaptive} strategy where the agents can still share the belief on a single hypothesis, but this hypothesis can change over time, for example, following a Markov chain model.  It would be particularly interesting to formulate the social learning problem inspired by the existing literature on hidden-Markov-model filtering~\cite{Krishnamurthy}, and determine whether the occurrence of mislearning can be prevented in this case. 

Finally, another extension would be to consider a MAP-inspired rule where, at each instant, the hypothesis shared by each agent is the one corresponding to the maximum belief component, i.e., replacing \eqref{eq:justone} by:
\begin{equation}
\widehat{\bm{\psi}}_{k,i}(\theta)=\begin{cases}
\bm{\psi}_{k,i}(\bm{\theta}_{k,i}^{\sf max} ), &\text{ if }\theta = \bm{\theta}_{k,i}^{\sf max}\\
\frac{1}{H-1}(1-\bm{\psi}_{k,i}(\bm{\theta}_{k,i}^{\sf max} )), &\text{ otherwise, }
\end{cases}\label{eq:maxtheta}
\end{equation}
with
\begin{equation}
\bm{\theta}_{k,i}^{\sf max} = \arg\max_{\theta\in\Theta}\bm{\psi}_{k,i}(\theta),
\end{equation} 
where the change of the transmitted hypothesis depends on the learning process itself.

\begin{figure}[ht]
	\centering
	\includegraphics[width=3.4in]{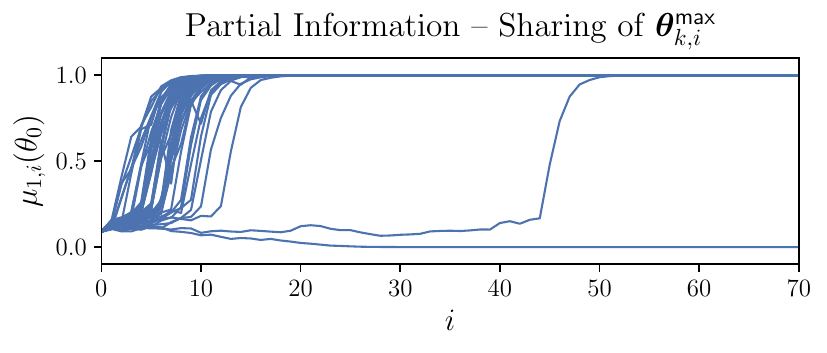}
	\caption{Alternative partial information strategy where the transmitted hypothesis at each instant is the most relevant one.}\label{fig:maxtheta}
\end{figure}

In Fig.~\ref{fig:maxtheta}, we see the evolution of the true belief component relative to agent 1 over time using the alternative approach in \eqref{eq:maxtheta}. The simulation takes place under the same network setup and family of Gaussian likelihoods described in section~\ref{sec:gaussiansim},  and the algorithm was run for 50 independent realizations of the signals observed by agents. 

We can see that the agent is able to learn the true hypothesis in 49 out of the 50 simulations. Future research efforts should aim at characterizing the probability of correct learning as a function of the network structure and data setup.


\appendices
\section{Proof of Proposition~\ref{prop:binary}}\label{ap:bin}
Let the belief vector $\bm{\mu}_{k,i}$ be split into two components for every agent $k$: $\bm{\mu}_{k,i}(\thetashare)$ and $\bm{\mu}_{k,i}(\thetasharec)$, the latter defined as
\begin{equation}
\bm{\mu}_{k,i}(\thetasharec)=\sum_{\tau\neq\thetashare}{\bm{\mu}_{k,i}(\tau)}. \label{eq:muthetac}
\end{equation}
Similarly, for the intermediate belief vector $\bm{\psi}_{k, i}$, we define:
\begin{equation}
\bm{\psi}_{k,i}(\thetasharec)=\sum_{\tau\neq\thetashare}{\bm{\psi}_{k,i}(\tau)}.\label{eq:psithetac}
\end{equation}
Remember, from \eqref{eq:1}, that all non-transmitted components of $\bm{\mu}_{k,i}$ evolve equally according to:
\begin{align}
\bm{\mu}_{k,i}(\tau)=\frac{\bm{\mu}_{k,i}(\thetasharec)}{H-1}
\end{align}
for any $\tau\neq\thetashare$. Replace \eqref{eq:Bayesupdate} into \eqref{eq:psithetac}:
\begin{align}
&\bm{\psi}_{k,i}(\thetasharec)=\frac{\sum\limits_{\tau\neq\thetashare}\bm{\mu}_{k,i-1}(\tau)L_{k}(\bm{\xi}_{k,i}|\tau)}{\sum\limits_{\theta'\in\Theta}\hspace{-4pt}\bm{\mu}_{k,i-1}(\theta')L_{k}(\bm{\xi}_{k,i}|\theta')}\nonumber\\\quad
&\stackrel{\text{(a)}}{=}\frac{\sum\limits_{\tau\neq\thetashare}\frac{\bm{\mu}_{k,i-1}(\thetasharec)}{H-1}L_{k}(\bm{\xi}_{k,i}|\tau)}{\bm{\mu}_{k,i-1}(\thetashare)L_{k}(\bm{\xi}_{k,i}|\thetashare)+\hspace{-4pt}\sum\limits_{\theta'\neq\thetashare}\hspace{-4pt}\frac{\bm{\mu}_{k,i-1}(\thetasharec)}{H-1}L_{k}(\bm{\xi}_{k,i}|\theta')}\nonumber\\
&\stackrel{\text{(b)}}{=}\frac{\bm{\mu}_{k,i-1}(\thetasharec)L_{k}(\bm{\xi}_{k,i}|\thetasharec)}{\bm{\mu}_{k,i-1}(\thetashare)L_{k}(\bm{\xi}_{k,i}|\thetashare)+\bm{\mu}_{k,i-1}(\thetasharec)L_{k}(\bm{\xi}_{k,i}|\thetasharec)},\label{eq:propbinnary}
\end{align}
where in (a) the non-transmitted components are replaced by \eqref{eq:1}, and in (b) the likelihood function corresponding to the complementary hypothesis $\thetasharec$ is replaced by \eqref{eq:likelihoodthetac}. Note that \eqref{eq:propbinnary} corresponds to the Bayesian update for the complementary hypothesis $\thetasharec$ under the set of two hypotheses $\Theta_b=\{\thetashare,\thetasharec\}$, and with the fictitious likelihood in \eqref{eq:likelihoodthetac}.

Similarly to the belief vectors $\bm{\mu}_{k,i}$, we now show that the non-transmitted components of the modified belief $\widehat{\bm{\psi}}_{k,i}$ evolve equally over time. From \eqref{eq:justone} and \eqref{eq:psithetac} we have, for any non-transmitted hypothesis $\tau\neq\thetashare$:
\begin{equation}
\widehat{\bm{\psi}}_{\ell,i}(\tau)=\frac{\bm{\psi}_{\ell,i}(\thetasharec)}{H-1}. \label{eq:psihatthetac}
\end{equation}
Consider now the combination step. Replacing \eqref{eq:combine} into \eqref{eq:muthetac} results in:
\begin{align}
&\bm{\mu}_{k,i}(\thetasharec)=
\frac{\sum\limits_{\tau\neq\thetashare}\hspace{-4pt}
	\exp\left(
	\sum\limits_{\ell=1}^K a_{\ell k} \log \widehat{\bm{\psi}}_{\ell,i}(\tau)
	\right)
}
{
	\sum\limits_{\theta^{\prime}\in\Theta}\hspace{-4pt} \exp\left(\sum\limits_{\ell=1}^K a_{\ell k} \log \widehat{\bm{\psi}}_{\ell,i}(\theta^{\prime})
	\right)
}\nonumber\\
&\stackrel{\text{(a)}}{=}
\frac{
	\sum\limits_{\tau\neq\thetashare}\hspace{-5pt}\exp\hspace{-2pt}\left(
	\sum\limits_{\ell=1}^K a_{\ell k} \log \frac{\bm{\psi}_{\ell,i}(\thetasharec)}{H-1}
	\right)
}
{
	\exp\hspace{-2pt}\left(\sum\limits_{\ell=1}^K \hspace{-1pt}a_{\ell k} \log \bm{\psi}_{\ell,i}(\thetashare)
	\hspace{-2pt}\right)\hspace{-2pt}+\hspace{-7pt}\sum\limits_{\theta^{\prime}\neq\thetashare}\hspace{-6pt}\exp\hspace{-2pt}\left(\sum\limits_{\ell=1}^K \hspace{-1pt}a_{\ell k} \log \frac{\bm{\psi}_{\ell,i}(\thetasharec)}{H-1}
	\hspace{-2pt}\right)
}\nonumber\\
&=
\frac{
	\exp\hspace{-2pt}\left(
	\sum\limits_{\ell=1}^K a_{\ell k} \log \bm{\psi}_{\ell,i}(\thetasharec)
	\right)
}
{
	\exp\hspace{-2pt}\left(\sum\limits_{\ell=1}^K \hspace{-1pt}a_{\ell k} \log \bm{\psi}_{\ell,i}(\thetashare)
	\hspace{-2pt}\right)\hspace{-2pt}+\hspace{-1pt}\exp\hspace{-2pt}\left(\sum\limits_{\ell=1}^K \hspace{-1pt}a_{\ell k} \log \bm{\psi}_{\ell,i}(\thetasharec)
	\hspace{-2pt}\right)
},\label{eq:combbinary}
\end{align}
where in (a), the non-transmitted components of the modified belief vector are replaced by $\eqref{eq:psihatthetac}$. Note that \eqref{eq:combbinary} is equivalent to writing a (log-linear) combination step for the binary set of hypotheses $\Theta_{b}=\{\thetashare,\thetasharec\}$.

Since $\bm{\mu}_{k,i}(\thetashare)$ and $\bm{\mu}_{k,i}(\thetasharec)$ (and similarly $\bm{\psi}_{k,i}(\thetashare)$ and $\bm{\psi}_{k,i}(\thetasharec)$) sum up to one, we have that for every $\theta\in\Theta_b$, the partial information algorithm enunciated in \eqref{eq:Bayesupdate}--\eqref{eq:combine} behaves in the same manner as if each agent $k$ performed the two steps in the traditional social learning algorithm seen in \eqref{eq:bayesup}--\eqref{eq:combsl} for the two hypotheses in $\Theta_b$, which agrees with the claim in Proposition~\ref{prop:binary}.
\section{Proof of Theorem~\ref{the:truthpartial}}
\label{ap:truthpartial}
We first introduce an intermediate result, where we show that for each agent the log-ratio between any non-transmitted and transmitted belief components will have an asymptotic exponential behavior. In order to avoid misunderstanding, we remark that this result is already known in social learning theory~\cite{nedic2017fast}. Nevertheless, we deem it useful to report here a proof for this result to make the article self-contained, and to make useful connections of this particular proof with other results that we prove relying on the recursive inequalities in Lemmas 8 and 9 further ahead.
\begin{lemma}[\textbf{Asymptotic Rate of Convergence}]\label{lemma:rate}
	Under Assumptions~\ref{as:integ} and~\ref{as:posit}, for all $\theta \in \Theta\setminus\{\thetashare\}$ and every agent $k=1,2,\dots,K$, we have that:
	\begin{equation}
	\frac{1}{i}\log\frac{\bm{\mu}_{k,i}(\theta)}{\bm{\mu}_{k,i}(\thetashare)}\stackrel{\textnormal{ a.s.}}{\longrightarrow} d_{\sf ave}(\thetashare)-d_{\sf ave}(\thetasharec).\label{eq:lemrate}
	\end{equation}
\end{lemma}
\begin{IEEEproof}
	We know from \eqref{eq:allmuequal} that for any non-transmitted hypotheses $\tau,\theta\neq\thetashare$:
	\begin{align}
	\bm{\mu}_{k,i}(\tau)=\bm{\mu}_{k,i}(\theta).\label{eq:equalmu}
	\end{align}
	Moreover, from \eqref{eq:justone}:
	\begin{align}
	\log \frac{\widehat{\bm{\psi}}_{\ell,i}(\theta)}{\widehat{\bm{\psi}}_{\ell,i}(\thetashare)}&=\log \frac{\left(1-\bm{\psi}_{\ell,i}(\thetashare)\right)/(H-1)}{\bm{\psi}_{\ell,i}(\thetashare)}
	\nonumber\\&=\log \frac{\sum_{\tau\neq\thetashare}\bm{\psi}_{\ell,i}(\tau)/(H-1)}{\bm{\psi}_{\ell,i}(\thetashare)}.\label{eq:lem1a}
	\end{align}
	Substituting \eqref{eq:Bayesupdate} into \eqref{eq:lem1a}, we obtain:
	\begin{equation}
	\log \frac{\widehat{\bm{\psi}}_{\ell,i}(\theta)}{\widehat{\bm{\psi}}_{\ell,i}(\thetashare)}=\log \frac{\sum_{\tau\neq\thetashare}\bm{\mu}_{\ell,i-1}(\tau)L_\ell(\bm{\xi}_{\ell,i}|\tau)/(H-1)}{\bm{\mu}_{\ell,i-1}(\thetashare)L_\ell(\bm{\xi}_{\ell,i}|\thetashare)}.\label{eq:lem1b}
	\end{equation}
	Using \eqref{eq:equalmu} in \eqref{eq:lem1b} yields:
	\begin{align}
	&\log \frac{\widehat{\bm{\psi}}_{\ell,i}(\theta)}{\widehat{\bm{\psi}}_{\ell,i}(\thetashare)}=\log \frac{\bm{\mu}_{\ell,i-1}(\theta)\sum_{\tau\neq\thetashare}L_\ell(\bm{\xi}_{\ell,i}|\tau)/(H-1) }{\bm{\mu}_{\ell,i-1}(\thetashare)L_\ell(\bm{\xi}_{\ell,i}|\thetashare)}\nonumber\\
	&=
	\log\frac{\bm{\mu}_{\ell,i-1}(\theta)}{\bm{\mu}_{\ell,i-1}(\thetashare)}
	+
	\log\frac{
		\sum_{\tau\neq\thetashare}L_\ell(\bm{\xi}_{\ell,i}|\tau)/(H-1)}
	{L_{\ell}(\bm{\xi}_{\ell,i} | \thetashare)}\nonumber\\
	&\stackrel{\text{(a)}}{=} \log\frac{\bm{\mu}_{\ell,i-1}(\theta)}{\bm{\mu}_{\ell,i-1}(\thetashare)}
	+
	\log\frac{
		L_\ell(\bm{\xi}_{\ell,i}|\thetasharec)}
	{L_{\ell}(\bm{\xi}_{\ell,i} | \thetashare)},\label{eq:thetathetatxnosa}
	\end{align}
	where in (a) we used the definition of the likelihood for the non-transmitted hypotheses found in \eqref{eq:likelihoodthetac}. Using \eqref{eq:combine_2}, we obtain the following recursion: 
	\begin{align}
	\log \frac{\bm{\mu}_{k,i}(\theta)}{\bm{\mu}_{k,i}(\thetashare)} &=\sum_{\ell=1}^Ka_{\ell k}\log\frac{\bm{\mu}_{\ell,i-1}(\theta)}{\bm{\mu}_{\ell,i-1}(\thetashare)}\nonumber
	\\&+
	\sum_{\ell=1}^Ka_{\ell k}\log\frac{
		L_{\ell}(\bm{\xi}_{\ell,i}|\thetasharec)}
	{L_{\ell}(\bm{\xi}_{\ell,i} | \thetashare)}.\label{eq:rec1}
	\end{align}
	Define the vectors:
	\begin{align}
	&\bm{y}_{i}(\theta)\triangleq\mathrm{col}\left\{\log \frac{\bm{\mu}_{k,i}(\theta)}{\bm{\mu}_{k,i}(\thetashare)}\right\}^K_{k=1},\\
	&\bm{x}_{i}\triangleq A\T\mathrm{col}\left\{\log\frac{L_{k}(\bm{\xi}_{k,i}|\thetasharec)}
	{L_{k}(\bm{\xi}_{k,i} | \thetashare)}\right\}^K_{k=1},
	\end{align}
	where the $\mathrm{col}$ operator concatenates a sequence of variables into a column vector. We can then rewrite \eqref{eq:rec1} in vector form for all $\theta\in\Theta\setminus\{ \thetashare\}$:
	\begin{equation}
	\bm{y}_i(\theta)=A\T\bm{y}_{i-1}(\theta)+\bm{x}_{i}.\label{eq:recvec1}
	\end{equation}
First, note that the recursion in \eqref{eq:recvec1} takes the form of the sequence of random vectors seen in auxiliary Lemma~\ref{lem:aux1} (see Appendix~\ref{ap:aux}). Since the random vectors $\bm{x}_i$ are i.i.d. across time and have finite expectation\footnote{The i.i.d. property across time is inherited from variables $\bm{\xi}_{k,i}$ for all $k=1,2,\dots,K$. Note that for each element of $\bm{x}_i$, $\E(\bm{x}_{k, i})$ can easily be rewritten as a function of two KL divergences: 
			\begin{equation}
			\E(\bm{x}_{k, i})=\sum_{\ell =1}^Ka_{\ell k}\E\hspace{-1pt}\left(\hspace{-1pt}\log\frac{L_{\ell}(\bm{\xi}_{\ell,i}|\thetasharec)}{L_{\ell}(\bm{\xi}_{\ell,i}|\thetashare)}\hspace{-1pt}\right)\hspace{-1pt}=\hspace{-1pt}\sum_{\ell =1}^K\hspace{-1pt}a_{\ell k}\hspace{-1pt}\left(d_\ell(\thetashare)-d_\ell(\thetasharec)\right)\hspace{-1pt}.
			\end{equation}
			The first term on the RHS is finite from Assumption~\ref{as:integ}, whereas the second term is finite from Assumption~\ref{as:integ} and the inequality in \eqref{eq:klineq}.}, Property~\ref{prop:int} (also found in Appendix~\ref{ap:aux}) can be applied and shows that $\bm{x}_i$ satisfies the conditions \eqref{eq:aslim1} and \eqref{eq:aslim2} required by Lemma~\ref{lem:aux1}. Particularly, in view of Property~\ref{prop:int}, the vector $\bar{x}$ takes the form of the expectation vector $\E(\bm{x}_i)$. Since $A$ is left-stochastic, all conditions in Lemma~\ref{lem:aux1} are satisfied and we can therefore apply its result as follows. For each $\theta\neq \thetashare$, we have that
	\begin{align}
	&\frac{1}{i}\log\frac{\bm{\mu}_{k,i}(\theta)}{\bm{\mu}_{k,i}(\thetashare)}\stackrel{\textnormal{a.s.}}{\longrightarrow}
	\sum_{k=1}^Kv_{k}\sum_{\ell=1}^Ka_{\ell k}\mathbb{E}\left(\log\frac{
		L_{\ell}(\thetasharec)}
	{L_{\ell}(\thetashare)}\right)\nonumber\\
	&= \sum_{\ell=1}^Kv_{\ell}\mathbb{E}\left(\log\frac{
		L_{\ell}(\thetatrue)}
	{L_{\ell}(\thetashare)}\right)-\sum_{\ell=1}^Kv_{\ell}\mathbb{E}\left(\log\frac{
		L_{\ell}(\thetatrue)}
	{L_{\ell}(\thetasharec)}\right)\nonumber\\
	&=d_{\sf ave}(\thetashare)-d_{\sf ave}(\thetasharec),\label{eq:limthetathetax}
	\end{align}
	where we recall that $\sum_{k=1}^K v_k a_{\ell k}=v_\ell$ since $v$ is the Perron eigenvector.
\end{IEEEproof}
\begin{IEEEproof}[Proof of Theorem~\ref{the:truthpartial}]
Note that the RHS of \eqref{eq:lemrate} represents a key quantity in the algorithm: conditionally on its sign, we have that the log-ratio of belief components on the LHS of \eqref{eq:lemrate} will increase or decrease indefinitely. 

If $\thetashare=\thetatrue$, we have that 
\begin{equation}
d_{\sf ave}(\thetatrue)=0.
\end{equation}
Under Assumption~\ref{as:clear}, there exists at least one clear-sighted agent in the network, say agent $\kappagood$, for which 
\begin{equation}
d_{\kappagood}(\thetatruec)>0.
\end{equation}
From the positivity of the Perron eigenvector, we have that
\begin{equation}
d_{\sf ave}(\thetatruec)> 0.
\end{equation}
Finally, from \eqref{eq:limthetathetax} with $\thetashare=\thetatrue$, we obtain
\begin{align}
&\frac{1}{i}\log\frac{\bm{\mu}_{k,i}(\theta)}{\bm{\mu}_{k,i}(\thetatrue)}\stackrel{\textnormal{a.s.}}{\longrightarrow}  d_{\sf ave}(\thetatrue)-d_{\sf ave}(\thetatruec)<0\nonumber\\&
\implies 
\log\frac{\bm{\mu}_{k,i}(\theta)}{\bm{\mu}_{k,i}(\thetatrue)}\stackrel{\textnormal{a.s.}}{\longrightarrow}-\infty\nonumber\\&\implies \bm{\mu}_{k,i}(\theta)\stackrel{\textnormal{a.s.}}{\longrightarrow}0,
\end{align}
which holds for all $\theta\in\Theta\setminus\{\thetashare\}$. This, in turn, implies that
\begin{equation}
\bm{\mu}_{k,i}(\thetatrue)\stackrel{\textnormal{a.s.}}{\longrightarrow}1,
\end{equation}
thus concluding the proof of Theorem~\ref{the:truthpartial}.
\end{IEEEproof}

\section{Proof of Theorem~\ref{the:selfaware}}\label{ap:selfaware}
In order to reach the conclusion in Theorem~\ref{the:selfaware}, we need first to establish some intermediate results (see Lemmas~\ref{lem:mumart},~\ref{lem:oneagentlearns} and~\ref{lem:oneall} enunciated next), which depend on auxiliary results found in Appendix~\ref{ap:alsa} (these results are stated in Lemmas~\ref{lem:unshared},~\ref{lem:discardonediscardall} and~\ref{lem:boundconv}). We resort moreover to two auxiliary lemmas (see Lemmas~\ref{lem:aux1} and~\ref{lem:scalarevolut} in Appendix~\ref{ap:aux}) which refer to statistical properties of more general recursions. 

Consider the truth sharing case, for which $\thetashare=\thetatrue$. The first key result, which can be seen in Lemma~\ref{lem:mumart} enunciated next, is that the random sequence 
\begin{equation}
\bm{m}_{i}\triangleq\sum_{k=1}^Kv_k\log \bm{\mu}_{k,i}(\thetashare)\label{eq:mi}
\end{equation}
is a submartingale. To lighten the notation we will denote the KL divergence between the true likelihood function $L_k(\xi|\thetatrue)$ and a convex combination of the likelihoods $L_k(\xi|\theta)$ by:
\begin{equation}
\delta_k(\alpha)\triangleq \E \left(\log\ \frac{L_k(\bm{\xi}_{k,i}|\thetatrue)}{\sum\limits_{\theta\in\Theta}\alpha(\theta) L_k(\bm{\xi}_{k,i}|\theta)}\right),\label{eq:klconvex}
\end{equation}
where $\alpha$ is the convex combination vector, i.e., $\alpha$ is a vector belonging to the $H-$simplex $\Delta^H$. 

We define the signal profile at each instant $i$ as $\bm{\xi}_i\triangleq \{\bm{\xi}_{1,i},\bm{\xi}_{2,i},\dots, \bm{\xi}_{K,i}\}$. We also define the {\em filtration} over the past observations as $\mathcal{F}_j$ for $j=1,2,\dots$, where $\mathcal{F}_j$ is the sub-$\sigma$-field generated by the observations up to instant $j$, namely,
	\begin{equation}
	\mathcal{F}_j\triangleq \sigma \left(\bm{\xi}_1,\bm{\xi}_2,\dots,\bm{\xi}_j\right),
	\end{equation}
	which satisfies $\mathcal{F}_1\subseteq\mathcal{F}_2\subseteq\dots\subseteq\mathcal{F}_\infty\triangleq \sigma \left(\bm{\xi}_1,\bm{\xi}_2,\dots\right)$. 
\begin{lemma}[\textbf{Submartingale Sequence}]
	\label{lem:mumart}
	Let $\thetashare=\thetatrue$, and consider the random sequence $\{\bm{m}_i\}$ in \eqref{eq:mi}. This sequence has the following properties. 
	\begin{enumerate}
		\item\begin{equation}
		\E(\bm{m}_{i}|\mathcal{F}_{i-1})
		\geq \bm{m}_{i-1} +\sum_{k=1}^K v_k \delta_k(\bm{\mu}_{k,i-1}).
		\label{eq:submartingale}
		\end{equation}
		\item
		The sequence $\bm{m}_i$ is a nonpositive submartingale with respect to the process $\{\bm{\xi}_{i}\}$.
		\item
		There exists a random variable $\bm{m}_{\infty}$ such that:
		\begin{equation}
		\bm{m}_{i} \stackrel{\textnormal{a.s.}}{\longrightarrow} \bm{m}_{\infty}.
		\label{eq:mumartasconv}
		\end{equation}
		\item
		The expectation $\E(\bm{m}_i)$ converges to a finite limit.
	\end{enumerate}
\end{lemma}
\begin{IEEEproof}
Consider \eqref{eq:combsa} with $\theta=\thetashare$. In view of \eqref{eq:justone}, we have:
\begin{equation}
\bm{\mu}_{k,i}(\thetashare)=\frac{\exp\left(\sum\limits_{\ell=1}^Ka_{\ell k}\log \bm{\psi}_{\ell,i}(\thetashare) \right)}{\sum\limits_{\theta'\in\Theta}\hspace{-3pt}\exp\hspace{-2pt}\left(\hspace{-2pt}a_{kk}\log \bm{\psi}_{k,i}(\theta')\hspace{-2pt}+\hspace{-3pt}\sum\limits_{\substack{\ell=1\\\ell\neq k}}^K\hspace{-1pt}a_{\ell k}\log \widehat{\bm{\psi}}_{\ell,i}(\theta') \hspace{-2pt}\right)}.\label{eq:combsa1}
\end{equation}
To simplify the notation, we define the following auxiliary variables:
\begin{align}
\bm{\mathcal{Y}}_{k,i}(\thetashare)&\triangleq\exp\left(\sum\limits_{\ell=1}^Ka_{\ell k}\log \bm{\psi}_{\ell,i}(\thetashare) \right),\label{eq:defY}\\
\bm{\mathcal{Z}}_{k,i}(\thetasharec)&\triangleq \hspace{-4pt}\sum_{\tau\neq\thetashare}\hspace{-4pt}\exp\hspace{-2pt}\left(\hspace{-1pt}a_{kk}\log \bm{\psi}_{k,i}(\tau)+\hspace{-2pt}\sum_{\substack{\ell=1\\\ell\neq k}}^Ka_{\ell k}\log \widehat{\bm{\psi}}_{\ell,i}(\tau) \hspace{-1pt}\right),
\end{align}
from which we can rewrite the combination step in \eqref{eq:combsa1} as:
\begin{equation}
\bm{\mu}_{k,i}(\thetashare)=\frac{\bm{\mathcal{Y}}_{k,i}(\thetashare)}{\bm{\mathcal{Y}}_{k,i}(\thetashare)+\bm{\mathcal{Z}}_{k,i}(\thetasharec)}.\label{eq:the2aux1}
\end{equation}	
Using \eqref{eq:justone}, we can develop the expression for $\bm{\mathcal{Z}}_{k,i}(\thetasharec)$ as:
\begin{align}
&\bm{\mathcal{Z}}_{k,i}(\thetasharec)\nonumber\\&=
\sum_{\tau\neq \thetashare}\hspace{-4pt}
\exp\left(
\displaystyle{a_{kk}\log\bm{\psi}_{k,i}(\tau)+ \sum_{\substack{\ell=1\\\ell\neq k}}^K a_{\ell k} 
	\log \frac{1-\bm{\psi}_{\ell,i}(\thetashare)}{H-1}
}
\right)\nonumber\\
%	&=&
%	\sum_{\tau\neq \thetashare}
%	\exp\left\{
%	\displaystyle{a_{kk}\log\bm{\psi}_{k,i}(\tau)+ \sum_{\substack{\ell=1\\\ell\neq k}}^K a_{\ell k} 
%		\log [1-\bm{\psi}_{\ell,i}(\thetashare)]+(1-a_{kk})\log\frac{1}{(H-1)}
%	}
%	\right\}\\
&\stackrel{\text{(a)}}{=}
\sum_{\tau\neq \thetashare}\hspace{-4pt}
\exp\left(\log\frac{\bm{\psi}_{k,i}(\tau)^{a_{kk}}}{(H-1)^{1-a_{kk}}}
+
\displaystyle{\sum_{\substack{\ell=1\\\ell\neq k}}^K a_{\ell k} 
	\log \bm{\psi}_{\ell,i}(\thetasharec)}
\right)\nonumber\\
&=
\left[\exp\left(
\displaystyle{\sum_{\substack{\ell=1\\\ell\neq k}}^K a_{\ell k} 
	\log \bm{\psi}_{\ell,i}(\thetasharec)
}
\right)\right]
\frac{\sum_{\tau\neq \thetashare}
	\displaystyle{\bm{\psi}_{k,i}(\tau)^{a_{kk}}}}
{(H-1)^{1-a_{kk}}},
\label{eq:mainchain}
\end{align}
where in (a) we introduced $\bm{\psi}_{\ell,i}(\thetasharec)\triangleq1-\bm{\psi}_{\ell,i}(\thetashare)$.
Now, applying the sum-of-powers inequality\footnote{For $r,s\neq 0$ with $r<s$, and for positive values $x_i$, we have that~\cite{bullen2013handbook} :
	$$	\left(
	\frac{1}{n}\sum_{i=1}^n x_i^{r}\right)^{1/r}\leq \left(\frac{1}{n}\sum_{i=1}^n x_i^{s}\right)^{1/s}.$$
	In particular, with the choice $s=1$ we can write the following inequality:
	$$	\frac{1}{n^{1-r}}\sum_{i=1}^n x_i^{r}\leq 
	\left(\sum_{i=1}^n x_i\right)^r.$$
} to the rightmost term in \eqref{eq:mainchain} results in:
\begin{align}
\frac{\sum_{\tau\neq \thetashare}
	\displaystyle{\bm{\psi}_{k,i}(\tau)^{a_{kk}}}}
{(H-1)^{1-a_{kk}}}
&\leq
\left(\sum_{\tau\neq \thetashare}
\displaystyle{\bm{\psi}_{k,i}(\tau)}\right)^{a_{kk}}\nonumber\\&=
\exp\left(
a_{kk}\log\bm{\psi}_{k,i}(\thetasharec)
\right).
\label{eq:psineq}
\end{align}
Replacing \eqref{eq:psineq} to \eqref{eq:mainchain} we get:
\begin{align}
\bm{\mathcal{Z}}_{k,i}(\thetasharec)
&\leq
\exp\left(
\displaystyle{\sum_{\ell=1}^K a_{\ell k} \log\bm{\psi}_{\ell,i}(\thetasharec)}
\right)\triangleq \bm{\mathcal{Y}}_{k,i}(\thetasharec).
\label{eq:musumthetanotshare}
\end{align}
In view of \eqref{eq:musumthetanotshare}, we can lower bound the expression in \eqref{eq:the2aux1}:
\begin{equation}
\bm{\mu}_{k,i}(\thetashare)\geq \frac{\bm{\mathcal{Y}}_{k,i}(\thetashare)}{\bm{\mathcal{Y}}_{k,i}(\thetashare)+\bm{\mathcal{Y}}_{k,i}(\thetasharec)}=\frac{1}{1+\frac{\bm{\mathcal{Y}}_{k,i}(\thetasharec)}{\bm{\mathcal{Y}}_{k,i}(\thetashare)}}.\label{eq:the2aux2}
\end{equation}
Applying $\log(\cdot)$ to both sides of \eqref{eq:the2aux2}, and replacing back the definitions of $\bm{\mathcal{Y}}_{k,i}(\thetashare)$ and $\bm{\mathcal{Y}}_{k,i}(\thetasharec)$ from \eqref{eq:defY} and \eqref{eq:musumthetanotshare} respectively, we can write the following inequality:
\begin{align}
\log \bm{\mu}_{k,i}(\thetashare)&\geq \log \frac{1}{1+\exp\left(\sum\limits_{\ell=1}^Ka_{\ell k}\log \frac{\bm{\psi}_{\ell, i}(\thetasharec)}{\bm{\psi}_{\ell, i}(\thetashare)}\right)}\nonumber\\&\triangleq f\left(\sum\limits_{\ell=1}^Ka_{\ell k}\log \frac{\bm{\psi}_{\ell, i}(\thetasharec)}{\bm{\psi}_{\ell, i}(\thetashare)}\right),
\end{align}
where we defined the concave function\footnote{The concavity of the function $f(x)$ can be seen from its second derivative: $$\frac{d^2f(x)}{dx^2}=\frac{-e^x}{\left[1+e^x\right]^2}< 0,$$
	for any $x\in \mathbb{R}$.}
\begin{equation}
f(x)\triangleq\log\frac{1}{1+e^x}.
\end{equation}
Using Jensen's inequality, we have that
\begin{align}
&\log \bm{\mu}_{k,i}(\thetashare)\geq \sum_{\ell=1}^Ka_{\ell k}\log \frac{1}{1+e^{\log \frac{\bm{\psi}_{\ell, i}(\thetasharec)}{\bm{\psi}_{\ell, i}(\thetashare)}}}\nonumber\\
&=\sum_{\ell=1}^Ka_{\ell k}\log \frac{1}{1+\frac{\bm{\psi}_{\ell, i}(\thetasharec)}{\bm{\psi}_{\ell, i}(\thetashare)}}\nonumber\\
&=\sum_{\ell=1}^Ka_{\ell k}\log \frac{\bm{\psi}_{\ell, i}(\thetashare)}{\bm{\psi}_{\ell, i}(\thetashare)+\bm{\psi}_{\ell, i}(\thetasharec)}\nonumber\\
&=\sum_{\ell=1}^Ka_{\ell k}\log \bm{\psi}_{\ell, i}(\thetashare)\nonumber\\
&\stackrel{\text{(a)}}{=}
\sum_{\ell=1}^Ka_{\ell k}\log \bm{\mu}_{\ell,i-1}(\thetashare)\nonumber\\&+\sum_{\ell=1}^Ka_{\ell k}\log \frac{L_\ell(\bm{\xi}_{\ell, i}|\thetashare)}{\sum\limits_{\theta'\in\Theta}\bm{\mu}_{\ell, i-1}(\theta')L_\ell(\bm{\xi}_{\ell, i}|\theta')},\label{eq:lem4aux}
\end{align}
where in (a), we replaced $\bm{\psi}_{\ell,i}(\thetashare)$ using the Bayesian update seen in \eqref{eq:Bayesupdate}.

Taking the expectation of both sides of \eqref{eq:lem4aux} conditioned on $\mathcal{F}_{i-1}$, we have that:
\begin{align}
&\E\left(\log \bm{\mu}_{k,i}(\thetashare)\Big| \mathcal{F}_{i-1}\right)\geq
\sum_{\ell=1}^K a_{\ell k}\log \bm{\mu}_{\ell,i-1}(\thetashare)\nonumber\\
&+
\sum_{\ell=1}^K a_{\ell k}
\E
\left(\log \frac{L_\ell(\bm{\xi}_{\ell, i}|\thetashare)}{\sum\limits_{\theta'\in\Theta}\bm{\mu}_{\ell, i-1}(\theta')L_\ell(\bm{\xi}_{\ell, i}|\theta')}\Big| \mathcal{F}_{i-1}\right). 
\label{eq:mainmarteq}
\end{align}
Since the current signal profile $\bm{\xi}_{i}$ is independent from the past data vectors (and, hence, is independent from the past belief vector $\bm{\mu}_{\ell,i-1}$), we see that the second term on the RHS of \eqref{eq:mainmarteq} is the following KL divergence, as defined in \eqref{eq:klconvex} (we recall that we are considering the case $\thetashare=\thetatrue$):
\begin{equation}
\E
\left(\log \frac{L_\ell(\bm{\xi}_{\ell, i}|\thetashare)}{\sum\limits_{\theta'\in\Theta}\bm{\mu}_{\ell, i-1}(\theta')L_\ell(\bm{\xi}_{\ell, i}|\theta')}\Big| \mathcal{F}_{i-1}\right)
=
\delta_\ell(\bm{\mu}_{\ell,i-1}).
\label{eq:KLdivspec}
\end{equation}
Multiplying both sides of \eqref{eq:mainmarteq} by $v_{k}$, summing over $k$, and recalling that $\sum_{k=1}^K v_k a_{\ell k}=v_\ell$ because $v$ is the Perron eigenvector, Eqs. \eqref{eq:mainmarteq} and \eqref{eq:KLdivspec} imply part $1)$ of the lemma. 

Part $2)$ follows from part $1)$. In fact, $\bm{m}_i$ is nonpositive because $\bm{\mu}_{k,i}\leq 1$, and $\bm{m}_i$ is a submartingale because the KL divergence is nonnegative, and, hence, Eq. \eqref{eq:submartingale} implies:  
\begin{equation}
\E(\bm{m}_i | \mathcal{F}_{i-1})\geq \bm{m}_{i-1}.
\label{eq:mimiminus1}
\end{equation}
Part $3)$ follows from the martingale convergence theorem~\cite{billingsley2008probability}. 

Finally, part $4)$ follows by taking the total expectation in \eqref{eq:mimiminus1}, which yields:
\begin{equation}
0\geq \E(\bm{m}_i)\geq \E(\bm{m}_{i-1})\geq \ldots\geq m_0=\sum_{k=1}^K v_k \log\mu_{k,0}(\thetashare),
\label{eq:meanseq}
\end{equation}
which implies that the sequence of expectations is a (monotonically) convergent sequence.
\end{IEEEproof}

Using part 3) of Lemma~\ref{lem:mumart}, we can establish the following technical corollary which will be useful later in the analysis.
\begin{corollary}[\textbf{Expectation of Log-Beliefs $\bm{\psi}_{k,i}$}]
	Let $\thetashare=\thetatrue$. For all $i\geq 1$ we have that:
	\label{cor:psimart}
	\begin{equation}
	\E\left(\log\frac{1}{\bm{\psi}_{k,i}(\thetashare)}\right)
	\leq 
	\frac{1}{v_k}\,
	\sum_{\ell=1}^K v_{\ell} \log\frac{1}{\mu_{\ell,0}(\thetashare)}.
	\label{eq:publicbelTX}
	\end{equation}
\end{corollary}
\begin{IEEEproof}
	Using the Bayesian update in \eqref{eq:Bayesupdate} 
	we can write:
	\begin{align}
	&\E\left(\log\frac{1}{\bm{\psi}_{k,i}(\thetashare)}\right)\nonumber\\
	&=\E\hspace{-2pt}\left(\log\frac{1}{\bm{\mu}_{k,i-1}(\thetashare)}\right)
	-
	\E\hspace{-2pt}\left(
	\log\frac{L_{k}(\bm{\xi}_{k,i}|\thetashare)}
	{\hspace{-2pt}\sum\limits_{\theta\in\Theta}\hspace{-2pt}\bm{\mu}_{k,i-1}(\theta)L_{k}(\bm{\xi}_{k,i}|\theta)}
	\right)\nonumber\\
	&=
	\E\left(\log\frac{1}{\bm{\mu}_{k,i-1}(\thetashare)}\right)
	-
	\E\left(
	\delta_k(\bm{\mu}_{k,i-1})\right)\nonumber\\
	&\leq
	\E\left(\log\frac{1}{\bm{\mu}_{k,i-1}(\thetashare)}\right).
	\label{eq:ineqchainbounded}
	\end{align}
	On the other hand, using \eqref{eq:meanseq} we can write:
	\begin{align}
	v_k\log\bm{\mu}_{k,i-1}(\thetashare)\geq
	\sum_{\ell=1}^K v_{\ell}\log\bm{\mu}_{{\ell},i-1}(\thetashare)
	=\bm{m}_{i-1}\nonumber\\
	\implies
	\E\left(
	\log\frac{1}{\bm{\mu}_{k,i-1}(\thetashare)}
	\right)
	\leq
	\frac{1}{v_k}\,\sum_{\ell=1}^K v_{\ell}\log\frac{1}{\mu_{\ell,0}(\thetashare)},
	\end{align}
	which combined with \eqref{eq:ineqchainbounded} yields the desired claim.
	
\end{IEEEproof}

\begin{lemma}[\textbf{The Clear-Sighted Agent Learns the Truth}]
	\label{lem:oneagentlearns}
	Let $\thetashare=\thetatrue$. Under Assumptions~\ref{as:integ},~\ref{as:posit} and~\ref{as:strongclear} we have that:
	\begin{equation}
		\bm{\mu}_{\kappagood,i}(\thetashare)\stackrel{\textnormal{p}}{\longrightarrow} 1
	\end{equation}
\end{lemma}
\begin{IEEEproof}
Taking the total expectation in \eqref{eq:submartingale} we get:
\begin{equation}
\E(\bm{m}_i)\geq \E(\bm{m}_{i-1})+ \sum_{k=1}^K v_k \E \left(\delta_k(\bm{\mu}_{k,i-1})\right).
\label{eq:meanmi}
\end{equation}
First of all, we remark that the last expectation in \eqref{eq:meanmi} is computed with respect to the only random quantity that appears within brackets, that is $\bm{\mu}_{k,i-1}$. 
Using \eqref{eq:meanmi} along with the fact that the KL divergence is nonnegative, we see that:
\begin{equation}
0\leq
\sum_{k=1}^K v_k\E \left(\delta_k(\bm{\mu}_{k,i-1})\right)
\leq 
\E(\bm{m}_i)- \E(\bm{m}_{i-1}), 
\end{equation} 
which, in view of part $4)$ of Lemma~\ref{lem:mumart} implies that \cite{rudin1964principles}:
\begin{equation}
\lim_{i\rightarrow\infty}
\sum_{k=1}^K v_k\E \left(\delta_k(\bm{\mu}_{k,i-1})\right)=0
\end{equation}
Recalling that $v_k>0$, we conclude that $\delta_k(\bm{\mu}_{k,i-1})$ converges to zero in mean. This implies in particular that $\delta_k(\bm{\mu}_{k,i-1})$ converges to zero in probability, namely,
\begin{equation}
\delta_k(\bm{\mu}_{k,i-1})\stackrel{\textnormal{p}}{\longrightarrow} 0.
\label{eq:KLgoestozero}
\end{equation}
Recalling that $\delta_k(\bm{\mu}_{k,i-1})$ is the KL divergence between $ L_k(\thetashare)$ and $\sum_{\theta\in\Theta}\bm{\mu}_{k,i-1}(\theta) L_k(\theta)$ as defined in \eqref{eq:klconvex}, from Pinsker's inequality \cite{cover2012elements} we can write:
\begin{equation}
\delta_k(\bm{\mu}_{k,i-1})\geq
\frac 1 2\Big\| L_k(\thetashare) - \sum\limits_{\theta\in\Theta}\bm{\mu}_{k,i-1}(\theta) L_k(\theta)\Big\|^2,\label{eq:Pinsker}
\end{equation}
where $\|\cdot\|$ denotes the total variation norm. 

Let us now specialize the analysis to the clear-sighted agent $\kappagood$. From Assumption~\ref{as:strongclear}, the set of distinguishable hypotheses $\bar{\Theta}_{\kappagood}$ is non-empty. Thus, we have that:
\begin{align}
&L_{\kappagood}(\thetashare) - \sum_{\theta\in\Theta} \bm{\mu}_{\kappagood,i-1}(\theta) L_{\kappagood}(\theta)\nonumber\\
&=
\Bigg(1-\sum_{\theta\in\Theta_{\kappagood}}\bm{\mu}_{\kappagood,i-1}(\theta)\Bigg)L_{\kappagood}(\thetashare) \nonumber\\& -\sum_{\theta\in\bar{\Theta}_{\kappagood}}\bm{\mu}_{\kappagood,i-1}(\theta) L_{\kappagood}(\theta)\nonumber\\
&=
\sum_{\theta\in\bar{\Theta}_{\kappagood}}\bm{\mu}_{\kappagood,i-1}(\theta)
\Bigg(
L_{\kappagood}(\thetashare) - 
\sum_{\tau\in\bar{\Theta}_{\kappagood}}
\alpha(\tau)L_{\kappagood}(\tau)
\Bigg),
\label{eq:convexcomb}
\end{align}
where we defined:
\begin{equation}
\alpha(\tau)=\frac{\bm{\mu}_{k^\star,i-1}(\tau)}{\sum\limits_{\theta\in\bar{\Theta}_{k^\star}}\bm{\mu}_{k^\star,i-1}(\theta)}.
\end{equation}
Assumption~\ref{as:strongclear} establishes a lower bound $c$ on the KL divergence between the true likelihood and any convex combination of the distinguishable likelihoods, which implies that the true likelihood is not in the convex hull of distinguishable likelihoods. This further implies that there exists some $c'>0$, for which
\begin{equation}
\Big\|
L_{\kappagood}(\thetashare) - 
\sum_{\tau\in\bar{\Theta}_{\kappagood}}
\alpha(\tau)L_{\kappagood}(\tau)
\Big\|\geq c',\label{eq:totalv}
\end{equation}
where $\|\cdot\|$ represents the total variation norm~\cite{billingsley2008probability}. From \eqref{eq:convexcomb} and \eqref{eq:totalv} we can write:
\begin{align}
&\Big\|L_{\kappagood}(\thetashare) - \sum\limits_{\theta\in\Theta}\bm{\mu}_{\kappagood,i-1}(\theta) L_{\kappagood}(\theta)\Big\|\nonumber\\&=
\Big|\sum_{\theta\in\bar{\Theta}_{\kappagood}}\bm{\mu}_{\kappagood,i-1}(\theta)\Big|
\; 
\Big\|
L_{\kappagood}(\thetashare) - 
\sum_{\tau\in\bar{\Theta}_{\kappagood}}
\alpha(\tau)L_{\kappagood}(\tau)
\Big\|\nonumber\\
&\geq c'\Big|\sum_{\theta\in\bar{\Theta}_{\kappagood}}\bm{\mu}_{{\kappagood},i-1}(\theta)\Big|.
\end{align}
Joining the latter inequality with \eqref{eq:Pinsker} we get:
\begin{equation}
\delta_{\kappagood}(\bm{\mu}_{\kappagood,i-1})\geq
\frac{c'{}^2}{2} \, \Big|\sum_{\theta\in\bar{\Theta}_{\kappagood}}\bm{\mu}_{\kappagood,i-1}(\theta)\Big|^2.
\end{equation}
Since $c'$ is strictly positive, we conclude from \eqref{eq:KLgoestozero} that, for every $\theta\in\bar{\Theta}_{\kappagood}$:
\begin{equation}
\bm{\mu}_{\kappagood,i}(\theta)\stackrel{\textnormal{p}}{\longrightarrow} 0.
\label{eq:mukithetabar}
\end{equation}
It remains to show that the same result holds for the indistinguishable non-transmitted hypotheses, i.e., for $\theta\in\Theta_{\kappagood}\setminus\{\thetashare\}$. 
But this result comes directly from Lemma~\ref{lem:discardonediscardall}, under Assumptions~\ref{as:integ} and~\ref{as:posit}. 
We have therefore shown that, for the clear-sighted agent $\kappagood$, the beliefs for all $\theta\in\Theta\setminus\{\thetashare\}$ vanish in probability, which finally yields the claim since the sum of the beliefs over $\Theta$ is equal to $1$.
\end{IEEEproof}
\begin{lemma}[\textbf{Influence of a Learning Agent}]
	\label{lem:oneall}
	Let $\thetashare=\thetatrue$. Under Assumptions~\ref{as:integ},~\ref{as:posit} and~\ref{as:strongclear}, if, for a certain agent $h$,
	\begin{equation}
	\bm{\mu}_{h,i}(\thetashare)\stackrel{\textnormal{p}}{\longrightarrow}1,
	\label{eq:hlearns}
	\end{equation}
	then the same result holds for all agents $k\neq h$. 
\end{lemma}
\begin{IEEEproof}
	Let $h$ be an agent that fulfills \eqref{eq:hlearns}. Consider that the combination weight $a_{hk}$ is {\em strictly} positive. 
	From \eqref{eq:combsa_2} we can therefore write:
	\begin{align}
	&	\log\frac{\bm{\mu}_{k,i}(\theta)}{\bm{\mu}_{k,i}(\thetashare)}=a_{kk}\log\bm{\psi}_{k,i}(\theta)\nonumber\\
	&+
	\sum_{\ell\neq k}a_{\ell k}\log\frac{1-\bm{\psi}_{\ell,i}(\thetashare)}{H-1}+
	\sum_{\ell=1}^K
	a_{\ell k}\log\frac{1}{\bm{\psi}_{\ell,i}(\thetashare)}\nonumber\\
	&\leq
	a_{hk}\log(1-\bm{\psi}_{h,i}(\thetashare))
	+
	\sum_{\ell=1}^K 
	a_{\ell k}\log\frac{1}{\bm{\psi}_{\ell,i}(\thetashare)}.
	\label{eq:logmusum}
	\end{align}
	By exponentiating \eqref{eq:logmusum} we can write: 
	%(we assume $a_{hk}<1$, the other case being even easier to deal with):
	\begin{equation}
	\bm{\mu}_{k,i}(\theta)\leq
	\underbrace{
		\left(1-\bm{\psi}_{h,i}(\thetashare)\right)^{a_{hk}}
	}_{\triangleq\bm{x}_i}
	\,
	\underbrace{
		e^{
			\sum_{\ell=1}^K 
			a_{\ell k}\log\frac{1}{\bm{\psi}_{\ell,i}(\thetashare)}
		}
	}_{\triangleq\bm{y}_i}.
	\label{eq:musum}
	\end{equation}
	First, we prove that the term $\bm{x}_i$ in \eqref{eq:musum} goes to zero in probability. 
	To this end, we observe that:
	\begin{align}
	&1-\bm{\psi}_{h,i}(\thetashare)\nonumber\\&=
	\frac{
		\sum\limits_{\theta\neq\thetashare} \bm{\mu}_{h,i-1}(\theta) L_h(\bm{\xi}_{h,i}|\theta)
	}{\bm{\mu}_{h,i-1}(\thetashare) L_h(\bm{\xi}_{h,i}|\thetashare) + \sum\limits_{\theta\neq\thetashare} \bm{\mu}_{h,i-1}(\theta) L_h(\bm{\xi}_{h,i}|\theta)}
	\nonumber\\&\leq
	\sum_{\theta\neq\thetashare} \frac{\bm{\mu}_{h,i-1}(\theta)}{\bm{\mu}_{h,i-1}(\thetashare)} \frac{L_h(\bm{\xi}_{h,i}|\theta)}{L_h(\bm{\xi}_{h,i}|\thetashare)}.
	\end{align}
	We now show that each individual term of the summation,
	\begin{equation}
	\underbrace{
		\frac{\bm{\mu}_{h,i-1}(\theta)}{\bm{\mu}_{h,i-1}(\thetashare)}
	}_{\triangleq \bm{s}_i} 
	\,
	\underbrace{\frac{L_h(\bm{\xi}_{h,i}|\theta)}{L_h(\bm{\xi}_{h,i}|\thetashare)}
	}_{\triangleq \bm{t}_i},
	\label{eq:Slutsky2}
	\end{equation}
	vanishes in probability as $i\rightarrow\infty$. 
	Indeed, the term $\bm{s}_i$ in \eqref{eq:Slutsky2} vanishes in probability as $i\rightarrow\infty$ in view of Lemma~\ref{lem:oneagentlearns}. 
	On the other hand, the random variables $\bm{t}_i$ are identically distributed.\footnote{We remark that the random variables $\bm{t}_i$ are well-behaved since $\frac{L_h(\bm{\xi}_{h,i}|\theta)}{L_h(\bm{\xi}_{h,i}|\thetashare)}$ is a (nonnegative) random variable with finite expectation equal to $1$.} 
	By application of Slutsky's theorem~\cite{ShaoBook}, we conclude that the product $\bm{s}_i \bm{t}_i$ converges to $0$ in distribution (and, hence, in probability). 
	
	Second, we show that $\bm{y}_i$ matches the conditions in \eqref{eq:modelemma} (see Lemma~\ref{lem:boundconv} in Appendix~\ref{ap:alsa}). 
	By application of Markov's inequality we conclude that, for any $M>0$:
	\begin{align}
	\P(\bm{y}_i>M)
	&=
	\P\left(
	\sum_{\ell=1}^K 
	a_{\ell k}\log\frac{1}{\bm{\psi}_{\ell,i}(\thetashare)}>\log M
	\right)\nonumber\\&
	\leq
	\frac{1}{\log M}\,\sum_{\ell=1}^K
	a_{\ell k}
	\E\left(
	\log\frac{1}{\bm{\psi}_{\ell,i}(\thetashare)}
	\right)\nonumber\\
	&\leq
	\frac{1}{\log M}\sum_{\ell=1}^K 
	\frac{a_{\ell k}}{v_{\ell}}
	\sum_{m=1}^K v_m \log\frac{1}{\mu_{m,0}(\thetashare)},
	\label{eq:meanbound}
	\end{align}
	where the latter inequality follows by Corollary~\ref{cor:psimart}. Since the final upper bound in \eqref{eq:meanbound} does not depend on $i$, we see that $\bm{y}_i$ fulfills \eqref{eq:modelemma} with the choice $g(M)=C/\log M$ for some finite positive constant $C$.
	
	Therefore, we conclude from Lemma~\ref{lem:boundconv} that the product $\bm{x}_i \bm{y}_i$ appearing in the upper bound in \eqref{eq:musum} goes to zero in probability and, hence, that:
	\begin{equation}
	\bm{\mu}_{k,i}(\theta)\stackrel{\textnormal{p}}{\longrightarrow} 0,
	\label{eq:claimclaim}
	\end{equation} 
	for any agent $k$ for which $a_{hk}>0$. 
	Since the network is strongly connected, given an agent $h$ that fulfills \eqref{eq:hlearns}, and an arbitrary agent $k$ (not necessarily a neighbor of $h$), there will always be a path connecting $h$ to $k$. Iterating the above reasoning along this path implies the desired result.
\end{IEEEproof}

We can now conclude the proof of Theorem~\ref{the:selfaware}. Under Assumption~\ref{as:strongclear}, there exists at least one clear-sighted agent $\kappagood$. Lemma~\ref{lem:oneagentlearns} guarantees that agent $\kappagood$ learns the truth in probability, whereas Lemma~\ref{lem:oneall} ensures that learning propagates across the network. It is therefore legitimate to write:
	\begin{equation}
	\sum_{k=1}^K v_k \log\bm{\mu}_{k,i}(\thetashare) \stackrel{\textnormal{p}}{\longrightarrow} 0.
	\label{eq:mumartconvprob}
	\end{equation}
	Using part $3)$ of Lemma~\ref{lem:mumart} (and since almost-sure convergence implies convergence in probability), from \eqref{eq:mumartconvprob} we conclude that:
	\begin{equation}
	\sum_{k=1}^K v_k \log\bm{\mu}_{k,i}(\thetashare) \stackrel{\textnormal{a.s.}}{\longrightarrow} 0.
	\label{eq:asconvtozero}
	\end{equation}
	On the other hand, since $v_k>0$ and $\log\bm{\mu}_{k,i}(\thetashare)\leq 0$, the convergence in \eqref{eq:asconvtozero} implies that:
	\begin{equation}
	\log\bm{\mu}_{k,i}(\thetashare) \stackrel{\textnormal{a.s.}}{\longrightarrow} 0\implies \bm{\mu}_{k,i}(\thetashare) \stackrel{\textnormal{a.s.}}{\longrightarrow} 1,
	\end{equation}
	for all agents $k=1,2,\dots,K$.

\section{Auxiliary Lemmas for the Approach with Self-Awareness and their Proofs}
\label{ap:alsa}
	\begin{lemma}[\textbf{Convergence for Non-Transmitted Hypotheses}]
	\label{lem:unshared}
	Let $\theta,\theta'\in\Theta\setminus\{\thetashare\}$, and define:
	\begin{equation}
	\bm{q}_{k,i}(\theta,\theta')\triangleq\log \frac{\bm{\mu}_{k,i}(\theta)}{\bm{\mu}_{k,i}(\theta')}.
	\label{eq:qkdef}
	\end{equation}
	For every $k=1,2,\dots,K$, under Assumptions~\ref{as:integ} and~\ref{as:posit}, there exists a random variable $\bm{q}_{k,\infty}(\theta,\theta')$ ensuring the following convergence in distribution:
	\begin{equation}
	\bm{q}_{k,i}(\theta,\theta')\stackrel{\textnormal{d}}{\longrightarrow} \bm{q}_{k,\infty}(\theta,\theta').
	\label{eq:qkconv}
	\end{equation}
\end{lemma}
\begin{IEEEproof}
	Since $\theta$ and $\theta'$ are distinct from $\thetashare$, using \eqref{eq:Bayesupdate}, \eqref{eq:justone} and \eqref{eq:combsa} we can write:
	\begin{equation}
	\log \frac{\bm{\mu}_{k,i}(\theta)}{\bm{\mu}_{k,i}(\theta')}=
	a_{kk}\log \frac{\bm{\mu}_{k,i-1}(\theta)}{\bm{\mu}_{k,i-1}(\theta')}+
	a_{kk}\log \frac{L_k(\bm{\xi}_{k,i}|\theta)}{L_k(\bm{\xi}_{k,i}|\theta')}.
	\label{eq:unsharedevolut}
	\end{equation}
	The result in \eqref{eq:qkconv} follows from part 1) of auxiliary Lemma~\ref{lem:scalarevolut} by setting:
	\begin{equation}
	a=a_{kk},\quad
	\bm{y}_i=\log \frac{\bm{\mu}_{k,i}(\theta)}{\bm{\mu}_{k,i}(\theta')}, 
	\quad
	\bm{x}_i=\log \frac{L_k(\bm{\xi}_{k,i}|\theta)}{L_k(\bm{\xi}_{k,i}|\theta')},
	\end{equation}
	where Assumption~\ref{as:integ} guarantees that $\bm{x}_i$ satisfies the conditions in Lemma~\ref{lem:scalarevolut}, and Assumption~\ref{as:posit} guarantees that $y_0$ assumes a finite value.
\end{IEEEproof}

From Lemma~\ref{lem:unshared}, we see that the log-ratio of belief components concerning non-transmitted hypotheses converges in distribution to a random variable $\bm{q}_{k,\infty}$. Investigating this limiting random variable in more detail, thanks to part 1) of Lemma~\ref{lem:scalarevolut}, we are able to write it as
\begin{equation}
\bm{q}_{k,\infty}(\theta,\theta')=\sum_{i=1}^{\infty}a_{kk}^i\log \frac{L_k(\bm{\xi}_{k,i}|\theta)}{L_k(\bm{\xi}_{k,i}|\theta')}.\label{eq:summrv}
\end{equation}
For each realization of signal profiles $\left(\bm{\xi}_{1}, \bm{\xi}_{2},\dots\right)$, the infinite summation in \eqref{eq:summrv} will converge almost surely to a random value. The distribution with which these random values are generated will be the same distribution that governs the oscillatory behavior of $\bm{q}_{k,i}(\theta,\theta')$ as $i\rightarrow \infty$.

Although the characterization of the limiting random variable $\bm{q}_{k,\infty}(\theta,\theta')$, described in \eqref{eq:summrv}, does not appear intuitive, its mere existence will enable other (stronger) convergence results starting from the one presented in Lemma~\ref{lem:discardonediscardall}. We see now that if a certain agent $k$ discards any non-transmitted hypothesis $\theta\in\Theta\setminus\{\thetashare\}$, then the existence of the limiting random variable $\bm{q}_{k,\infty}(\theta,\theta')$ will allow it to discard all other non-transmitted hypotheses.

\begin{lemma}[\textbf{Rejection of Non-Transmitted Hypotheses}]
	\label{lem:discardonediscardall}
	Assume, for a given agent $k$, and for one non-transmitted hypothesis $\theta'\in\Theta\setminus\{\thetashare\}$:
	\begin{equation}
 \bm{\mu}_{k,i}(\theta')\stackrel{\textnormal{p}}{\longrightarrow} 0,
	\label{eq:thetadistinguish}
	\end{equation}
	and that Assumptions~\ref{as:integ} and~\ref{as:posit} hold. Then the same convergence holds for all hypotheses $\theta\in\Theta\setminus\{\theta',\thetashare\}$ for the same agent.
\end{lemma}

\begin{IEEEproof}
	Let $\theta\neq\thetashare$ be a non-transmitted hypothesis that fulfills \eqref{eq:thetadistinguish}. In view of \eqref{eq:qkdef}, for any $\theta'\in \Theta\setminus \{\theta,\thetashare\}$ we can write:
	\begin{equation}
	\bm{\mu}_{k,i}(\theta)=\bm{\mu}_{k,i}(\theta') e^{\bm{q}_{k,i}(\theta,\theta')}.
	\label{eq:muktautheta}
	\end{equation}
	Now, under Assumptions~\ref{as:integ} and~\ref{as:posit}, Lemma~\ref{lem:unshared} reveals that $\bm{q}_{k,i}(\theta,\theta')$ converges in distribution to a certain random variable $\bm{q}_{k,\infty}(\theta,\theta')$. In view of the continuous mapping theorem~\cite{ShaoBook}, we conclude that:
	\begin{equation}
	e^{\bm{q}_{k,i}(\theta,\theta')}\stackrel{\textnormal{d}}{\longrightarrow} e^{\bm{q}_{k,\infty}(\theta,\theta')}.
	\label{eq:eqki}
	\end{equation}
	Examining \eqref{eq:muktautheta}, we see that $\bm{\mu}_{k,i}(\theta)$ is given by the product of two random sequences: $i)$ the first sequence, $\{\bm{\mu}_{k,i}(\theta')\}$, vanishes in probability as $i\rightarrow\infty$ in view of \eqref{eq:thetadistinguish}; $ii)$ the second sequence, $\{e^{\bm{q}_{k,i}(\theta,\theta')}\}$, converges in distribution as $i\rightarrow\infty$ in view of \eqref{eq:eqki}. 
	Using Slutsky's Theorem~\cite{ShaoBook}, we conclude that $\bm{\mu}_{k,i}(\theta)$ converges to zero in distribution, and, hence, in probability.  
\end{IEEEproof}
In other words, whenever an agent discards a non-transmitted hypothesis, it will automatically discard all other non-transmitted hypotheses. This result will bind together the evolution of the non-transmitted hypotheses in the case when the respective beliefs components are converging in probability to zero, which we refer to as \emph{parallel rejection} of non-transmitted hypotheses.

Finally we introduce a technical result, which is used in the proof of Lemma~\ref{lem:oneall}.
\begin{lemma}[\textbf{Useful Convergence Result}]
	\label{lem:boundconv}
	Let $\bm{z}_i=\bm{x}_i\bm{y}_i$, where $\{\bm{x}_i\}$ and $\{\bm{y}_i\}$ are two sequences of nonnegative random variables such that $\bm{x}_i$ vanishes in probability, and:
	\begin{equation}
	\P(\bm{y}_i>M)\leq g(M),\textnormal{ with }{\lim_{M\rightarrow \infty}}g(M)=0.
	\label{eq:modelemma}
	\end{equation}
	Then, we have that:
	\begin{equation}
	\bm{z}_i\stackrel{\textnormal{p}}{\longrightarrow} 0.
	\end{equation}
\end{lemma}
\begin{IEEEproof}
	Let us consider the following implication of events, for any positive values $M$ and $\gamma$:
	\begin{equation}
	\Big\{\bm{x}_i\leq \frac{\gamma}{M}\Big\}
	\bigcap 
	\Big\{\bm{y}_i\leq M\Big\}
	\implies
	\Big\{
	\bm{x}_i\bm{y}_i\leq \gamma
	\Big\},
	\end{equation}
	which, using De Morgan's laws \cite{billingsley2008probability}, is equivalent to:
	\begin{equation}
	\Big\{
	\bm{x}_i\bm{y}_i> \gamma
	\Big\}
	\implies
	\Big\{\bm{x}_i> \frac{\gamma}{M}\Big\}
	\bigcup
	\Big\{\bm{y}_i> M\Big\}.
	\label{eq:eventsimp}
	\end{equation}
	Since, for any two events $\mathcal{A}, \mathcal{B}$, the condition $\mathcal{A}\implies\mathcal{B}$ implies that $\P(\mathcal{A})\leq\P(\mathcal{B})$, from \eqref{eq:eventsimp}, and using the union bound, we conclude  that:
	\begin{align}
	\P(\bm{z}_i>\gamma)&\leq
	\P(\bm{x}_i> \gamma/M)+ 
	\P(\bm{y}_i>M)
	\nonumber\\
	&\leq
	\P(\bm{x}_i> \gamma/M)
	+
	g(M),
	\label{eq:ineqchain}
	\end{align}
	where the latter inequality follows by the upper bound in \eqref{eq:modelemma}.
	Now, let us fix a value $\varepsilon>0$. For sufficiently large $M$, we have that $g(M)\leq\varepsilon/2$ in view of the limit appearing in \eqref{eq:modelemma}. 
	On the other hand, since by assumption $\bm{x}_i$ converges to zero in probability, for given values of $M$ and $\gamma$ there exists certainly a sufficiently large $i_0$ such that, for every $i\geq i_0$, also the quantity $\P(\bm{x}_i>\gamma/M)$ is upper bounded by $\varepsilon/2$, which implies, for $i\geq i_0$:
	\begin{equation}
	\P(\bm{z}_i>\gamma)\leq\varepsilon,
	\end{equation}
	and the claim of the lemma is proved.
\end{IEEEproof}

\section{Proof of Theorem~\ref{the:belcoll}}\label{ap:thebelcoll}
From Lemma~\ref{lemma:rate} (see Appendix~\ref{ap:truthpartial}), we see that the sign of the quantity on the RHS of \eqref{eq:lemrate} will dictate different convergence behaviors. Note that KL divergences are finite from Assumption~\ref{as:integ}. First, consider the case when
\begin{equation}
d_{\sf ave}(\thetasharec)> d_{\sf ave}(\thetashare),
\end{equation}
which implies that the asymptotic rate of convergence seen in \eqref{eq:lemrate} is strictly negative. Since $\bm{\mu}_{k,i}(\theta)$ is bounded by 1 for any hypothesis $\theta$, then
\begin{align}
&\frac{1}{i}\log\frac{\bm{\mu}_{k,i}(\theta)}{\bm{\mu}_{k,i}(\thetashare)}\stackrel{\textnormal{a.s.}}{\longrightarrow}  d_{\sf ave}(\thetashare)-d_{\sf ave}(\thetasharec)<0\nonumber\\&
\implies 
\log\frac{\bm{\mu}_{k,i}(\theta)}{\bm{\mu}_{k,i}(\thetashare)}\stackrel{\textnormal{a.s.}}{\longrightarrow}-\infty\nonumber\\&\implies \bm{\mu}_{k,i}(\theta)\stackrel{\textnormal{a.s.}}{\longrightarrow}0,
\end{align}
which holds for all $\theta\in\Theta\setminus\{\thetashare\}$. This, in turn, implies that
\begin{equation}
\bm{\mu}_{k,i}(\thetashare)\stackrel{\textnormal{a.s.}}{\longrightarrow}1.
\end{equation}
Next, consider the case:
\begin{equation}
d_{\sf ave}(\thetasharec)< d_{\sf ave}(\thetashare),
\end{equation}
implying that the asymptotic rate of convergence in \eqref{eq:lemrate} is strictly positive. In this case, since again $\bm{\mu}_{k,i}(\theta)$ is bounded, we have that 
\begin{align}
&\frac{1}{i}\log\frac{\bm{\mu}_{k,i}(\theta)}{\bm{\mu}_{k,i}(\thetashare)}\stackrel{\textnormal{a.s.}}{\longrightarrow} d_{\sf ave}(\thetashare)-d_{\sf ave}(\thetasharec)>0\nonumber\\&
\implies 
\log\frac{\bm{\mu}_{k,i}(\theta)}{\bm{\mu}_{k,i}(\thetashare)}\stackrel{\textnormal{a.s.}}{\longrightarrow}+\infty\nonumber\\&\implies\bm{\mu}_{k,i}(\thetashare)\stackrel{\textnormal{a.s.}}{\longrightarrow}0,
\end{align}
which, in view of \eqref{eq:1}, implies that, for every $\theta\in\Theta\setminus\{\thetashare\}$,
\begin{equation}
\bm{\mu}_{k,i}(\theta)\stackrel{\textnormal{a.s.}}{\longrightarrow}\frac{1}{H-1}.
\end{equation}

\section{Proof of Theorem~\ref{the:altsa}}\label{ap:alt_sa}
We will start by addressing the first part of Theorem~\ref{the:altsa}. Let us develop the recursion in \eqref{eq:combsa_2} with $\theta = \thetashare$ and $\theta^{\prime}=\thetatrue$.
\begin{align}
&\log \frac{\bm{\mu}_{k,i}(\thetashare)}{\bm{\mu}_{k,i}(\thetatrue)}=a_{kk}\log \frac{\bm{\mu}_{k,i-1}(\thetashare)}{\bm{\mu}_{k,i-1}(\thetatrue)}+a_{kk}\log \frac{L_k(\bm{\xi}_{k,i}|\thetashare)}{L_k(\bm{\xi}_{k,i}|\thetatrue)}\nonumber\\
&+\sum_{\substack{\ell=1\\\ell\neq k}}^Ka_{\ell k}\log \frac{\bm{\mu}_{\ell,i-1}(\thetashare)L_\ell(\bm{\xi}_{\ell,i}|\thetashare)}{\sum\limits_{\tau\neq \thetashare}\frac{1}{H-1}\bm{\mu}_{\ell,i-1}(\tau)L_\ell(\bm{\xi}_{\ell,i}|\tau)}\nonumber\\
&\stackrel{\text{(a)}}{\leq}\sum_{\ell =1}^Ka_{\ell k}\log \frac{\bm{\mu}_{\ell,i-1}(\thetashare)}{\bm{\mu}_{\ell,i-1}(\thetatrue)}+\sum_{\ell =1}^Ka_{\ell k}\log \frac{L_\ell(\bm{\xi}_{\ell,i}|\thetashare)}{L_\ell(\bm{\xi}_{\ell,i}|\thetatrue)}\nonumber\\
&+\sum_{\substack{\ell=1\\\ell\neq k}}^Ka_{\ell k}\log\left( \bm{\mu}_{\ell,i-1}(\thetatrue)L_\ell(\bm{\xi}_{\ell,i}|\thetatrue)\right)\nonumber\\
&-\sum_{\substack{\ell=1\\\ell\neq k}}^Ka_{\ell k} \sum_{\tau\neq \thetashare}\frac{\log \left(\bm{\mu}_{\ell,i-1}(\tau)L_\ell(\bm{\xi}_{\ell,i}|\tau)\right)}{H-1}\nonumber\\
&=\sum_{\ell =1}^Ka_{\ell k}\log \frac{\bm{\mu}_{\ell,i-1}(\thetashare)}{\bm{\mu}_{\ell,i-1}(\thetatrue)}+\sum_{\ell =1}^Ka_{\ell k}\log \frac{L_\ell(\bm{\xi}_{\ell,i}|\thetashare)}{L_\ell(\bm{\xi}_{\ell,i}|\thetatrue)}\nonumber\\
&-\sum_{\substack{\ell=1\\\ell\neq k}}^K \frac{a_{\ell k}}{H-1}\sum_{\tau\neq \thetashare}\left(\log\frac{ L_\ell(\bm{\xi}_{\ell,i}|\tau)}{L_\ell(\bm{\xi}_{\ell,i}|\thetatrue)}+\log \frac{\bm{\mu}_{\ell,i-1}(\tau)}{\bm{\mu}_{\ell,i-1}(\thetatrue)}\right),\label{eq:recurbound}
\end{align} 
where (a) follows from Jensen's inequality applied as follows:
\begin{align}
\log \left(\sum\limits_{\tau\neq \thetashare}\frac{1}{H-1}\bm{\mu}_{\ell,i-1}(\tau)L_\ell(\bm{\xi}_{\ell,i}|\tau)\right)\nonumber\\\geq\sum\limits_{\tau\neq \thetashare}\frac{1}{H-1}\log \left(\bm{\mu}_{\ell,i-1}(\tau)L_\ell(\bm{\xi}_{\ell,i}|\tau)\right).
\end{align}
Setting $\bm{y}_{k,i}=\log \frac{\bm{\mu}_{k,i}(\thetashare)}{\bm{\mu}_{k,i}(\thetatrue)}$ and
\begin{align}
\bm{x}_{k,i}&=\sum_{\ell =1}^Ka_{\ell k}\log \frac{L_\ell(\bm{\xi}_{\ell,i}|\thetashare)}{L_\ell(\bm{\xi}_{\ell,i}|\thetatrue)}\nonumber\\
&-\sum_{\substack{\ell=1\\\ell\neq k}}^K\frac{a_{\ell k} }{H-1}\sum_{\tau\neq \thetashare}\log\frac{ L_\ell(\bm{\xi}_{\ell,i}|\tau)}{L_\ell(\bm{\xi}_{\ell,i}|\thetatrue)}\nonumber\\
&-\sum_{\substack{\ell=1\\\ell\neq k}}^K\frac{a_{\ell k} }{H-1}\sum_{\tau\neq \thetashare}\log \frac{\bm{\mu}_{\ell,i-1}(\tau)}{\bm{\mu}_{\ell,i-1}(\thetatrue)},
\label{eq:xchoice}
\end{align}
we can write \eqref{eq:recurbound} in vector form as:
\begin{equation}
\bm{y}_i\preccurlyeq A\T \bm{y}_{i-1}+\bm{x}_{i},\label{eq:recurbound1}
\end{equation}
where the symbol $\preccurlyeq$ denotes element-wise inequality.
Therefore, the recursion in \eqref{eq:recurbound1} matches the model in \eqref{eq:lemrecurs}, but for the fact that we have an inequality in place of an equality. Since the matrix $A$ has nonnegative entries, we can still develop the recursion preserving the inequality, allowing us to use the results from Lemma~\ref{lem:aux1} (Appendix~\ref{ap:aux}) in the form of an inequality.

We need now to show that $\bm{x}_i$, as defined in \eqref{eq:xchoice}, satisfies the conditions \eqref{eq:aslim1} and \eqref{eq:aslim2} in Lemma~\ref{lem:aux1}. Regarding the log-likelihood ratio terms, i.e., the first two terms on the RHS of \eqref{eq:xchoice}, since these terms satisfy Assumption~\ref{as:integ} and since the observations $\bm{\xi}_{\ell,i}$ are i.i.d. across time, the result of Lemma~\ref{lem:aux1} can be applied to these terms, in view of Property~\ref{prop:int} (Appendix~\ref{ap:aux}). For these two terms, we have that
\begin{align}
\bar{x}_k&=\sum_{\ell =1}^Ka_{\ell k}\E\left(\log \frac{L_\ell(\thetashare)}{L_\ell(\thetatrue)}\right)\nonumber\\&-\sum_{\substack{\ell=1\\\ell\neq k}}^K\frac{a_{\ell k} }{H-1}\sum_{\tau\neq \thetashare}\E\left(\log\frac{ L_\ell(\tau)}{L_\ell(\thetatrue)}\right).
\end{align}
For what concerns the log-belief ratio, i.e., the third term on the RHS of \eqref{eq:xchoice}, we have that this term behaves like the recursion seen in \eqref{eq:unsharedevolut}, which reveals that the log-belief ratio for the non-transmitted hypotheses matches the model in \eqref{eq:recurlemmascalar}. As a result, conditions \eqref{eq:aslim1} and \eqref{eq:aslim2} are automatically satisfied in view of Lemma~\ref{lem:scalarevolut}.

From Lemma~\ref{lem:aux1}, we have that
\begin{align}
&\limsup_{i\rightarrow \infty}\frac{1}{i}\log \frac{\bm{\mu}_{k,i}(\thetashare)}{\bm{\mu}_{k,i}(\thetatrue)} \nonumber\\
&\stackrel{\text{a.s.}}{\leq}-\sum_{\ell =1}^Kv_{\ell}d_{\ell}(\thetashare)+\sum_{\ell =1}^Kv_{\ell}\sum_{\substack{n=1\\n\neq \ell}}^K \frac{a_{n \ell}}{H-1}\sum_{\tau\neq \thetashare}d_n(\tau)\nonumber\\
&-\sum_{\ell =1}^Kv_{\ell}\sum_{\substack{n=1\\n\neq \ell}}^K \frac{a_{n \ell}}{H-1}\sum_{\tau\neq \thetashare}\lim_{i\rightarrow \infty}\frac{1}{i}\sum_{j=1}^{i}\log \frac{\bm{\mu}_{n,j-1}(\tau)}{\bm{\mu}_{n,j-1}(\thetatrue)}.\label{eq:th4aux}
\end{align}
We recall that for $\tau, \thetatrue \neq \thetashare$ and $\tau\in \bar{\Theta}_n$, according to Lemma~\ref{lem:scalarevolut} (Appendix~\ref{ap:aux}),
\begin{align}
\frac{1}{i}\sum_{j=1}^{i}\log\frac{ \bm{\mu}_{n,j}(\tau)}{\bm{\mu}_{n,j}(\theta_0)} &\stackrel{\textnormal{a.s.}}{\longrightarrow} \frac{a_{n n}}{1-a_{n n}}\mathbb{E}\left(\log \frac{L_n(\tau)}{L_n(\thetatrue)}\right)\nonumber\\
 &= -\frac{a_{n n}}{1-a_{n n}}d_n(\tau).\label{eq:th4aux2}
\end{align}
Thus replacing \eqref{eq:th4aux2} into \eqref{eq:th4aux}, yields
\begin{align}
&\limsup_{i\rightarrow \infty}\frac{1}{i}\log \frac{\bm{\mu}_{k,i}(\thetashare)}{\bm{\mu}_{k,i}(\thetatrue)}\stackrel{\textnormal{a.s.}}{\leq}-\sum_{\ell =1}^Kv_{\ell}d_{\ell}(\thetashare)\nonumber\\&+\sum_{\ell =1}^Kv_{\ell}\sum_{\substack{n=1\\n\neq \ell}}^Ka_{n \ell} \left(\frac{a_{n n}}{1-a_{n n}}+1\right) \frac{1}{H-1}\hspace{-4pt}\sum_{\tau\neq\thetashare}d_n(\tau)\nonumber\\
&=-\sum_{\ell =1}^Kv_{\ell}d_\ell(\theta_{\sf TX})+\frac{1}{H-1}\hspace{-4pt}\sum_{\tau\neq\thetashare}\sum_{\ell =1}^Kv_{\ell}\sum_{\substack{n=1\\n\neq \ell}}^Ka_{n \ell}\frac{1}{1-a_{n n}}d_n(\tau)\nonumber\\
&=-\sum_{\ell =1}^Kv_{\ell}d_\ell(\theta_{\sf TX})+\frac{1}{H-1}\hspace{-4pt}\sum_{\tau\neq\thetashare}\sum_{\ell =1}^Kv_{\ell}d_\ell(\tau), \label{eq:auxequ}
\end{align}
where \eqref{eq:auxequ} follows from algebraic manipulations, taking into account the left stochasticity of matrix $A$ and the definition of the Perron eigenvector $v$. 
As long as the RHS of \eqref{eq:auxequ} assumes a negative value, this implies that 
\begin{equation}
\bm{\mu}_{k,i}(\thetashare)\stackrel{\textnormal{a.s.}}{\longrightarrow}0.
\end{equation}
The proof for the first part of Theorem~\ref{the:altsa} is complete. We proceed now to examine the second part. Considering Assumption~\ref{as:boundedlik} and developing the recursion in \eqref{eq:unsharedevolut} for $\theta=\tau$ and $\theta'=\tau'$, the boundedness of log-likelihood ratios is inherited by the ratio of the log-beliefs for any non-transmitted hypotheses $\tau,\tau'\in\Theta\setminus\{\thetashare\}$. In fact, exploiting \eqref{eq:unsharedevolut} and the upper bound in \eqref{eq:assboundedlike}, and iterating over $i$, we can write:
\begin{align}
\log\frac{\bm{\mu}_{k,i}(\tau)}{\bm{\mu}_{k,i}(\tau')}
&\leq
a_{kk}^i\log\frac{\bm{\mu}_{k,0}(\tau)}{\bm{\mu}_{k,0}(\tau')}+
B\sum_{j=1}^i a_{kk}^{i-j+1}\nonumber\\&=
a_{kk}^i\log\frac{\mu_{k,0}(\tau)}{\mu_{k,0}(\tau')}+
a_{kk}\frac{1-a_{kk}^i}{1-a_{kk}} B.
\end{align}
We know that $a_{kk}^i$ converges to zero as $i\rightarrow \infty$. For an arbitrarily small $\varepsilon>0$, there exists an instant $i_0$ such that for $i>i_0$ we have that:
\begin{align}
&\log\frac{\bm{\mu}_{k,i}(\tau)}{\bm{\mu}_{k,i}(\tau')}
\leq
\frac{a_{kk}}{1-a_{kk}} B+ \varepsilon\log\frac{\mu_{k,0}(\tau)}{\mu_{k,0}(\tau')}\nonumber\\&\implies 
\bm{\mu}_{k,i}(\tau)\leq\bm{\mu}_{k,i}(\tau')e^{\frac{a_{kk}}{1-a_{kk}} B + \epsilon}.\label{eq:recursionbound}
\end{align}
where we defined:
\begin{equation}
\epsilon \triangleq   \varepsilon\log\frac{\mu_{k,0}(\tau)}{\mu_{k,0}(\tau')}.\label{eq:epsi}
\end{equation}
Note that if $\varepsilon$ is arbitrarily small, $\epsilon$ will also be arbitrarily close to zero due to Assumption~\ref{as:posit}.

Developing the recursion in \eqref{eq:combsa_2} with $\theta\in\Theta\setminus\{\thetashare\}$ and $\theta^{\prime}=\thetashare$, we have that:
\begin{align}
&\log \frac{\bm{\mu}_{k,i}(\theta)}{\bm{\mu}_{k,i}(\thetashare)}=a_{kk}\log\frac{\bm{\psi}_{k,i}(\theta)}{\bm{\psi}_{k,i}(\thetashare)}\nonumber\\
&+\sum_{\substack{\ell=1\\\ell\neq k}}^Ka_{\ell k}\log \frac{\sum_{\tau \neq \thetashare}L_\ell(\bm{\xi}_{\ell,i}|\tau)\bm{\mu}_{\ell, i-1}(\tau)}{L_\ell(\bm{\xi}_{\ell,i}|\thetashare)\bm{\mu}_{\ell, i-1}(\thetashare)(H-1)}\nonumber\\
&= \sum_{\ell=1}^Ka_{\ell k}\log\frac{\bm{\psi}_{\ell,i}(\theta)}{\bm{\psi}_{\ell,i}(\thetashare)}\nonumber\\
&+\sum_{\substack{\ell=1\\\ell\neq k}}^Ka_{\ell k}\log \frac{\sum_{\tau \neq \thetashare}L_\ell(\bm{\xi}_{\ell,i}|\tau)\bm{\mu}_{\ell, i-1}(\tau)}{L_\ell(\bm{\xi}_{\ell,i}|\theta)\bm{\mu}_{\ell, i-1}(\theta)(H-1)}\nonumber\\
&\stackrel{\text{(a)}}{\leq} \sum_{\ell=1}^Ka_{\ell k}\log\frac{\bm{\psi}_{\ell,i}(\theta)}{\bm{\psi}_{\ell,i}(\thetashare)}+\sum_{\substack{\ell=1\\\ell\neq k}}^Ka_{\ell k}\log \frac{\sum_{\tau \neq \thetashare}L_\ell(\bm{\xi}_{\ell,i}|\tau)}{L_\ell(\bm{\xi}_{\ell,i}|\theta)(H-1)}\nonumber\\
&+\sum_{\substack{\ell=1\\\ell\neq k}}^Ka_{\ell k}\left(\frac{a_{\ell \ell }}{1-a_{\ell \ell }}B+\epsilon\right)\nonumber\\
&\stackrel{\text{(b)}}{=} \sum_{\ell=1}^Ka_{\ell k}\log\frac{\bm{\mu}_{\ell,i-1}(\theta)}{\bm{\mu}_{\ell,i-1}(\thetashare)}+\sum_{\ell=1}^Ka_{\ell k}\log\frac{L_\ell(\bm{\xi}_{\ell,i}|\theta)}{L_\ell(\bm{\xi}_{\ell,i}|\thetashare)}\nonumber\\
&+\sum_{\substack{\ell=1\\\ell\neq k}}^Ka_{\ell k}\log \frac{L_{\ell} (\bm{\xi}_{\ell,i}|\thetasharec)}{L_\ell(\bm{\xi}_{\ell,i}|\theta)}+\sum_{\substack{\ell=1\\\ell\neq k}}^Ka_{\ell k}\left(\frac{a_{\ell \ell }}{1-a_{\ell \ell }}B+\epsilon\right)\nonumber\\
&= \sum_{\ell=1}^Ka_{\ell k}\log\frac{\bm{\mu}_{\ell,i-1}(\theta)}{\bm{\mu}_{\ell,i-1}(\thetashare)}+\sum_{\ell=1}^Ka_{\ell k}\log\frac{L_\ell(\bm{\xi}_{\ell,i}|\theta)}{L_\ell(\bm{\xi}_{\ell,i}|\thetashare)}\nonumber\\&+\sum_{\ell=1}^Ka_{\ell k}\log \frac{L_{\ell} (\bm{\xi}_{\ell,i}|\thetasharec)}{L_\ell(\bm{\xi}_{\ell,i}|\theta)}-a_{kk}\log \frac{L_{k} (\bm{\xi}_{k,i}|\thetasharec)}{L_k(\bm{\xi}_{k,i}|\theta)}\nonumber\\&+\sum_{\substack{\ell=1\\\ell\neq k}}^Ka_{\ell k}\left(\frac{a_{\ell \ell }}{1-a_{\ell \ell }}B+\epsilon\right)\nonumber\\
&= \sum_{\ell=1}^Ka_{\ell k}\log\frac{\bm{\mu}_{\ell,i-1}(\theta)}{\bm{\mu}_{\ell,i-1}(\thetashare)}+\sum_{\ell=1}^Ka_{\ell k}\log\frac{L_\ell(\bm{\xi}_{\ell,i}|\thetasharec)}{L_\ell(\bm{\xi}_{\ell,i}|\thetashare)}\nonumber\\&-a_{kk}\log \frac{L_{k} (\bm{\xi}_{k,i}|\thetasharec)}{L_k(\bm{\xi}_{k,i}|\theta)}+\sum_{\substack{\ell=1\\\ell\neq k}}^Ka_{\ell k}\left(\frac{a_{\ell \ell }}{1-a_{\ell \ell }}B+\epsilon\right),\label{eq:lastrecurs}
\end{align}
where in (a) we used the bound in \eqref{eq:recursionbound} for $\tau'=\theta$, that is:
\begin{equation}
\frac{\bm{\mu}_{\ell,i-1}(\tau)}{\bm{\mu}_{\ell,i-1}(\theta)}\leq e^{\frac{a_{\ell\ell}}{1-a_{\ell\ell}} B + \epsilon}
\end{equation}
and in (b) we used the definition of $L_\ell(\thetasharec)$ seen in \eqref{eq:likelihoodthetac}. Setting $\bm{y}_{k,i}=\log\frac{\bm{\mu}_{k,i}(\theta)}{\bm{\mu}_{k,i}(\thetashare)}$ and:
\begin{align}
\bm{x}_{k,i}&=\sum_{\ell=1}^Ka_{\ell k}\log\frac{L_\ell(\bm{\xi}_{\ell,i}|\thetasharec)}{L_\ell(\bm{\xi}_{\ell,i}|\thetashare)}-a_{kk}\log \frac{L_{k} (\bm{\xi}_{k,i}|\thetasharec)}{L_k(\bm{\xi}_{k,i}|\theta)}\nonumber\\&+\sum_{\substack{\ell=1\\\ell\neq k}}^Ka_{\ell k}\left(\frac{a_{\ell \ell }}{1-a_{\ell \ell }}B+\epsilon\right),\label{eq:xlemma8}
\end{align}
we can rewrite \eqref{eq:lastrecurs} in vector form as:
\begin{equation}
\bm{y}_i\preccurlyeq A\T\bm{y}_{i-1}+\bm{x}_i,\label{eq:lastrecurs1}
\end{equation}
for all $i>i_0$. Accordingly, the recursion in \eqref{eq:lastrecurs1} satisfies the model in \eqref{eq:lemrecurs} with inequality, and with initial state $\bm{y}_{k,i_0}$. As we develop the recursion, the inequality in \eqref{eq:lastrecurs1} is preserved. Regarding the conditions on $\bm{x}_i$ for applying Lemma~\ref{lem:aux1}, the first two terms on the RHS of \eqref{eq:xlemma8} inherit the i.i.d. property of the observations $\bm{\xi}_i$ and have finite expectation.
 The third term on the RHS of \eqref{eq:xlemma8} is deterministic and bounded. Applying Property~\ref{prop:int}, we see that $\bm{x}_i$ satisfies the conditions \eqref{eq:aslim1} and \eqref{eq:aslim2} in Lemma~\ref{lem:aux1}. From Lemma~\ref{lem:aux1} and for each $\theta\neq\thetashare$, we have that
\begin{align}
&\limsup_{i\rightarrow\infty}\frac{1}{i}
\log\frac{\bm{\mu}_{k,i}(\theta)}{\bm{\mu}_{k,i}(\thetashare)}
\stackrel{\textnormal{a.s.}}{\leq}
\sum_{k=1}^K v_{k}\sum_{\ell=1}^Ka_{\ell k}\E\hspace{-2pt} \left(\log\frac{L_\ell(\thetasharec)}{L_\ell(\thetashare)}\right)\nonumber\\
&-\hspace{-2pt}\sum_{k=1}^Kv_k a_{kk}\E \hspace{-2pt}\left(\hspace{-2pt}\log \frac{L_{k} (\thetasharec)}{L_k(\theta)}\hspace{-2pt}\right)\hspace{-2pt}+\hspace{-2pt}
\sum_{k=1}^K v_{k}\sum_{\substack{\ell=1\\\ell\neq k}}^Ka_{\ell k}\hspace{-2pt}\left(\hspace{-2pt}\frac{ a_{\ell \ell}}{1-a_{\ell\ell}}B\hspace{-2pt}+\hspace{-2pt}\epsilon\hspace{-2pt}\right)\hspace{-2pt}.
\end{align}
Taking into account the arbitrariness of $\epsilon$, we end up with the following result:
\begin{align}
&\limsup_{i\rightarrow\infty}\frac{1}{i}
\log\frac{\bm{\mu}_{k,i}(\theta)}{\bm{\mu}_{k,i}(\thetashare)}
\stackrel{\textnormal{a.s.}}{\leq}
 -\sum_{k=1}^K v_{k}d_{k} (\thetasharec)+\sum_{k=1}^K v_{k}d_{k} (\thetashare)\nonumber\\
 &+ \sum_{k=1}^K v_{k}a_{kk}d_{k} (\thetasharec)-\sum_{k=1}^K v_{k}a_{kk}d_{k} (\theta)\nonumber\\&+
B\sum_{k=1}^K v_{k}\sum_{\substack{\ell=1\\\ell\neq k}}^Ka_{\ell k}\frac{ a_{\ell \ell}}{1-a_{\ell\ell}}\nonumber\\
&\stackrel{\text{(a)}}{\leq}
-\sum_{k=1}^K v_{k}d_{k} (\thetasharec)+\sum_{k=1}^K v_{k}d_{k} (\thetashare)+ \sum_{k=1}^K v_{k}a_{kk}d_{k} (\thetasharec)\nonumber\\
&+B\sum_{k=1}^K v_{k}\sum_{\substack{\ell=1\\\ell\neq k}}^Ka_{\ell k}\frac{ a_{\ell \ell}}{1-a_{\ell\ell}}\nonumber\\
&\stackrel{\text{(b)}}{=}\sum_{k=1}^K v_{k}d_{k} (\thetashare)-\sum_{k=1}^K v_{k}(1-a_{kk})d_{k} (\thetasharec)+B\sum_{k=1}^Kv_ka_{kk},\label{eq:auxequ2}
\end{align}
where in (a) we considered that $\sum_{k=1}^K v_{k}a_{kk}d_{k} (\theta)\geq0$ from the nonnegativity of the KL divergences and of terms $a_{kk}$ and the positivity of the Perron eigenvector. In (b), we considered the  left stochasticity of matrix $A$ and the definition of the Perron eigenvector. As long as the RHS of \eqref{eq:auxequ2} assumes a negative value, it implies that for all $\theta\in\Theta\setminus\{\thetashare\}$:
\begin{equation}
\bm{\mu}_{k,i}(\theta)\stackrel{\textnormal{a.s.}}{\longrightarrow}0\implies \bm{\mu}_{k,i}(\thetashare)\stackrel{\textnormal{a.s.}}{\longrightarrow}1.
\end{equation}

\section{Auxiliary Results and their Proofs}\label{ap:aux}
\begin{lemma}[\textbf{Main Vector Recursion}]\label{lem:aux1}
	Let a sequence of random vectors $\bm{y}_i$ with dimension $K \times 1$ be defined through the following recursion, for $i=1,2,\dots$
	\begin{equation}
	\bm{y}_i=A\T \bm{y}_{i-1}+\bm{x}_i.\label{eq:lemrecurs}
	\end{equation}
	where $\bm{y}_0$ is an initial (a.s. finite) random vector, and $A$ is a primitive left-stochastic (deterministic) matrix satisfying:
	\begin{equation}
	\lim_{i\rightarrow \infty}{A^i}=v\mathbbm{1}\T, \label{eq:Acomb}
	\end{equation}
	for some vector $v$ with positive entries such that $\mathbbm{1}\T v=1$. Moreover ${\bm{x}_i}$ is a sequence of random vectors (with entries $\{\bm{x}_{\ell,i}\}$) possessing the following properties, for a certain deterministic vector $\bar{x}$:
		\begin{align}
		\frac{1}{i}\sum_{j =1}^i\bm{x}_{\ell, j}&\stackrel{\textnormal{a.s.}}{\longrightarrow}\bar{x}_\ell,\label{eq:aslim1}\\
		\lim\sup_{i\rightarrow \infty}\frac{1}{i}\sum_{j =1}^i|\bm{x}_{\ell, j}|\,&\stackrel{\textnormal{a.s}}{=} \bm{M}_\ell,\label{eq:aslim2}
%		 \frac{\bm{x}_{\ell, i}}{i}& \stackrel{\textnormal{a.s.}}{\longrightarrow}0,\label{eq:aslim3}
		\end{align}
for some nonnegative (a.s. finite) random variables $\bm{M}_{\ell}$, with $\ell=1,2,\ldots, K$. Then we have that,
	\begin{equation}
		\frac {1}{ i} \, \bm{y}_{i}\stackrel{\textnormal{a.s.}}{\longrightarrow}\mathbbm{1}v\T \bar{x}.
	\end{equation}
\end{lemma}
\begin{IEEEproof}
	Iterating the recursion in \eqref{eq:lemrecurs}, we get:
\begin{equation}
\bm{y}_i=(A^i)\T \bm{y}_0 + \sum_{j=0}^{i-1} (A^{j})\T \bm{x}_{i-j}.
\label{eq:recursdevelop}
\end{equation}
Once scaled by $i$, the first term on the RHS vanishes almost surely when $i$ tends to infinity in view of the properties of $A$. We focus on the second term. It is useful to rewrite the summation in \eqref{eq:recursdevelop} as follows:
\begin{equation}
\frac{1}{i}\sum_{j=0}^{i-1} (A^j)\T \bm{x}_{i-j}=
\frac{1}{i}\sum_{j=0}^{i-1} (A^j-v\mathbbm{1}\T)\T \bm{x}_{i-j}+
\frac{1}{i}\sum_{j=0}^{i-1} \mathbbm{1}v\T \bm{x}_{i-j}.\label{eq:t123}
\end{equation}
Regarding the last term on the RHS of \eqref{eq:t123}, in view of \eqref{eq:aslim1}, we have that:
\begin{equation}
\frac{1}{i}\sum_{j=0}^{i-1} \mathbbm{1}v\T  \bm{x}_{i-j}=\mathbbm{1}v\T \frac{1}{i}\sum_{j=1}^{i} \bm{x}_j\stackrel{\text{a.s.}}{\longrightarrow}\mathbbm{1}v\T  \bar{x}.
\end{equation}
Accordingly, the claim of the lemma will be proved if we show that the first term on the RHS of \eqref{eq:t123} vanishes with probability one. From \eqref{eq:Acomb}, for some $\varepsilon>0$, there exists an index $i_0$ such that, for all $j>i_0$:
\begin{equation}
\left|[A^j]_{\ell k} - v_{\ell }\right|<\varepsilon.\label{eq:vareps}
\end{equation}
Let us therefore split the term of interest as:
\begin{align}
\frac{1}{i}\sum_{j=0}^{i-1} (A^j-v\mathbbm{1}\T)\T \bm{x}_{i-j}=\frac{1}{i}\sum_{j=0}^{i_0} (A^j-v\mathbbm{1}\T)\T \bm{x}_{i-j}\nonumber\\+\frac{1}{i}\sum_{j=i_0+1}^{i-1} (A^j-v\mathbbm{1}\T)\T \bm{x}_{i-j}.\label{eq:2ndterm}
\end{align}
Regarding the first term on the RHS of \eqref{eq:2ndterm}, we can write the absolute value of its $k-$th component as:
\begin{align}
&\frac{1}{i}
\left|
\sum_{j=0}^{i_0} \sum_{\ell=1}^K\Big([A^j]_{\ell k}-v_k\Big)\,\bm{x}_{\ell,i-j}
\right|\nonumber\\&
\leq
\frac{1}{i}\sum_{j=0}^{i_0} \sum_{\ell=1}^K\Big|[A^j]_{\ell k}-v_k\Big| |\bm{x}_{\ell,i-j}|\nonumber\\& \stackrel{\text{(a)}}{\leq}
\sum_{j=0}^{i_0} \sum_{\ell=1}^K\frac{|\bm{x}_{\ell,i-j}|}{i}\stackrel{\textnormal{a.s.}}{\longrightarrow}0,\label{eq:lasteq}
\end{align}
where the inequality in (a) follows because $A$ is left-stochastic and $v$ is the Perron eigenvector, while almost sure convergence to 0 is due to the fact that, in view of \eqref{eq:aslim1}, we have $ \frac{\bm{x}_{\ell, i}}{i} \stackrel{\textnormal{a.s.}}{\longrightarrow}0$.
 
Let us address the second term on the RHS of \eqref{eq:2ndterm}. Considering its $k$-th component, we can write its absolute value as:
\begin{align}
&\frac{1}{i}
\left|
\sum_{j=i_0+1}^{i-1} \sum_{\ell=1}^K
\left([A^j]_{\ell k}-v_{\ell}\right) \bm{x}_{\ell,i-j}
\right|
\nonumber\\&\stackrel{\text{(a)}}{\leq}
\varepsilon\sum_{\ell=1}^K\frac{1}{i}\sum_{j=i_0+1}^{i-1} |\bm{x}_{\ell,i-j}|=
\varepsilon\sum_{\ell=1}^K\frac{1}{i}\sum_{j=1}^{i-i_0-1} |\bm{x}_{\ell,j}|,\label{eq:lemma8aux}
\end{align}
where in (a) we used the bound in \eqref{eq:vareps}. From \eqref{eq:lemma8aux}, in view of \eqref{eq:aslim2}, it follows that
\begin{equation}
\lim\sup_{i\rightarrow \infty} \frac{1}{i}
\left|
\sum_{j=i_0+1}^{i-1} \sum_{\ell=1}^K
\left([A^j]_{\ell k}-v_{\ell}\right) \bm{x}_{\ell,i-j}
\right|\stackrel{\textnormal{a.s.}}{\leq}\varepsilon \sum_{\ell=1}^K\bm{M}_\ell.\label{eq:lemmaaux8}
\end{equation}
Finally, in view of \eqref{eq:lasteq} and \eqref{eq:lemmaaux8} we can write the absolute value of the $k-$th component of \eqref{eq:2ndterm} as:
\begin{equation}
\lim\sup_{i\rightarrow \infty} \frac{1}{i}
\left|
\sum_{j=0}^{i} \sum_{\ell=1}^K\Big([A^j]_{\ell k}-v_k\Big)\,\bm{x}_{\ell,i-j}
\right|\stackrel{\textnormal{a.s.}}{\leq}\varepsilon \sum_{\ell=1}^K\bm{M}_\ell.\label{eq:lemma8final}
\end{equation}
From \eqref{eq:lemma8final}, due to the arbitrariness of $\varepsilon$, the term on the LHS of \eqref{eq:2ndterm} vanishes and the proof is complete.
\end{IEEEproof}
The following property shows that conditions \eqref{eq:aslim1} and \eqref{eq:aslim2} in Lemma~\ref{lem:aux1} are satisfied for the particular case in which the random vectors $\{\bm{x}_i\}$ are i.i.d. and have finite expectation.

\begin{property}[\textbf{Convergence of i.i.d. Random Variables with Finite Expectation}]\label{prop:int}
Consider the sequence of i.i.d. integrable random variables $\{\bm{x}_{i}\}$ with $\E(\bm{x}_i)=\bar{x}$. Then, the following properties are satisfied:
	\begin{align}
	\frac{\bm{x}_{i}}{i}& \stackrel{\textnormal{a.s.}}{\longrightarrow}0,\label{eq:indexi0}\\
	\frac{1}{i}\sum_{j =1}^i\bm{x}_{j}&\stackrel{\textnormal{a.s.}}{\longrightarrow}\bar{x},\label{eq:abssl}\\
	\frac{1}{i}\sum_{j =1}^i|\bm{x}_{j}|&\stackrel{\textnormal{a.s.}}{\longrightarrow}\E(|\bm{x}_{j}|). \label{eq:absslln}
	\end{align}
	Equation \eqref{eq:indexi0} follows from integrability\footnote{For any integrable random variable $\bm{z}$, and any $\varepsilon > 0$, we have that \cite{KaiLaiChungBook}[Theorem 3.2.1]:
		\begin{equation}
		\varepsilon \sum_{i=1}^{\infty}\P(|\bm{z}|>\varepsilon i)\leq \E(|\bm{z}|)<\infty.\label{eq:bcbound}
		\end{equation}
		Since $\bm{x}_{i}$ are integrable and identically distributed, from \eqref{eq:bcbound}, we have $ \sum_{i=1}^{\infty}\P( |\bm{x}_i|>\varepsilon i)<\infty$. Therefore, condition \eqref{eq:indexi0} follows from the Borel-Cantelli lemma\cite{billingsley2008probability}.} and Eqs. \eqref{eq:abssl} and \eqref{eq:absslln} follow from the Strong Law of Large Numbers (SLLN)~\cite{billingsley2008probability}.
	\QED
\end{property}
\begin{lemma}[\textbf{Scalar Recursion}]
	\label{lem:scalarevolut}
	Let $\bm{y}_i$ be a (scalar) random variable satisfying, for $0<a<1$ and $i=1,2,\dots$:
	\begin{equation}
	\bm{y}_i=a \bm{y}_{i-1}+a\bm{x}_i,
	\label{eq:recurlemmascalar}
	\end{equation}
	where $\{\bm{x}_i\}$ are i.i.d. {\em integrable} random variables whose expectation is given by $\E(\bm{x}_i)={\sf m}_{\sf x}$, and $\bm{y}_0$ is an initial (a.s. finite) random variable.
We have that:
\begin{enumerate}
	\item 	$\bm{y}_i$ converges in distribution, as $i\rightarrow\infty$, to a random variable $\bm{y}_{\infty}$ that can be defined as:
		\begin{equation}
	\bm{y}_i\stackrel{\textnormal{d}}{\longrightarrow}\bm{y}_{\infty}\triangleq\sum_{j=1}^\infty a^j \bm{x}_j.
	\end{equation}
	\item The following conditions are satisfied:
	\begin{align}
	\frac{\bm{y}_i}{i}&\stackrel{\textnormal{a.s.}}{\longrightarrow}0,\label{eq:zerolim}\\
	\frac 1 i\sum_{j=1}^i \bm{y}_j&\stackrel{\textnormal{a.s.}}{\longrightarrow}\frac{a}{1-a}{\sf m}_{\sf x},\label{eq:slln}\\
	\limsup_{i\rightarrow\infty}\frac 1 i\sum_{j=1}^i |\bm{y}_j|& \stackrel{\textnormal{a.s.}}{\leq} \frac{a}{1-a}\E(|\bm{x}_j|)
.\label{eq:limsup}
	\end{align}
	
\end{enumerate}	

\end{lemma}
\begin{IEEEproof}
	For item $1)$, we develop the recursion in \eqref{eq:recurlemmascalar}:
	\begin{equation}
	\bm{y}_i=a^i\bm{y}_0 + \sum_{j=1}^i a^{j} \bm{x}_{i-j+1}.\label{eq:recursion}
	\end{equation}
	As $i$ goes to infinity, the first term on the RHS of \eqref{eq:recursion} vanishes almost surely.
	Regarding the second term on the RHS of \eqref{eq:recursion}, since $\bm{x}_i$ are i.i.d. across $i$, we can write the following equality in distribution for $i=1,2,\dots$:
	\begin{equation}
\sum_{j=1}^i a^{j} \bm{x}_{i-j+1}
	\stackrel{\textnormal{d}}{=}
	\sum_{j=1}^i a^j \bm{x}_j.
	\label{eq:eqind}
	\end{equation}
	The random series on the RHS of \eqref{eq:eqind} is the sum of independent random variables, with
	\begin{equation}
	\sum_{j=1}^\infty \E(|a^j \bm{x}_j|)=\E(|\bm{x}|)\sum_{j=1}^\infty a^j=\E(|\bm{x}|) \frac{a}{1-a}<\infty,
	\end{equation}
	where index $j$ was suppressed due to identical distribution across time. This condition is sufficient to conclude that the random series is almost-surely (and absolutely) convergent~\cite[Lemma 3.6$'$]{Loeve1951}. Denoting the value of the series by $\bm{y}_{\infty}$, we conclude that:
\begin{equation}
	\bm{y}_i\stackrel{\textnormal{d}}{\longrightarrow}\bm{y}_{\infty}.
\end{equation}
	For item $2)$, we will first show the result in \eqref{eq:zerolim}. To do that, consider again the recursion in \eqref{eq:recursion}:
\begin{align}
\bm{y}_i&=a^i\bm{y}_0 + \sum_{j=1}^i a^{j} \bm{x}_{i-j+1}\nonumber\\&\implies  
\frac{1}{i} \bm{y}_i=
\frac{1}{i}a^i\bm{y}_0 + 
\frac{1}{i}\sum_{j=1}^i a^{j} \bm{x}_{i-j+1}.
\label{eq:recursxscalar}
\end{align}
The first term on the RHS of \eqref{eq:recursxscalar} converges to zero almost surely as $i$ goes to infinity. Since $0<a<1$, we know that $\{a^j\}$ forms a converging sequence, which implies that for some $\varepsilon>0$, there exists an index $i_0$ such that, for all $j>i_0$:
\begin{equation}
|a^j|<\varepsilon.\label{eq:powerseq}
\end{equation}
We can therefore rewrite the second term on the RHS of \eqref{eq:recursxscalar} as
\begin{equation}
\frac{1}{i}\sum_{j=1}^i a^{j} \bm{x}_{i-j+1}=\frac{1}{i}\sum_{j=1}^{i_0} a^{j} \bm{x}_{i-j+1}+\frac{1}{i}\sum_{j=i_0+1}^i a^{j} \bm{x}_{i-j+1} \label{eq:recursxscalar2}.
\end{equation}
Let us address the first term on the RHS of \eqref{eq:recursxscalar2}, but considering its absolute value:
\begin{align}
\frac{1}{i}\left|\sum_{j=1}^{i_0} a^{j} \bm{x}_{i-j+1}\right|&\leq  \frac{1}{i}\sum_{j=1}^{i_0} \left|\bm{x}_{i-j+1}\right|
= \sum_{j=i-i_0+1}^{i} \frac{|\bm{x}_{j}|}{i}\stackrel{\textnormal{a.s.}}{\longrightarrow}0,\label{eq:lemma9van}
\end{align}
which vanishes almost surely in view of \eqref{eq:indexi0}. Now consider the second term on the RHS of \eqref{eq:recursxscalar2}. In view of \eqref{eq:powerseq}, the term can be bounded as:
\begin{align}
&\frac{1}{i}\left|\sum_{j=i_0+1}^i a^{j} \bm{x}_{i-j+1}\right|\leq \frac{1}{i}\,\varepsilon\sum_{j=i_0+1}^i \left|\bm{x}_{i-j+1}\right| = \frac{1}{i}\,\varepsilon\sum_{j=1}^{i-i_0} \left|\bm{x}_{j}\right|\label{eq:limsupeps}\\
&\implies \limsup_{i\rightarrow \infty}\frac{1}{i}\left|\sum_{j=i_0+1}^i a^{j} \bm{x}_{i-j+1}\right|\stackrel{\textnormal{a.s.}}{\leq}\varepsilon\E(\left|\bm{x}_{j}\right|),\label{eq:limsupeps1}
\end{align}
where the RHS of \eqref{eq:limsupeps} converges to the RHS of \eqref{eq:limsupeps1} in view of the SLLN (since $\bm{x}_j$ is integrable). 
Taking the absolute value of the LHS of \eqref{eq:recursxscalar2} and using \eqref{eq:lemma9van} and \eqref{eq:limsupeps1}, we can write:
\begin{equation}
\lim\sup_{i\rightarrow \infty}\frac{1}{i}\left|\sum_{j=1}^i a^{j}\bm{x}_{i-j+1}\right|\stackrel{\textnormal{a.s.}}{\leq}\varepsilon\E(\left|\bm{x}_{j}\right|).\label{eq:limsupeps2}
\end{equation}	
Due to the arbitrariness of $\varepsilon$ in \eqref{eq:limsupeps2}, we conclude that the limit superior in \eqref{eq:limsupeps1} vanishes, and therefore \eqref{eq:zerolim} holds.

Let us now show the result in \eqref{eq:slln}, but considering the original recursion in \eqref{eq:recurlemmascalar}:
\begin{align}
&\bm{y}_i=a \bm{y}_{i-1}+a\bm{x}_i\nonumber\\
&\implies \frac{1}{n}\sum_{i=1}^{n}\bm{y}_i= \frac{a}{n}\sum_{i=1}^{n}\bm{y}_{i-1}+\frac{a}{n}\sum_{i=1}^{n}\bm{x}_i.
\label{eq:sumrecur}
\end{align}
Note that the first term on the RHS of \eqref{eq:sumrecur} can be written as
\begin{equation}
\frac{1}{n}a\sum_{i=1}^{n}\bm{y}_{i-1}=\frac{a}{n}\sum_{k=0}^{n-1}\bm{y}_{k}=\frac{a}{n}\sum_{k=1}^{n}\bm{y}_{k}-a\frac{\bm{y}_n}{n}+a\frac{\bm{y}_0}{n},
\end{equation}
and therefore \eqref{eq:sumrecur} can be rewritten as
\begin{align}
(1-a)\frac{1}{n}\sum_{i=1}^{n}\bm{y}_i=-a\frac{\bm{y}_n}{n}+a\frac{\bm{y}_0}{n}+a\frac{1}{n}\sum_{i=1}^{n}\bm{x}_i\stackrel{\textnormal{a.s.}}{\longrightarrow}a{\sf m}_{\sf x},\label{eq:sumrecur2}
\end{align}
where the first term on the RHS vanishes almost surely in view of \eqref{eq:zerolim} and so does the second term, whereas the third term converges almost surely to $a{\sf m}_{\sf x}$ from the SLLN. It remains to verify condition \eqref{eq:limsup}. To this aim, it is useful to introduce the recursion:
\begin{equation}
\bm{s}_i=a \bm{s}_{i-1}+|\bm{x}_i|, \textnormal{ with initial condition } s_0=|\bm{y}_0|.
\label{eq:recursabs}
\end{equation}
From \eqref{eq:recursabs} we can write:
\begin{equation}
\bm{s}_i=a^i |\bm{y}_0| + \sum_{j=1}^i a^{j}|\bm{x}_{i-j+1}|.
\label{eq:absum}
\end{equation}
Comparing \eqref{eq:absum} against \eqref{eq:recursion}, by application of the triangle inequality we conclude that $|\bm{y}_i|\leq \bm{s}_i$.
On the other hand, $\bm{s}_i$ matches the model in \eqref{eq:recurlemmascalar} and, hence, in view of \eqref{eq:slln} we can write ($\E(|\bm{x}|)$ is the common mean of the random variables $|\bm{x}_j|$):
\begin{equation}
\lim_{i\rightarrow\infty} \frac{1}{i}\sum_{j=1}^i\bm{s}_j=\frac{a}{1-a}\E(|\bm{x}|).
\label{eq:convergv}
\end{equation}
Moreover, since $|\bm{y}_i|\leq \bm{s}_i$ we have that:
\begin{equation}
\limsup_{i\rightarrow\infty} \frac{1}{i}\sum_{j=1}^i|\bm{y}_j|\stackrel{\textnormal{a.s.}}{\leq}
\lim_{i\rightarrow\infty} \frac{1}{i}\sum_{j=1}^i\bm{s}_j\stackrel{\textnormal{a.s.}}{=}\frac{a}{1-a}\E(|\bm{x}|)<\infty,
\end{equation}
which reveals that condition \eqref{eq:limsup} holds. 
\end{IEEEproof}

\ifCLASSOPTIONcaptionsoff
  \newpage
\fi



\bibliographystyle{IEEEtran}

%\bibliography{bib/biblio.bib}

\begin{IEEEbiographynophoto}{Virginia Bordignon}
	received engineering degrees from Ecole Centrale de Lyon, France, and the Federal University of Rio Grande do Sul (UFRGS), Brazil, the M.S. degree in electrical engineering from UFRGS, and the Ph.D. degree in electrical engineering from \'Ecole Polytechnique F\'ed\'erale de Lausanne (EPFL), Switzerland. She is currently a Postdoctoral Researcher with the Adaptive Systems Laboratory, EPFL, Switzerland. Her research interests include statistical inference, distributed learning and information processing over networks.
\end{IEEEbiographynophoto}

\begin{IEEEbiographynophoto}{Vincenzo Matta} is currently a Full Professor in telecommunications at the Department of Information and Electrical Engineering and Applied Mathematics (DIEM), University of Salerno, Italy. He is the author of more than 130 articles published on international journals and proceedings of international conferences. His research interests include adaptation and learning over networks, social learning, statistical inference on graphs, and security in communication networks. He serves as an Associate Editor for the IEEE Open Journal of Signal Processing. He served as an Associate Editor for the IEEE Transactions on Aerospace and Electronic Systems, the IEEE Signal Processing Letters, and the IEEE Transactions on Signal and Information Processing over Networks, and a Senior Area Editor for the IEEE Signal Processing Letters. He is a member of the Sensor Array and Multichannel Technical Committee of the Signal Processing Society (SPS), and served as an IEEE SPS Representative to the IEEE Transactions on Signal and Information Processing over Networks.
\end{IEEEbiographynophoto}

\begin{IEEEbiographynophoto}{Ali H. Sayed} is Dean of Engineering at EPFL, Switzerland, where he also leads the Adaptive Systems Laboratory. He has served before as distinguished professor and chairman of electrical engineering at UCLA. He is a member of the US National Academy of Engineering (NAE) and The World Academy of Sciences (TWAS). He served as President of the IEEE Signal Processing Society in 2018 and 2019. His work has been recognized with several awards including more recently the 2022 IEEE Fourier Award, the 2020 IEEE Norbert Wiener Society Award, and several Best Paper Awards. He is a Fellow of IEEE, EURASIP, and AAAS.
\end{IEEEbiographynophoto}

\end{document}